\documentclass{article}
\usepackage[utf8]{inputenc}

\usepackage{xcolor,soul}
\usepackage{cite}
\usepackage{graphicx}
\usepackage{booktabs}
\usepackage{array}
\usepackage{amsfonts, amsmath, amsthm, amssymb}
\usepackage{graphicx}
\usepackage{changepage}
\usepackage{color,soul}
\usepackage{enumitem}   
\usepackage{framed,multirow}
\usepackage{latexsym}
\usepackage{url}
\usepackage{bm}
\usepackage{relsize}

\usepackage[ruled,vlined]{algorithm2e}

\newcolumntype{L}[1]{>{\raggedright\let\newline\\\arraybackslash\hspace{0pt}}m{#1}}
\newcolumntype{C}[1]{>{\centering\let\newline\\\arraybackslash\hspace{0pt}}m{#1}}
\newcolumntype{R}[1]{>{\raggedleft\let\newline\\\arraybackslash\hspace{0pt}}m{#1}}

\usepackage{pifont}

\usepackage{tikz}

\makeatletter
\newcommand{\thickhline}{%
    \noalign {\ifnum 0=`}\fi \hrule height 1pt
    \futurelet \reserved@a \@xhline
}
\newcolumntype{"}{@{\hskip\tabcolsep\vrule width 1pt\hskip\tabcolsep}}
\makeatother

\usepackage{geometry}
\usepackage{siunitx}

\title{\textbf{HAITCH: A Framework for Distortion and Motion Correction in Fetal Multi-Shell Diffusion-Weighted MRI}}

\author{Haykel Snoussi, Davood Karimi, Onur Afacan, Mustafa Utkur, Ali Gholipour\\
\\
\textit{Boston Children's Hospital, and Harvard Medical School, Boston, MA 02115 USA.}}

\begin{document}

\maketitle

\begin{abstract}
Diffusion magnetic resonance imaging (dMRI) is pivotal for probing the microstructure of the rapidly-developing fetal brain. However, fetal motion during scans and its interaction with magnetic field inhomogeneities result in artifacts and data scattering across spatial and angular domains. The effects of those artifacts are more pronounced in high-angular resolution fetal dMRI, where signal-to-noise ratio is very low. Those effects lead to biased estimates and compromise the consistency and reliability of dMRI analysis. This work presents HAITCH, the first and the only publicly available tool to correct and reconstruct multi-shell high-angular resolution fetal dMRI data. HAITCH offers several technical advances that include a blip-reversed dual-echo acquisition for dynamic distortion correction, advanced motion correction for \textit{model-free} and robust reconstruction, optimized multi-shell design for enhanced information capture and increased tolerance to motion, and outlier detection for improved reconstruction fidelity. The framework is open-source, flexible, and can be used to process any type of fetal dMRI data including single-echo or single-shell acquisitions, but is most effective when used with multi-shell multi-echo fetal dMRI data that cannot be processed with any of the existing tools. Validation experiments on real fetal dMRI scans demonstrate significant improvements and accurate correction across diverse fetal ages and motion levels. HAITCH successfully removes artifacts and reconstructs high-fidelity fetal dMRI data suitable for advanced diffusion modeling, including fiber orientation distribution function estimation. These advancements pave the way for more reliable analysis of the fetal brain microstructure and tractography under challenging imaging conditions.
\end{abstract}


\section{Introduction}
\label{sec:introduction}
Fetal diffusion magnetic resonance imaging (dMRI) provides detailed insights into the development of brain connectivity and microstructure during the prenatal period~\cite{jakab2017utero,hutter2018slice,deprez2019higher,hunt2019challenges,jakab2019developmental,khan2019fetal,jaimes2020vivo,machado2021spatiotemporal,calixto2023characterizing,calixto2023population,calixto2024advances,calixto2024detailed,kebiri2024deep}. This powerful technique relies on a mathematical model that delves into the microscopic behavior of water molecules within brain tissue. This model begins by calculating the probability \( P(\mathbf{R_{\tau}} | \mathbf{R_{0}}, \tau) \), which represents the likelihood of a water molecule moving from position \( \mathbf{R_{0}} \) to \( \mathbf{R_{\tau}} \) over time \( \tau \). 
Since determining this probability for a single molecule is impossible, we calculate the Ensemble Average Propagator (EAP), \( P(\mathbf{R}) \), which represents the averaged probability across all molecules within a voxel. 
Building upon Stejskal and Tanner's work~\cite{stejskal1965spin} on the pulsed gradient spin-echo sequence, we can express the normalized dMRI signal, \( E(\mathbf{q}) \), as the Fourier transform of the EAP. When the diffusion-weighted gradient duration \( \delta \) is sufficiently shorter than the time between two gradient pulses \( \Delta \), \( E(\mathbf{q}) \) can be expressed as:
\begin{equation}
E(\mathbf{q}) = \int_{\mathbf{R} \in \mathbb{R}^{3}} P(\mathbf{R}) \exp(-2\pi i\mathbf{q} \cdot \mathbf{R}) \, d\mathbf{R}
\end{equation}
where \( \mathbf{q} \) is a 3D-vector representing the effective gradient direction (\( \mathbf{q} = q\mathbf{u} \) where \( \mathbf{u} \) is a 3D unit vector).
A standard dMRI scan consists of acquiring a reference insensitive to diffusion (b-value \( = 0\)) and a set of diffusion-weighted images (b-value \( > 0\)) in non-collinear directions. 
By employing a suitable diffusion model, this setup allows the estimation of the local diffusion properties and the brain's structural connectivity network. The diffusion model estimation assumes that the gathered dMRI signals originate from the same physical point. However, achieving this condition in fetal scans is challenging due to several factors such as maternal breathing and unpredictable fetal motion~\cite{gholipour2014fetal,deprez2019higher}. These issues cause misalignment in dMRI sequences and disrupt the assumption of a consistent relationship between image space and anatomy. Moreover, the inherent limitations of echo planar imaging (EPI) used in dMRI acquisition, such as geometric distortion and susceptibility to spin history effects, distort the dMRI signals~\cite{hutter2018slice,afacan2019fetal,andersson2021diffusion}. The problem is magnified in the context of fetal imaging due to the complex interaction between fetal motion and the susceptibility-induced inhomogeneities of the main magnetic field (B0), combined with the small size of the fetal brain and the low signal-to-noise ratio (SNR). Inconsistencies originating from these factors are a major impediment to accurately characterizing fetal brain connectivity and hinder the overall reproducibility of \textit{in-utero} dMRI studies~\cite{jakab2017utero,xiao2024reproducibility}.

To address these challenges, several fetal-specific retrospective motion correction methods have been developed. 
The initial approach of motion correction for fetal dMRI, introduced by~\cite{jiang2009diffusion}, extended slice-to-volume reconstruction (SVR) techniques originally developed for \textit{in-utero} structural MRI. 
By assuming that local diffusion properties can be represented by a rank-2 tensor model, each slice was registered to the simulated volume. A diffusion tensor matrix reconstruction was then employed to integrate the realigned slices into a regular grid to produce the final reconstructed volume. \cite{oubel2012reconstruction} proposed a groupwise registration method that aligns diffusion-weighted images collectively before utilizing a derived image for registration with the T2-weighted (T2w) reference image. Affine transformation matrices were applied to realign the original sequences, addressing both motion and eddy-current effects. Then, a dual radial basis function-based interpolation was used to reconstruct a consistent image. \cite{fogtmann2013unified} presented a super-resolution reconstruction of the diffusion tensors, with some similarities in the registration concept to~\cite{oubel2012reconstruction} and several key distinctions from ~\cite{scherrer2012super}. Notable differences include the implementation of a unified reconstruction-alignment formulation, yielding a diffusion-sensitive slice registration model. Additionally, Fogtmann \textit{et al.} incorporated point spread function deconvolution into the 3D image reconstruction process. This allowed the creation of 3D datasets with isotropic spatial resolution from multiple scattered slices acquired in different anatomical planes. \cite{marami2017temporal} proposed motion correction and reconstruction through a dynamic model of fetal head motion; where motion parameters are estimated through a Kalman filtering approach. Additionally, they reconstructed diffusion tensors using a weighted least squares fit. This approach notably contributed to the creation of the first spatio-temporal diffusion tensor fetal atlas, as detailed in~\cite{khan2019fetal}. \cite{deprez2019higher} advanced the field by correcting motion using super-resolution reconstruction with spherical harmonics (SH). This approach is particularly effective in identifying crossing fibers in the fetal brain. Building upon their previous work~\cite{kuklisova2012reconstruction}, this technique notably includes intensity correction within the reconstruction process.

The above-mentioned techniques have predominantly employed data representations grounded in diffusion tensor imaging and single-shell SH. However, these approaches have limitations in capturing the complexity of the developing brain.
Tensor model struggles to represent crossing fibers which is a prevalent feature in fetal brain white matter. A \textit{model-free} signal representation is crucial for such motion correction methodologies to guarantee a broad range of imaging analyses. Similarly, \cite{tournier2020data} highlight the insufficiency of single-shell SH in capturing the full spectrum of diffusion information needed for neonatal imaging, which is highly similar to fetal imaging. Consequently, these methodologies either restrict the type of diffusion information extracted or reduce the information content and impose constraints on the variety of input data that can be processed. This significantly hinders the scope of subsequent analytical endeavors aimed at comprehensively characterizing the developing fetal brain microstructure. Moreover, the effect of local geometric distortions and their interaction with fetal and maternal motion has been either ignored or not adequately addressed in those studies. While conventional methods for EPI distortion correction rely on static field mapping~\cite{jezzard1995correction}, which is only appropriate for motion-free data, methods that rely on dMRI volumes with reversed phase encoding are not effective for the continuous and large motion that typically affects fetal dMRI~\cite{andersson2003correct,andersson2017towards}. This warrants further investigation of dMRI acquisition and processing methods to address motion and geometric distortions.

In this work, we introduce the High Angular resolution diffusion Imaging reconsTruction and Correction approacH (HAITCH), a novel framework that tackles these identified challenges by integrating optimized acquisition and reconstruction strategies to mitigate the combined effects of fetal motion, geometric distortion, and their interaction in fetal dMRI. To this end, our objective was to develop a robust methodology and a toolbox for acquiring and reconstructing fetal dMRI data that ensures high fidelity and suitability for detailed \textit{in-vivo} and \textit{in-utero} analysis of the fetal brain microstructure development. 
\newpage
To achieve this, HAITCH focuses on three core technical contributions, represented by the bold-bordered boxes in Fig.~\ref{fig:flowchart}: 
\textit{i)} Optimized multi-shell sampling scheme for improved fetal dMRI, \textit{ii)} Dual-Echo sequence enables dynamic distortion correction, \textit{iii)} 
Advanced motion correction and model-free reconstruction techniques. We validate the accuracy of our framework using real fetal dMRI scans acquired at Boston Children's Hospital. The implementation of the HAITCH framework is publicly available as the first module of the Fetal and Neonatal Development Imaging (FEDI) toolbox: \url{https://fedi.readthedocs.io}.
\section{Material and methods}
\label{sec:methods}
\begin{figure}[ht]
\centering
\includegraphics[width=0.49\textwidth]{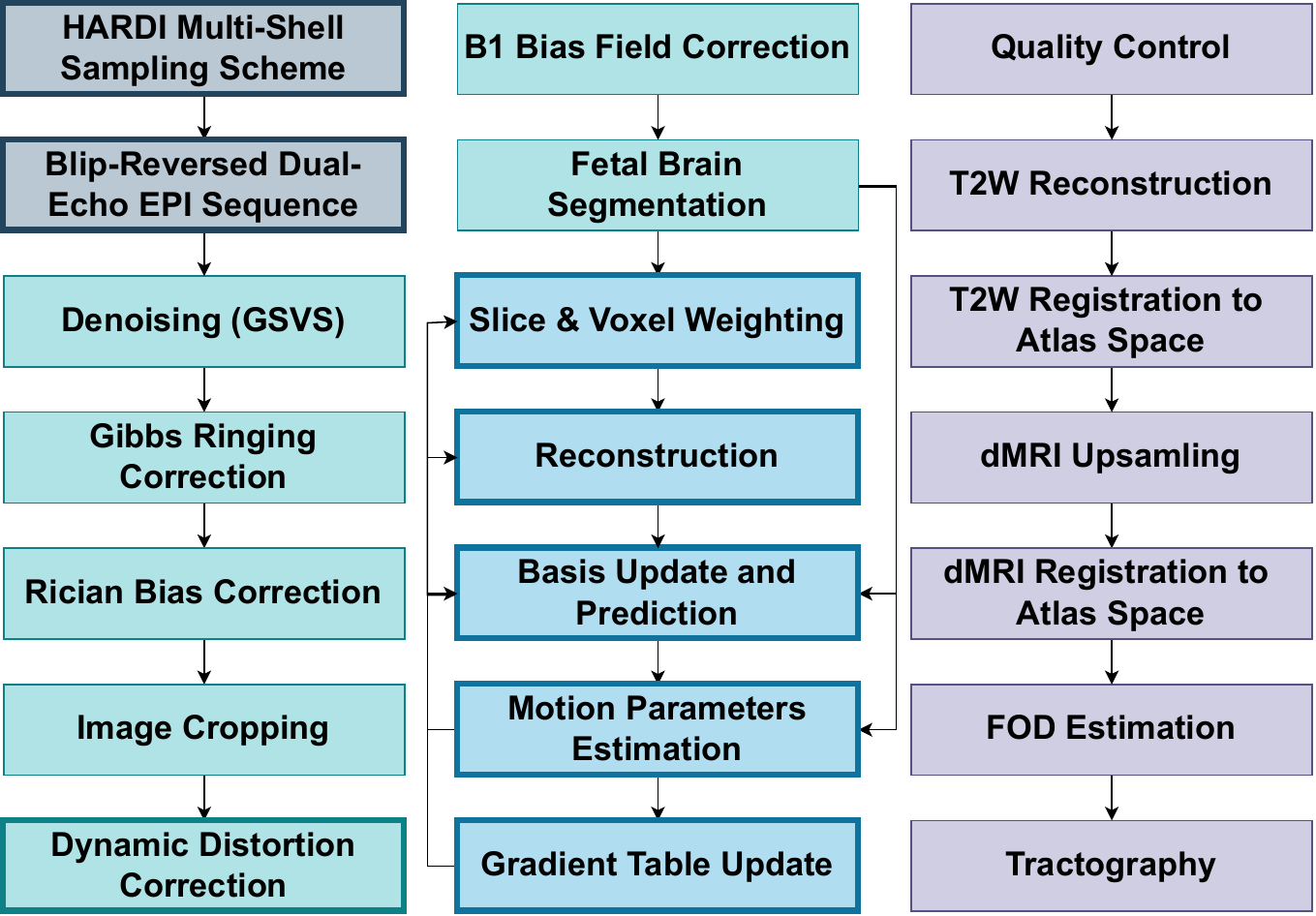}
\caption{HAITCH Framework Processing Stages. This flowchart illustrates the key steps involved in the HAITCH Framework. Boxes with bold borders refer to our main contributions. Grey boxes outline the acquisition stages, including the sampling scheme and the dual-echo sequence. Green boxes detail the pre-processing steps, distortion correction, B1 bias field correction, and fetal brain segmentation. Blue boxes refer to the iterative approach of motion correction and image reconstruction that includes slice \& voxel weighting, \textit{model-free} reconstruction, signal basis update, signal prediction, registration, and updating the gradient table. Purple boxes showcase the post-processing steps, encompassing T2-weighted image reconstruction, spatial normalization for atlas registration, and streamlined tractography. Refer to Section~\ref{sec:methods} for details on the sampling scheme, dual-echo sequence, motion correction, and dynamic distortion correction. Section~\ref{sec:implementation} discusses pre-processing, post-processing, and implementation specifics. Boxes with bold borders refer to our main contributions.}
\label{fig:flowchart}
\end{figure}

HAITCH employs a multi-stage approach to address the challenges associated with fetal dMRI acquisition. Initially, a specialized multi-shell high angular resolution diffusion imaging (HARDI) sampling scheme is designed to increase the overall dataset's tolerance to motion. Our recommended acquisition involves a modified dual-echo EPI sequence that allows dynamic correction of time-varying geometric distortions. Following acquisition, the data undergoes preprocessing including denoising, Gibbs ringing correction, and Rician bias correction. Geometric distortions are then addressed through non-static field map estimation. Subsequently, B1 bias field correction and fetal brain segmentation are performed. Motion correction is achieved through an iterative refinement process. A flowchart summarizing the various processing stages of HAITCH is shown in Fig.~\ref{fig:flowchart}. This approach progressively improves the data by repeatedly updating slice weights, transform coefficients, and motion parameters, which will be described in detail in the sections that follow. Through these iterative updates, the self-consistency of the data is enhanced, leading to progressively improved motion-corrected images.
In this section, we first explain the multi-shell sampling scheme, then we describe the signal representation that makes the foundation of our modeling, sampling, image reconstruction, outlier detection, slice weighting, basis updating, and motion parameters estimation. At the end, we explain the dynamic distortion correction component based on our recommended multi-echo EPI sequence. The details of the experiments conducted, pre-processing, implementation aspects of our methods, and post-processing are explained in Section~\ref{sec:implementation}. Section~\ref{sec:results} will then present the results obtained.

\subsection{A Multi-Shell HARDI Sampling Scheme for the Fetal Brain}
\label{subsection:hardi_scheme}
dMRI with multi-shell HARDI provides rich information about the tissue microstructure by varying b-values and directional sampling. Utilizing an extension of the electrostatic repulsion cost function~\cite{dubois2006optimized,caruyer2013design}, we have customized a sequence with HARDI, interlaced multi-shell, and an incremental sampling scheme for uniform angular coverage. This sampling scheme ensures a balance between the global angular distribution of all diffusion gradient directions and the angular distribution on each shell. The angular density of coverage for each shell increases uniformly as fetal dMRI acquisition proceeds. This optimization of the angular coverage and the temporal ordering aims to mitigate the effects of both sudden and slow fetal motion during acquisition. Acquired slices and volumes with close gradient directions are separated in acquisition time, thereby increasing the motion tolerance of the entire dataset. With this acquisition scheme, any part of the acquired data will contain uniformly distributed orientations, which is not possible with conventional schemes. 

Furthermore, to maximize the information content of the data within our multi-shell HARDI acquisition scheme, we determined the per-shell sampling density and the associated b-values based on the detectable number of SH, as detailed below. Previous studies have shown that the angular frequency in neonates is significantly lower compared to adults~\cite{tournier2020data,tournier2013determination}.~\cite{tournier2020data} demonstrated in a neonatal study that SH terms beyond order 8 are indistinguishable from noise for typical SNR values. SH terms of order 6 were small, suggesting a minimum requirement of 28 directions at high b-values. At lower b-values, only terms of orders 2 or 4 could be detected, implying minimum requirements of 6 and 15 directions, respectively. In practice, acquiring more than the minimum diffusion-weighted directions is highly recommended to account for potential imperfections in the diffusion-encoding gradient directions and ensure sufficient effective SNR for utilizing SH terms with order 6. Additionally,~\cite{tournier2020data} determined that [0, 400 s/mm\(^2\), 1800 s/mm\(^2\) ] and [0, 400 s/mm\(^2\), 1000 s/mm\(^2\), 2600 s/mm\(^2\)] b-value combinations as the optimal for 2-shells and 3-shells sampling schemes in their neonatal imaging study. Their corresponding number of directions expressed as a percentage of the total number of volumes acquired are as follows: [12\%, 29\%, 59\%] and [6\%, 19\%, 28\%, 47\%], respectively. 

Considering the potential differences and similarities between fetal and neonatal imaging, the dual-echo nature of our sequence, and achievable b-values due to inherent low SNR in fetal dMRI~\cite{karimi2021deep}, we propose a 2-shell HARDI scheme summarized in Table~\ref{tab:acquisition_parameters}. This scheme incorporates 11 interleaved \(b=0\) images, and diffusion-weighted images at \(b=400\) s/mm\(^2\) and \(b=900\) s/mm\(^2\) with 28 and 56 directions per shell, respectively. 
\begin{table}[ht]
\centering
\begin{tabular}{|c|c|c|c|}
\hline
b-value (s/mm\textsuperscript{2}) & 0 & 400 & 900 \\ \hline
Number of Directions & 11 & 28 & 56 \\ \hline
Percentage of Total Volumes & 12\% & 29\% & 59\% \\ \hline
\end{tabular}
\vspace{-0.1 cm}
\caption{Summary of the scheme: b-values and directions per shell}
\label{tab:acquisition_parameters}
\end{table}

\subsection{Signal Representation}
\subsubsection{Diffusion Signal Modeling}
To effectively capture the orientation-dependent nature of dMRI contrast, we employ the 3D Simple Harmonic Oscillator Reconstruction and Estimation (SHORE) technique~\cite{ozarslan2013mean,merlet2013continuous}. SHORE serves as a basis function representation that accurately models both the radial and angular properties of the dMRI signal by fitting a linear combination of orthogonal basis functions. SHORE allows us to represent the diffusion signal \( E(\mathit{q} \mathbf{u}) \) as a truncated linear combination of orthonormal basis functions $\Phi_{nlm}(\mathit{q}, \mathbf{u})$ :
\begin{equation}
E(\mathit{q} \mathbf{u}) = \sum_{l=0, \text{even}}^{N_{\text{max}}} \sum_{n=l}^{(N_{\text{max}}+l)/2} \sum_{m=-l}^{l} \mathbf{c}_{nlm} \Phi_{nlm}(\mathit{q}, \mathbf{u})
\label{eq:shorebasis}
\end{equation}
where the $\mathbf{c}_{n\ell m} = \langle E, \Phi_{n\ell m} \rangle$ are the transform coefficients that capture the contribution of each basis function, $\; \Phi_{nlm}(\mathbf{q}, \mathbf{u}) = X_{nl}(q, \zeta) Y_{l}^{m}(\mathbf{u})$ can be decomposed into (i) $X_{nl}(q, \zeta)$, the radial basis, is a generalized Laguerre polynomial modulated by exponential decay and scaling factor $\zeta$, capturing the radial behavior and ensuring orthonormality. $n$ represents the radial order. (ii) $Y^{m}_{\ell}(\mathbf{u})$, the angular basis, is a SH function of order $\ell$ and degree $m$, modeling the angular properties, $N_{\text{max}}$ is the maximal order of the functions in the truncated series. 
The sparsity inherent in the SHORE model, characterized by the concentration of significant information within a few coefficients, is crucial for effective diffusion signal modeling~\cite{merlet2013continuous}. Furthermore, the ability of SHORE to accurately detect multiple diffusion directions has been demonstrated in previous studies~\cite{fick2015using}. This representation translates the fetal dMRI scan into a compact 4D representation of SHORE coefficients associated with each voxel in the image grid. With this diffusion signal representation and motion parameters, we can predict the expected dMRI contrast using a forward model. This prediction allows us to compare it with the motion-corrupted acquired data. Subsequently, we formulate an inverse problem that optimizes the similarity between the predicted and acquired data while simultaneously estimating the fetal motion traces.


\subsubsection{Forward Modeling}
Fetal motion during dMRI acquisition disrupts the signal, introduces artifacts, and degrades the acquired data consistency. By assuming rigid motion, we can model this disruption using a forward model. This model allows us to predict the expected dMRI signal in the presence of motion. This involves transformations in the $q$-space domain to account for changes in spatial and angular coordinates due to motion. It essentially simulates how diffusion would appear under the influence of motion, providing a mathematical relationship between the acquired dMRI data, the underlying true signal, and the motion parameters. For a given set of fetal motion parameters $\boldsymbol{\mu}$, the predicted signal $\hat{\boldsymbol{E}}$ could be expressed as:
\begin{equation}
\hat{\boldsymbol{E}}(\boldsymbol{\mu}, \boldsymbol{q}) = \boldsymbol{C}(\boldsymbol{q}) ~ \boldsymbol{M}(\boldsymbol{\mu}) ~ \boldsymbol{D}(\boldsymbol{\mu}) ~ \boldsymbol{A} ~~ \mathbf{c}
\end{equation}
where $\boldsymbol{A}$ represents noise, $\boldsymbol{D}$ accounts for EPI distortion, $\boldsymbol{M}(\boldsymbol{\mu})$ represents the effects of rigid motion parameters on the signal, and $\boldsymbol{C}(\boldsymbol{q})$ encodes the SHORE basis for $q$-space. This equation essentially enables generating predictions for the signal in a scattered collection, each with their own motion state and dMRI encoding.

\subsection{Inverse Problem and Reconstruction}
Given the scattered data acquired during the dMRI scan, we aim to find the optimal reconstruction coefficients \( \boldsymbol{c}^* \) and rigid motion parameters \( \boldsymbol{\mu}^* \) that maximize the similarity between the acquired data and their prediction. We achieve this by solving an inverse problem. This involves finding the combination of reconstruction coefficients and motion parameters that best explain the acquired data. Mathematically, this is formulated as a large, sparse least-squares problem with an $\ell2$ regularization:
\begin{equation}
\mathbf{c}^{(k)} = \underset{\mathbf{c \in \mathbb{R}^{n_{c}} }}{\arg\min} \left\| W \left( \hat{\boldsymbol{E}}^{(k)}(\boldsymbol{\mu}^{(k)}, \boldsymbol{q}) - \mathbf{E} \right) \right\|_{\ell2} + \lambda_l \left\| Lc \right\|_{\ell2} + \lambda_n \left\| Nc \right\|_{\ell2}
\label{eq:argmin}
\end{equation}
where \( \hat{\boldsymbol{E}}^{(k)} \) is the predicted signal of all scattered data in the current iteration \( k \) using the coefficients \( \mathbf{c}^{(k)} \), \( W = \text{diag}\left( \cdots, \sqrt{w_s}, \cdots \right) \), \( w_s \) is a slice-wise or voxel-wise weight for outlier rejection. The weighting matrix \( W\) assigns higher importance to reliable slices or voxels and reduces the influence of outliers. \( N \) and \( L \in \mathbb{R}^{n_{q} \times n_{c}} \) are two diagonal matrices with \(\text{diag}(N) = n(n + 1)\) and \(\text{diag}(L) = l(l + 1)\), respectively. These matrices penalize the high frequencies of the radial and angular parts, respectively. The constants \( \lambda_n \) and \( \lambda_l \) are weights of the penalty terms. 

Equation~\ref{eq:argmin} minimizes a cost function that measures a weighted difference between the predicted signal \( \hat{\boldsymbol{E}}^{(k)} \) and the acquired signal \( \boldsymbol{E} \). We formulate the solution of this least squares problem as an iterative process that leverages a data-driven signal representation. We focus on optimizing the target image for maximum similarity to the acquired scattered data, while simultaneously identifying the unknown fetal motion traces that correspond to the fetal brain position. We utilize surrounding directional information to compensate for potential data loss and reconstruct scattered data across spatial and angular domains. The optimization process consists of repeatedly alternating between slice weighting, a reconstruction step optimizing the SHORE coefficients $\mathbf{c}$ given current motion parameters $\boldsymbol{\mu}$, and a registration step updating $\boldsymbol{\mu}$ for the current $\mathbf{c}$ throughout each epoch of our framework.

\subsection{Outlier Detection and Weighting}
\label{subsection:outliers}
Fetal dMRI is inherently susceptible to outliers and imaging irregularities, which can negatively impact image quality and the subsequent processing steps. Outliers are signal intensities that significantly deviate from the expected signal based on other measurements. Fetal motion and maternal breathing during diffusion encoding can induce severe signal dropouts. Moreover, rapid slice acquisition times combined with fetal motion can lead to spin-history artifacts, manifesting as signal hyper-intensities due to overlapping slice excitation. Hence, detecting and appropriately weighting outliers are essential to mitigate the impact of unreliable slices on the reconstruction. To examine the likelihood of a slice \( S_{i} \) or voxel \( V_{i} \) with gradient \(g\) being an outlier, we employ three complementary methods.

\subsubsection{Modified Z-Score}
We employ a modified \(z\)-score and median absolute deviation (MAD) to detect outliers within each b-value shell. MAD focuses on the median, which is less swayed by outliers, and is a better choice than the standard deviation for outlier detection in datasets that might have outliers. Given a fetal dMRI dataset, let \( \boldsymbol{E}[\hat{S}_{i,b,g}] \) represent the diffusion signal intensity average for the \(i\)-th slice with b-value \(b\) and gradient direction \(g\), and let \( \boldsymbol{E}[\widetilde{S_{i,b}}] \) present the median of those average across the same b-value \(b\). We compute the modified \(z\)-score for slice \(i\) with gradient direction \(g\) as follows:
\begin{equation}
Z_{i,g} = \frac{\mid\boldsymbol{E}[\hat{S}_{i,b,g}] - \boldsymbol{E}[\widetilde{S_{i,b}}]\mid}{1.4826 \cdot \text{MAD}(\boldsymbol{E}[\hat{S}_{i,b,g}] - \boldsymbol{E}[\widetilde{S_{i,b}}])}
\end{equation}
Subsequently, lower \(\eta_l\) and upper \(\eta_u\) thresholds for the modified \(z\)-scores are applied to assign a weight \(w_s\) to each slice as following:
\begin{equation}
w_s = 1 - \frac{Z_{i,g} - \eta_l}{\eta_u - \eta_l}
\end{equation}
These weights are incorporated within the initialization epoch to influence the initial predicted signal \( \hat{\boldsymbol{E}}^{(1)} \).

\subsubsection{Bayesian Gaussian Mixture Modeling}
The second method utilizes a probabilistic criterion to identify outliers based on the intensity difference between the acquired dMRI data and their corresponding predictions~\cite{christiaens2021scattered}. This approach identifies and excludes outlier slices exhibiting large residuals compared to other slices with similar b-values. We first calculate the root-mean-squared error (RMSE), denoted by \(\epsilon\), between the acquired data and the provided signal predictions. The RMSE values for inlier and outlier slices are expected to follow different probability distributions. Within each shell, a two-component Bayesian Gaussian Mixture Model (GMM) is employed to model the log-transformed RMSE values, thereby delineating inlier and outlier slices. The GMM assigns a probability, considered as \( w_s \), to each slice belonging to the inlier component.
\subsubsection{Voxel-Wise Weighting}
The previous two methods focused on slice-level weighting to address signal dropout caused by bulk motion during acquisition. However, local variations within slices can arise due to factors like spin history and physiological motion. To down-weight these localized outliers, we employ a voxel-wise weighting applied in the final processing epoch. This method builds upon the voxel-wise standardized residuals with the modified z-score which are then converted to weights for each voxel as follows:
\begin{equation}
w_v = \frac{1}{ (z_v^2 + 1)^2 } ~ \text{where} ~ z_v = \frac{E_v - \hat{E}_v}{1.4826 \cdot \text{MAD}(E_v - \hat{E}_v)}
\end{equation}
Here, \(E_v\) and \(\hat{E}_v\) represent the acquired and predicted signal intensity for a particular voxel, respectively.

\subsection{Registration}
In this step, we seek to compute and update the motion parameters by optimizing the spatial alignment between the predicted signal and the acquired scattered data. We achieve this by performing rigid image registration directly between the predicted and the acquired dMRI contrasts at each time point. This approach offers several advantages over conventional registration methods that often use the b0 image (unweighted image) as the reference for aligning the acquired dMRI data. Here, the predicted dMRI contrasts at each time point serve as the reference for registration with the acquired dMRI data. This way of registration is more accurate than the conventional registration to b0 images, as the fixed (predicted) and moving (acquired) images have the same diffusion sensitization.

\subsection{Dynamic Distortion Correction}
Fetal dMRI often relies on EPI for its efficient data acquisition. 
However, EPI inherits a limitation: the presence of local geometric distortions. These distortions are a result of the inhomogeneities of the main magnetic field B0, which are caused by the differences in magnetic susceptibility of tissue and air/gas in air/gas-tissue interfaces. A common approach to correct geometric distortions in adult dMRI involves acquiring scans with a single phase encoding direction and adding a \textit{b}=0 image with reversed phase encoding. A field map, assumed to be static, is then generated from the data and used for correcting distortions~\cite{andersson2003correct}. While effective for subjects with minimal motion, this method is inadequate for fetal imaging because fetal motion affects field inhomogeneities. The nearly-continuous fetal movements as well as gas movement in the maternal intestine, and maternal respiratory motion, can cause significant changes in B0 inhomogeneity. These changes would render static field map correction techniques inaccurate in fetal dMRI~\cite{kuklisova2017distortion,hutter2018slice,afacan2019fetal,afacan2020simultaneous}. 

\begin{figure}[ht]
\centering
\includegraphics[width=0.49\textwidth]{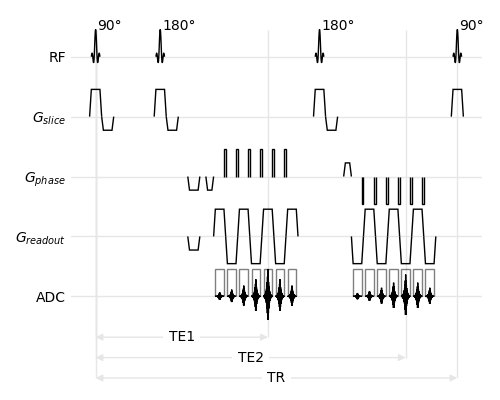}
\vspace{0.5cm}\\
\includegraphics[width=0.225\textwidth, trim={14cm 4cm 14cm 4cm}, clip]{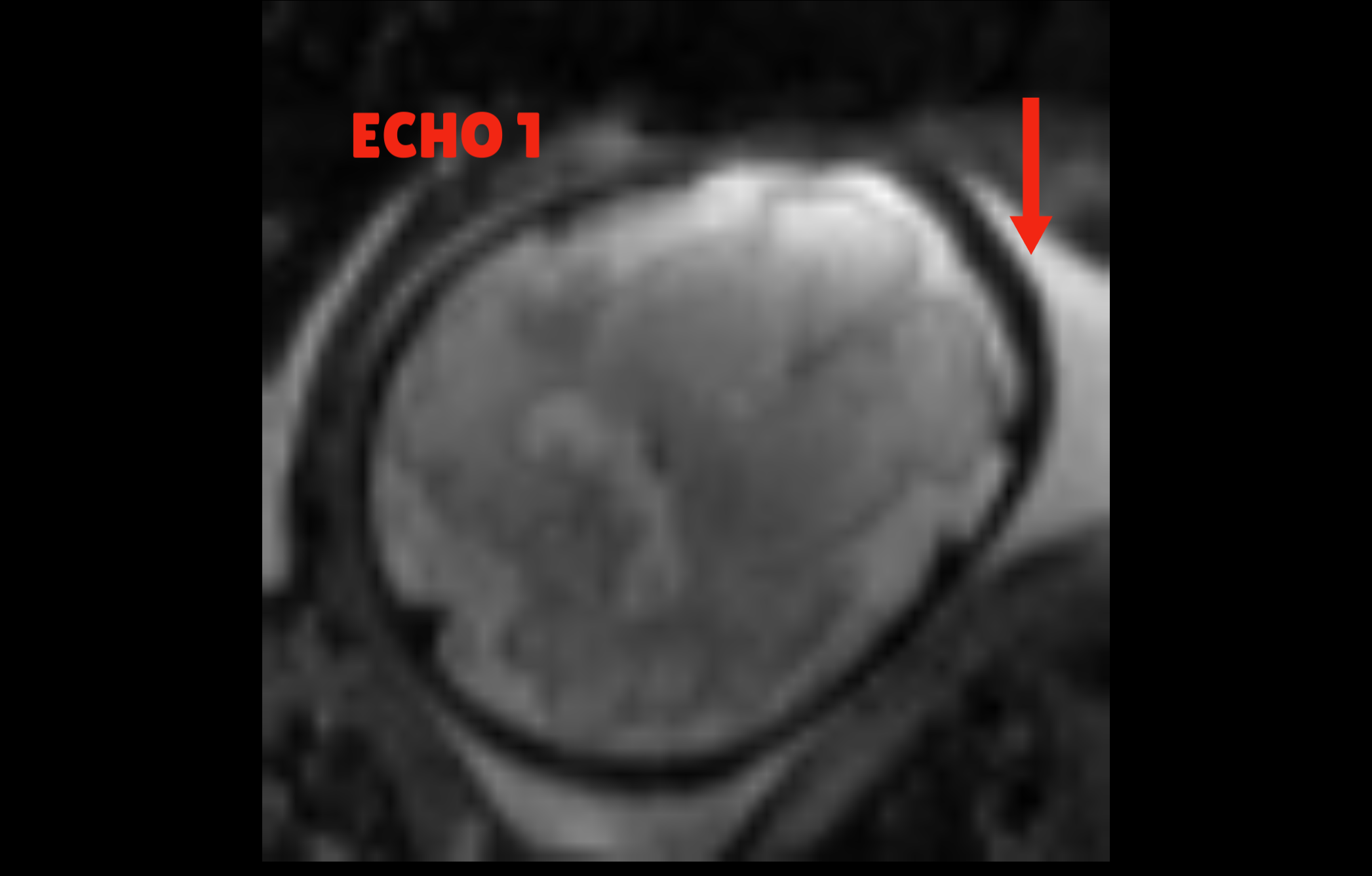}
\includegraphics[width=0.225\textwidth, trim={14cm 4cm 14cm 4cm}, clip]{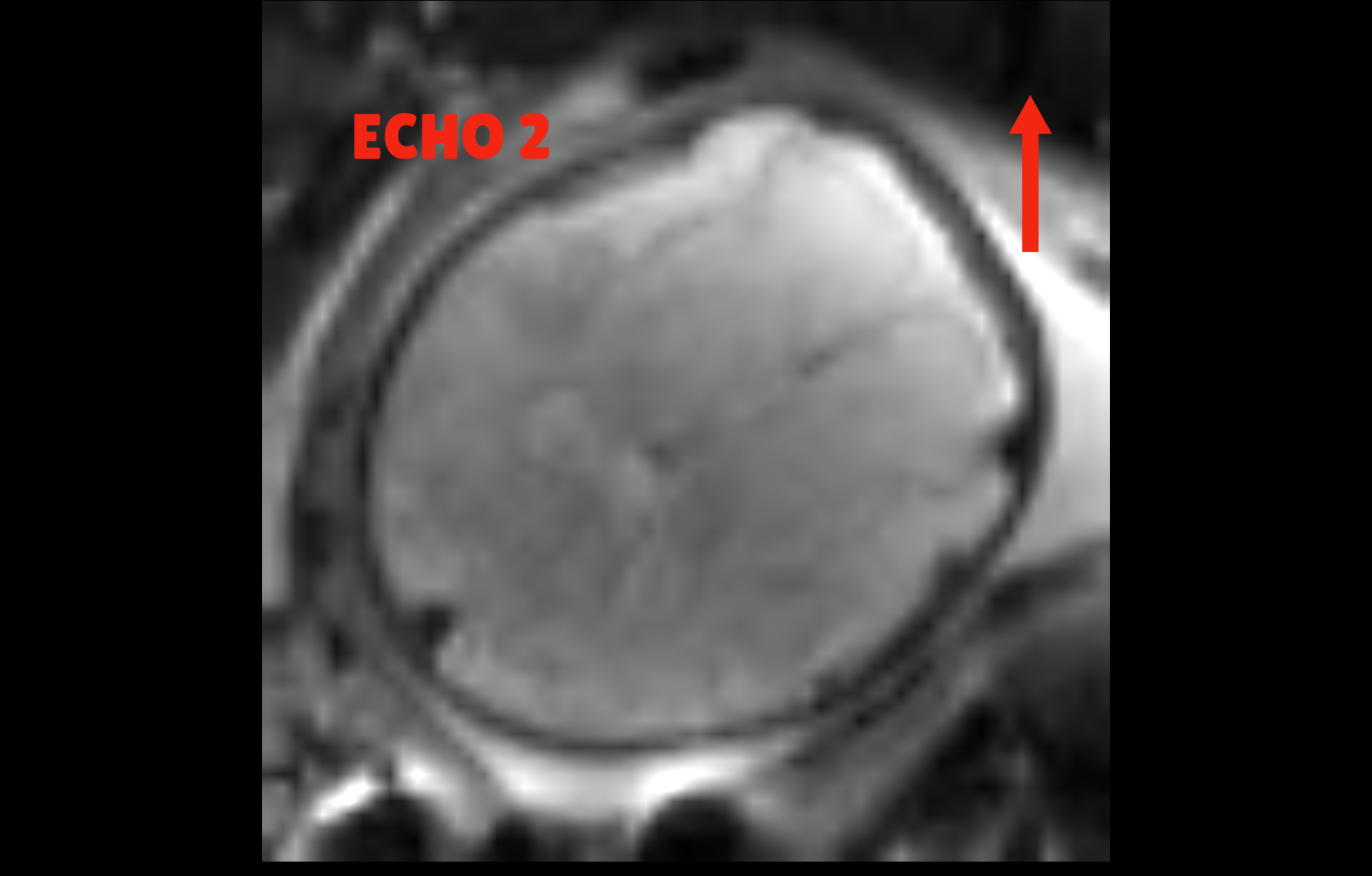}
\caption{(Top): Schematic diagram of the modified spin-echo EPI sequence.  This sequence acquires two echoes (TE1 and TE2) with reversed phase encoding directions.  Despite the extended readout, the dual-echo sequence maintains signal-to-noise ratio (SNR) efficiency comparable to a conventional EPI scan. Temporal proximity between the echoes (< 50 ms) ensures minimal motion artifacts between acquisitions. (Bottom): Illustration of the opposing susceptibility-induced distortions in the two echoes for an example.  The first echo (TE1) exhibits stretching (arrow), while the second echo (TE2) shows signal pile-up in the corresponding location.  This key feature allows for dynamic field map estimation and correction of geometric distortions even in the presence of fetal motion. Details of the implementation are discussed in Section~\ref{sec:implementation}.}
\label{fig:dualecho}
\end{figure}

To address this challenge, we propose using a modified spin-echo EPI sequence incorporating a second readout (echo). This second echo generates data that have the same resolution, field of view, and bandwidth of the first echo data but with the order of \textit{k}-space traversal reversed in the phase encoding direction. Fig.~\ref{fig:dualecho} illustrates the pulse sequence diagram. A 180$^{\circ}$ refocusing pulse and spoiler gradients were employed to generate the second spin-echo. The frequency encoding direction is reversed between the two echoes to ensure consistent temporal spacing between \textit{k}-space points.  Notably, the dual-echo sequence maintains the SNR efficiency comparable to the conventional EPI scan despite the extended readout train. The temporal proximity between the two echoes ($<$ 50 ms) ensures closely matched and nominally identical motion states for the two acquired slices. Importantly, the first and second echo-time data exhibit geometric distortions in opposite directions at each acquisition time point. Fig.~\ref{fig:dualecho} further illustrates this concept, where susceptibility-induced stretching in the first echo corresponds to signal pile-up in the second echo. This enables the application of the blip-reversed approach in dynamic environments such as fetal and neonatal imaging. This key advantage allows us to estimate a dynamic field map, thereby correcting for geometric distortions in the presence of motion. Implementation details are discussed in Section~\ref{sec:implementation}. 

\section{Implementation and Experiments}
\label{sec:implementation}
\subsection{Fetal Datasets}
To examine the efficacy of our proposed framework, we conducted 36 dMRI studies on 14 pregnant volunteers, scanned at a gestational age between 24 and 36 weeks. The studies were approved by the institutional review board, and written informed consent was obtained from all participants. Each scan session involved repeated single-shot fast spin echo T2w acquisitions, which were used to reconstruct a reference anatomical volume with an isotropic resolution of 0.8 mm\(^3\) using SVR~\cite{uus2020deformable,ebner2020automated}. 

Each scan session includes also a dMRI scan acquired in three different ways for testing, \textit{(i) Dual-Echo data}: Twenty-four dMRI scans with two echoes were acquired from 13 subjects. Echo times were 50 ms for TE1 and between 85 ms to 90 ms for TE2. These scans included various single and double-shell b-values (250, 400, 500, 600, 750, 800, 900, and 1000 s/mm\(^2\)) across 6-direction sets (13, 15, 20, 22, 26, 27, 44, 80 and 93 directions). \textit{(ii) Multi-Shell HARDI data}: Nine multi-shell HARDI dMRI scans were acquired from five volunteers. These scans were acquired following our optimized scheme described in subsection~\ref{subsection:hardi_scheme}. \textit{(iii) Dual-Echo Multi-Shell HARDI data}: An additional three scans from two volunteers possess also the sampling scheme described in subsection~\ref{subsection:hardi_scheme}. 

All scans are acquired using a 3T Siemens Prisma scanner with a 30-element body coil. Scans were performed with parallel imaging with GRAPPA factor 2,  minimum repetition times (TR) between 2.5 and 7 seconds, isotropic resolution settings of 2.0, 2.3, or 2.5 mm\(^3\), a field of view of 256 or 300 mm depending on the age and size of the fetus and the maternal body, and between 30-40 slices to cover the fetal brain. 



\subsection{Pre-processing}
dMRI data is susceptible to various artifacts arising from the limitations of the MRI hardware and techniques. To address these artifacts and ensure reliable estimation of diffusion parameters, we have established the following steps, as represented by green boxes in Fig.~\ref{fig:flowchart}:
\subsubsection{Denoising}
The data is initially denoised using the generalized singular value shrinkage (GSVS) method for noise estimation and reduction, as detailed in~\cite{cordero2019complex}. This method extends the random matrix theory-based approach of the Marchenko-Pastur Principal Component Analysis (MPPCA), introduced in~\cite{veraart2016diffusion,veraart2016denoising}. GSVS moves beyond the limitations of MPPCA, including the homoscedastic and uncorrelated noise assumptions and the hard separability of signal and noise. Instead, it utilizes the findings of~\cite{benaych2012singular} for optimal singular value shrinkage under a generalized Marchenko-Pastur law. This makes GSVS particularly suitable for motion-degraded and low SNR data, including fetal dMRI data with high motion artifacts. GSVS effectively removes only thermal noise without compromising anatomical features in approximately 2 minutes. 
\subsubsection{Gibbs Ringing Artifact Correction}
High-contrast boundaries, such as the edges between cerebrospinal fluid and gray matter or white matter in the fetal brain, may produce image artifacts known as Gibbs ringing. These artifacts arise from inadequate sampling of high frequencies, and may significantly impact the diffusion signal. To remove these artifacts, we employ a method that modulates the truncation in k-space as a convolution with a sinc-function in image space and interpolates the image through local subvoxel shifts~\cite{kellner2016gibbs}.

\subsubsection{Rician Bias Correction}
 dMRI images exhibit Rician or non-central \(\chi\) noise, especially in low-SNR settings like multi-shell HARDI fetal dMRI. Unlike Gaussian noise, Rician noise is non-additive and intensity-dependent, leading to Rician bias and reduced image contrast. To address this, we adapt the methodology introduced in~\cite{ades2018evaluation} with GSVS replacing MPPCA for Rician bias correction. GSVS denoising provides an unbiased estimate of the noise standard deviation $\sigma$ at each voxel based on data acquired with low b-values~\cite{veraart2016diffusion,cordero2019complex}. We then estimate the true signal intensity \(E_c\) through the following equation:
\begin{equation}
E_c^2 = \langle M \rangle^2 - \left( \xi(\theta) -2 \right) \sigma^2
\end{equation}
Here, $\langle M \rangle$ represents the expected value of the measured magnitude signal intensity, and $\xi(\theta)$ is a correction factor defined by $\theta = \frac{\eta}{\sigma}$, which is equivalent to the SNR~\cite{koay2006analytically}. For $\text{SNR}$ exceeding 2 dB, $\xi(\theta)$ approaches 1, allowing for accurate signal intensity estimation.

\subsection{Dynamic Distortion Correction}
Both the first echo time dMRI(TE1) scan and the second echo time dMRI(TE2) scan were used to correct the distortion following the reversed gradient polarity technique. To ascertain the sequence's efficiency and effectiveness, distortion correction was approached via two dynamic and one static correction methods: (i) Slice-wise correction: field map is estimated for each slice using the reversed echo time slice at every time point for low and high b-values images using the method introduced in Voss \textit{et al.}~\cite{voss2006fiber}. (ii) Volume-wise correction: field map is estimated for each volume using the reversed echo time volume at every time point for both low and high b-values images using FSL TOPUP~\cite{andersson2003correct}. (iii) Static map correction: the classic way of distortion correction with FSL TOPUP, employing the first b=0 image of TE1 and the last b=0 image of TE2 to calculate the static field map, which is then applied to the entire TE1 dataset. These three correction methods were applied reciprocally between dMRI(TE1) and dMRI(TE2), with outcomes evaluated via the Structural Similarity Index Measure (SSIM) and Peak signal-to-noise ratio (PSNR) metrics. The method with the highest SSIM and PSNR was chosen for subsequent processing.
\subsection{B1 Bias Field Correction}
Fetal dMRI data acquired with high-density coils suffers from signal variations due to B1 field inhomogeneity. This is particularly problematic in multi-shell acquisitions where complex anatomy and varying diffusion weightings magnify these effects. To address this, we employ the N4ITK algorithm~\cite{tustison2010n4itk} to estimate the corrected signal \(E_c(\mathbf{q})\) as follows:
\begin{equation}
E_c(\mathbf{q}) = E_{DC}(\mathbf{q}) \cdot \frac{\hat{E}_{DC(N4)}(\mathbf{b=0})}{\beta \cdot \hat{E}_{DC}(\mathbf{b=0})}
\end{equation}
\noindent
where \(E_{DC}(\mathbf{q})\) is the distortion corrected dMRI signal, \(\beta\) is the bias field estimated using the N4ITK from the masked average of all b=0 images, \(\hat{E}_{DC}(\mathbf{b=0})\) and \(\hat{E}_{DC(N4)}(\mathbf{b=0})\) are the estimated average intensity of b=0 images before and after applying the N4ITK bias correction respectively.
\subsection{Fetal Brain Segmentation}
Fetal dMRI scans encompass not only the fetal brain but also surrounding maternal structures. Efficient background removal is essential for enhancing the reliability of subsequent processing steps. However, fetal brain extraction is challenging due to factors like variable fetal head position, movement during scans, and the diverse appearance of the developing fetal brain with adjacent anatomy. To address these complexities, we leverage a deep learning-based method built upon a fully convolutional neural network architecture similar to U-Net++~\cite{karimi2020deep}. The model was trained following the established strategy recommended by nnU-Net~\cite{isensee2021nnu} and has been extensively validated on manually labeled data. This approach effectively removes background structures and performs image cropping to reduce processing time.

\subsection{Motion Correction and Reconstruction}
The HAITCH framework employs an iterative motion correction and reconstruction process that spans five epochs. Each epoch alternates between outlier reweighting, SHORE coefficients computation, basis updating, signal prediction, registration, and gradient table rotation steps.

The process begins with slice weighting using the modified \textit{Z}-score method, which does not require a pre-existing predicted signal. 
The reconstruction step then estimates the coefficients $\mathbf{c}$ by minimizing the cost function of Eq.~\ref{eq:argmin}. The computation of these coefficients involves the use of the SHORE basis of order 6 defined by Merlet and Deriche~\cite{merlet2013continuous}. It is important to note that this basis differs from the one available in the Dipy library~\cite{garyfallidis2014dipy}, resulting in more number of basis functions (72 in Merlet \& Deriche vs. 50 in Dipy for radial order 6). The scale parameter \(\zeta\) in Eq.~\ref{eq:shorebasis} is fixed to a value of 700, and the regularization parameters \(\lambda_l\) and \(\lambda_n\) in Eq.~\ref{eq:argmin} are set to $10^{-8}$ as recommended by~\cite{merlet2013continuous}. Using these coefficients and the pre-defined sampling scheme, we predict the dMRI signal which, in turn, is used as a reference for the registration step. Rigid registration is performed using ANTs~\cite{avants2009advanced} to compute motion parameters $\boldsymbol{\mu}$ from the acquired data. This involves volume-to-volume registration, where the fixed and moving images share the same diffusion contrast.
A global correlation metric is utilized to guide the registration process. Based on the estimated motion parameters, the gradient directions are subsequently rotated individually. A new epoch commences with slice weighting using the GMM method, which compares the registered data with the predicted signal. The reconstruction step then updates the SHORE basis, estimates new coefficients, and predicts a new dMRI signal in the initial basis functions for the next registration step. Therefore, motion parameters $\boldsymbol{\mu}$ are updated and refined over four epochs. The final epoch utilizes these refined parameters to obtain final registered data, followed by a voxel-wise outlier detection. 
Subsequently, SHORE coefficients are computed in the updated basis, and the process concludes by predicting the final dMRI signal in the original sampling scheme.
\subsection{Spatial Normalization to Atlas Space}
Following the motion correction and reconstruction steps, the HAITCH framework performs spatial normalization to a common reference space. This process involves two stages of registration. First, the motion-corrected and reconstructed dMRI data of the subject is registered to a reconstructed T2w image of the same subject in the world coordinates. The T2w image is obtained using a super-resolution SVR algorithm, either SVRTK~\cite{uus2020deformable} or NiftyMIC~\cite{ebner2020automated}. Second, the T2w image is registered to the publicly available CRL fetal brain atlas~\cite{gholipour2017normative}. This atlas provides a standardized anatomical reference for the fetal brain. 
Finally, the two transformation matrices obtained from these separate registrations are combined into a single transformation, which is then applied to the dMRI data to map it into the atlas space while upsampling it to an isotropic voxel resolution of 1.25 mm. These upsampled dMRI data are used in subsequent post-processing steps including fiber tractography.
\subsection{Diffusion Model Estimation and Tractography}
White matter tractography was conducted utilizing the MRtrix3 toolbox~\cite{tournier2019mrtrix3}. Initially, white matter, gray matter, and cerebrospinal fluid response functions were estimated using the unsupervised method. Subsequently, multi-shell multi-tissue constrained spherical deconvolution was performed to estimate fiber orientation distribution (FOD) maps~\cite{jeurissen2014multi}. Streamline tractography was then conducted using the iFOD2 algorithm, a probabilistic tracking method that utilizes second-order integration of the FOD maps~\cite{tournier2010improved}. Tracking parameters were carefully chosen to optimize the tractography and generate accurate white matter pathways from the fetal FOD maps. These included an angle threshold of 45$^{\circ}$, a cutoff value of 0.01, a specified number of streamlines (100,000), and an FOD power raised to 6 for enhanced specificity. Additionally, minimum and maximum track lengths were set to 10 mm and 120 mm, respectively, to target relevant white matter tracts.

\subsection{HAITCH: Open-Source Framework and Flexibility}
As mentioned earlier, HAITCH framework is publicly available as the first module of our FEDI toolbox, accessible at \url{https://fedi.readthedocs.io}, and \url{https://github.com/FEDIToolbox}. This toolbox includes both the HAITCH implementation and our optimized sampling scheme for fetal dMRI, making these resources readily available to the research community. HAITCH demonstrates its versatility by effectively handling various fetal dMRI data types, including single-shell, multi-shell, or single-echo acquisitions. It further empowers users with a range of options for customization. These include additional outlier detection methods and alternative registration methods and metrics, allowing researchers to tailor the framework to their specific needs. For dynamic distortion correction, besides the methods mentioned previously, users can easily switch to alternative approaches like Block-Matching~\cite{hedouin2017block} or EPIC~\cite{holland2010efficient} methods.

To take advantage of the full capacity of HAITCH, we recommend using a blip-reversed dual-echo EPI sequence. We have proactively shared our in-house built dual-echo spin echo EPI sequence (developed by co-authors OA and MO) on the Siemens R2C platform. This ensures that researchers will have all the necessary tools readily available for successful HAITCH implementation and application in their fetal brain imaging studies.

\section{Results}
\label{sec:results}
\subsection{Distortion Correction: Dynamic Field Map Estimation}
\begin{figure*}[!ht]
\centering
\vspace{0.3cm}
\includegraphics[width=0.49\textwidth, trim={0cm 0cm 0cm 0.9cm}, clip]{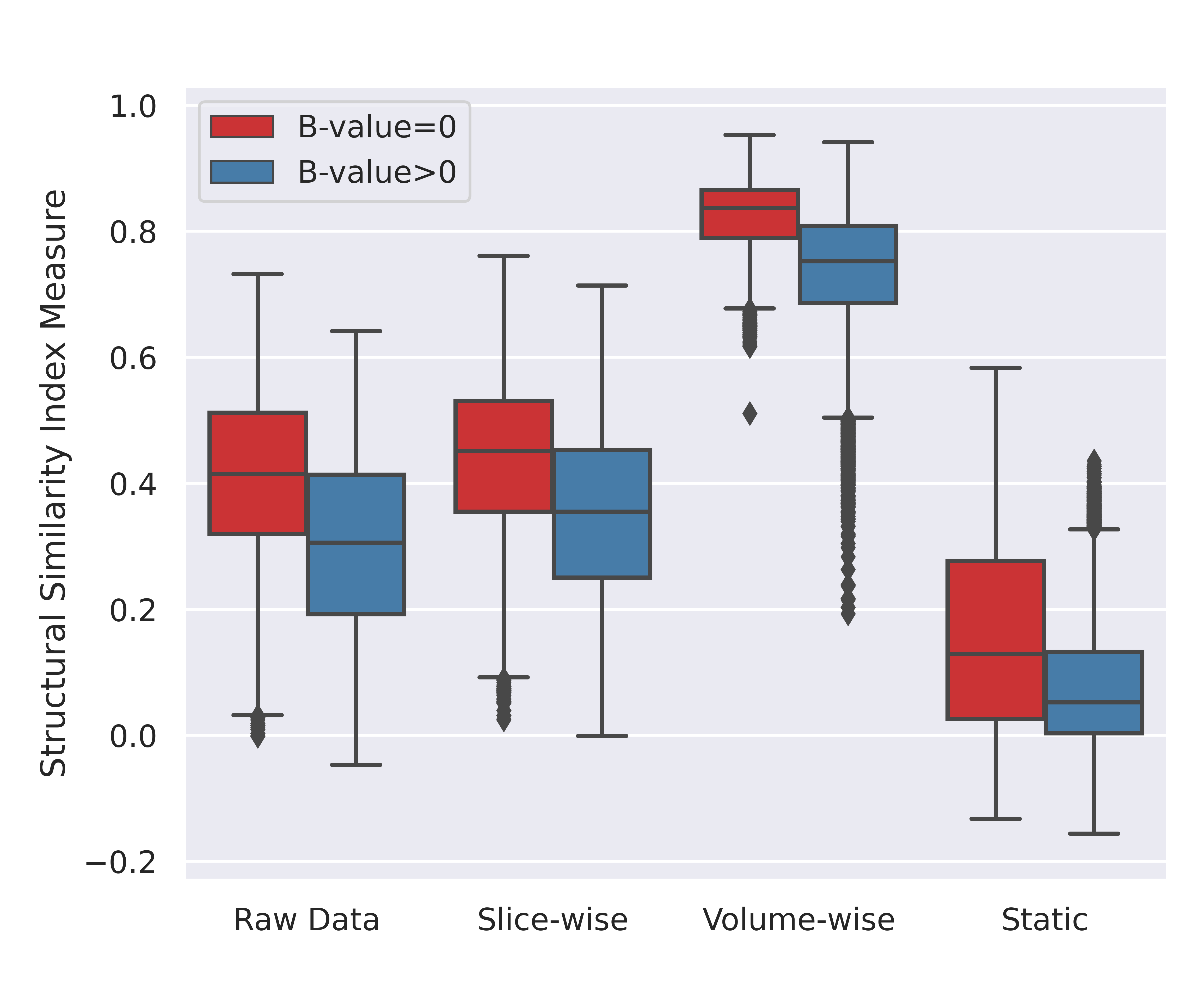}
\includegraphics[width=0.49\textwidth, trim={0cm 0cm 0cm 0.9cm}, clip]{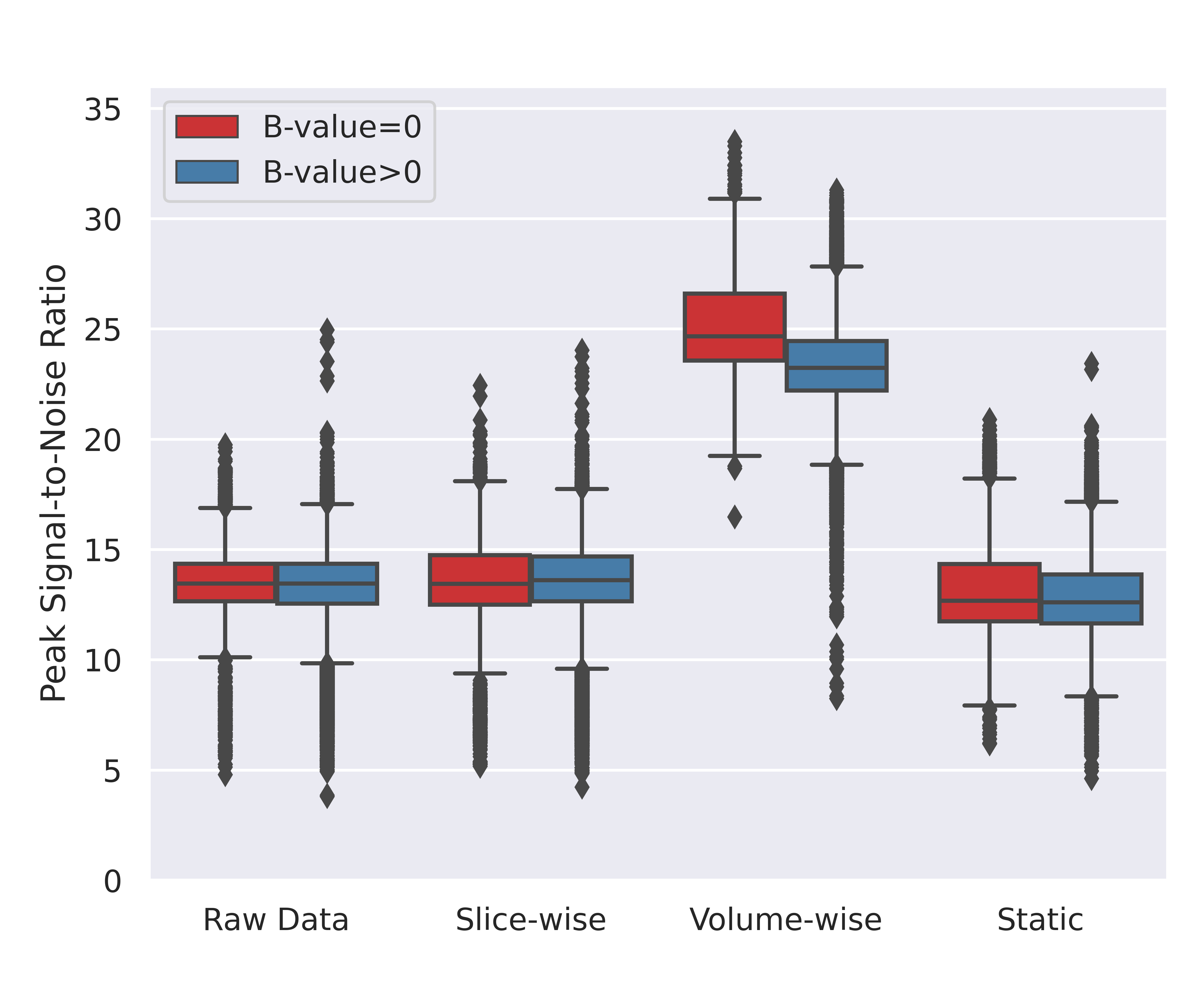}
\vspace{-0.3cm}
\caption{Quantitative Evaluation of Distortion Correction Techniques using SSIM and PSNR. Boxplots summarize the distribution of SSIM (left) and PSNR (right) values across 27 subjects. Results are presented for raw data and data processed with: Slice-wise correction, Volume-wise correction (HAITCH approach), and Static field map correction obtained from the three echo reversal correction methods. Red boxes represent b=0 images, while blue boxes display diffusion-weighted images ($b>0$). Higher SSIM and PSNR values indicate better image quality and reduced distortion. As evident from the plots, the volume-wise dynamic correction method achieved the best results across both metrics and image types.}
\label{fig:boxplots}
\end{figure*}

Field map estimation utilizing both dynamic and static echo reversal correction approaches was conducted on 27 dMRI scans. The effectiveness of each correction method was evaluated visually and quantitatively. Fig.~\ref{fig:dc_ssim} presents an illustrative example of distortion correction achieved for each method. Overall, visual inspection shows that volume-wise dynamic correction is superior in recovering anatomical brain shape compared to the raw data and other correction techniques. Corrected dMRI (TE1) and dMRI (TE2) appear to have greater visual similarity than those processed with other methods.To quantify these observations, Fig.~\ref{fig:boxplots} illustrates the SSIM and PSNR statistics for both b=0 and $b>0$ images across all slices for all scans. The analysis reveals that both dynamic field map estimation methods have significantly enhanced the SSIM and PSNR values compared to the raw data (uncorrected). The volume-wise approach yielded the highest SSIM and PSNR, indicating superior performance in correcting geometric distortions. Conversely, static field map-based correction resulted in lower SSIM and approximately similar PSNR compared to raw data. This highlights its ineffectiveness in addressing motion-induced field inhomogeneities.

\begin{table*}
\centering
\begin{tabular}{c@{\hspace{0.07cm}}c@{\hspace{0.07cm}}c@{\hspace{0.07cm}}c@{\hspace{0.07cm}}c@{\hspace{0.07cm}}c@{\hspace{0.07cm}}c@{\hspace{0.07cm}}c@{\hspace{0.07cm}}}
\includegraphics[width=0.115\textwidth,trim={52cm 12cm 16cm 13.5cm},clip]{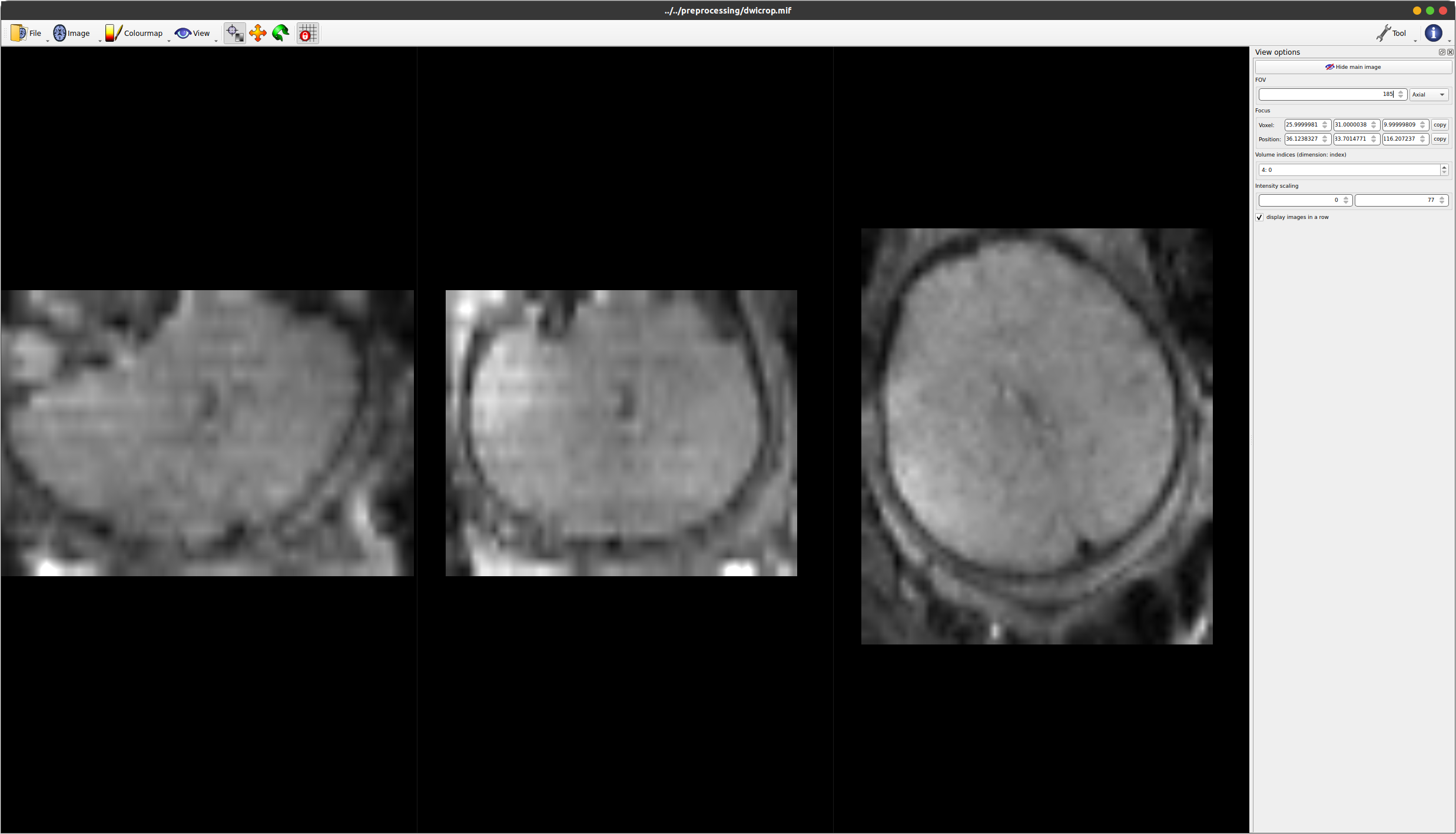} &
\includegraphics[width=0.115\textwidth,trim={52cm 12cm 16cm 13.5cm},clip]{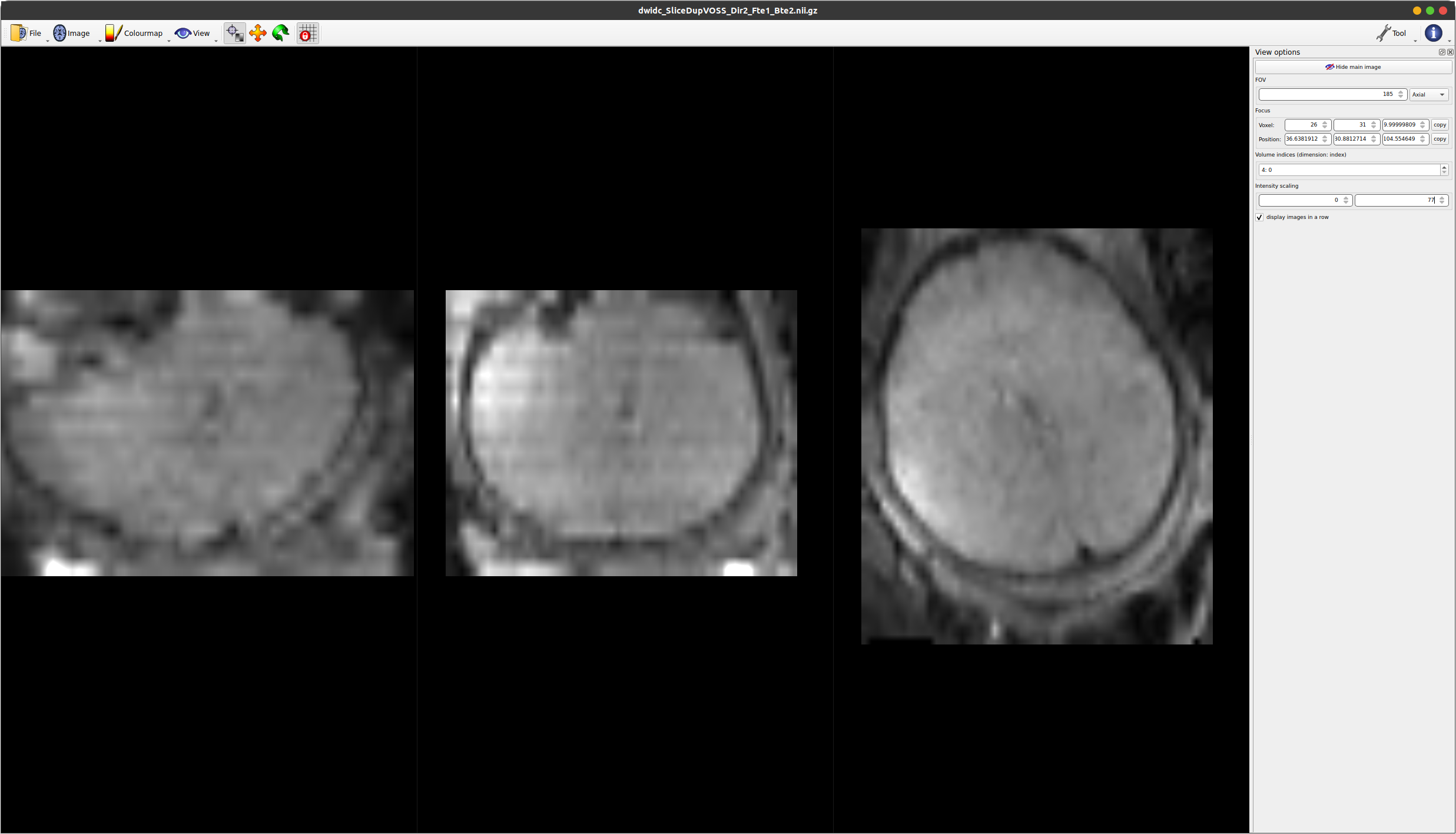} &
\includegraphics[width=0.115\textwidth,trim={52cm 12cm 16cm 13.5cm},clip]{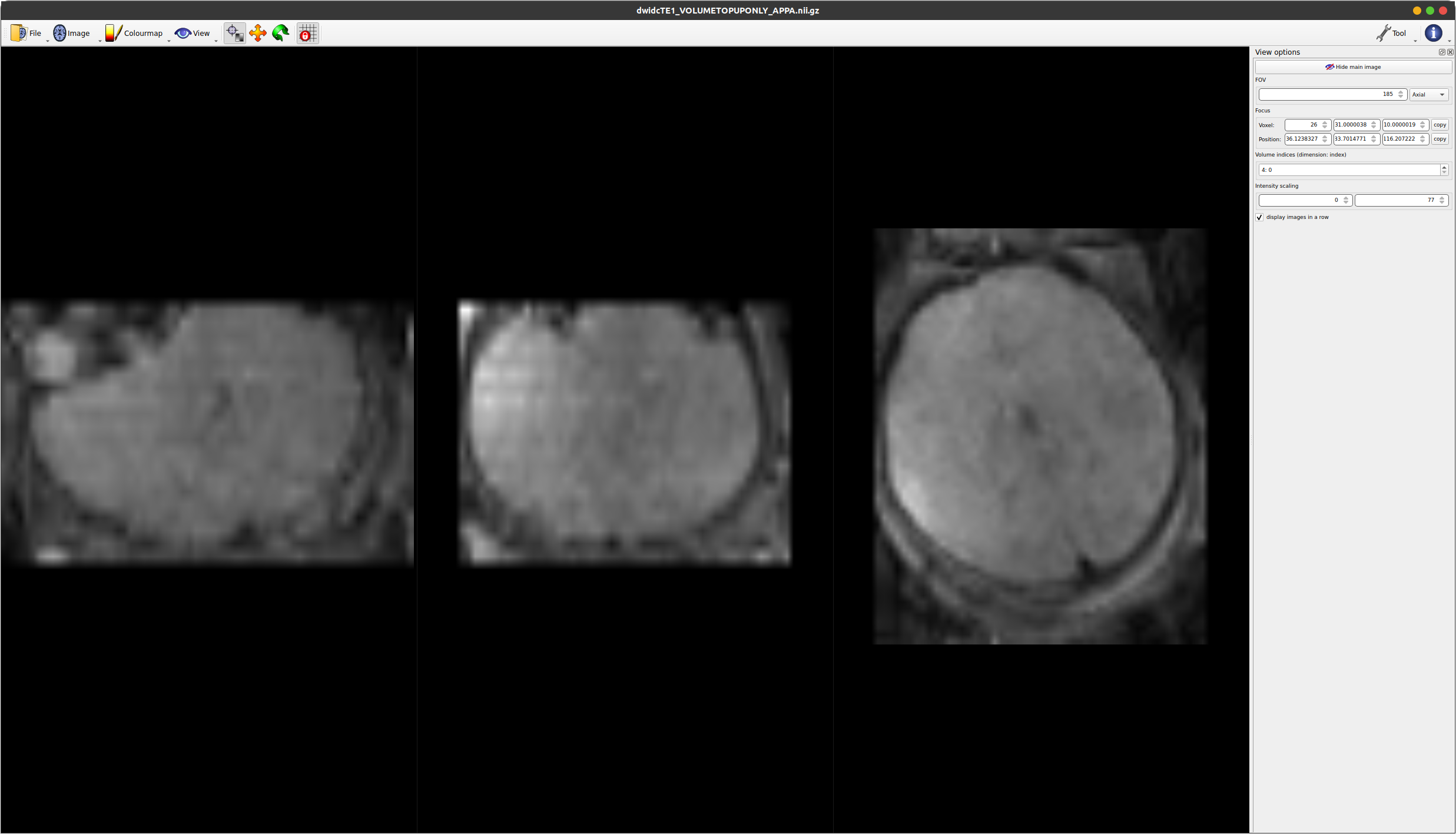} &
\includegraphics[width=0.115\textwidth,trim={52cm 12cm 16cm 13.5cm},clip]{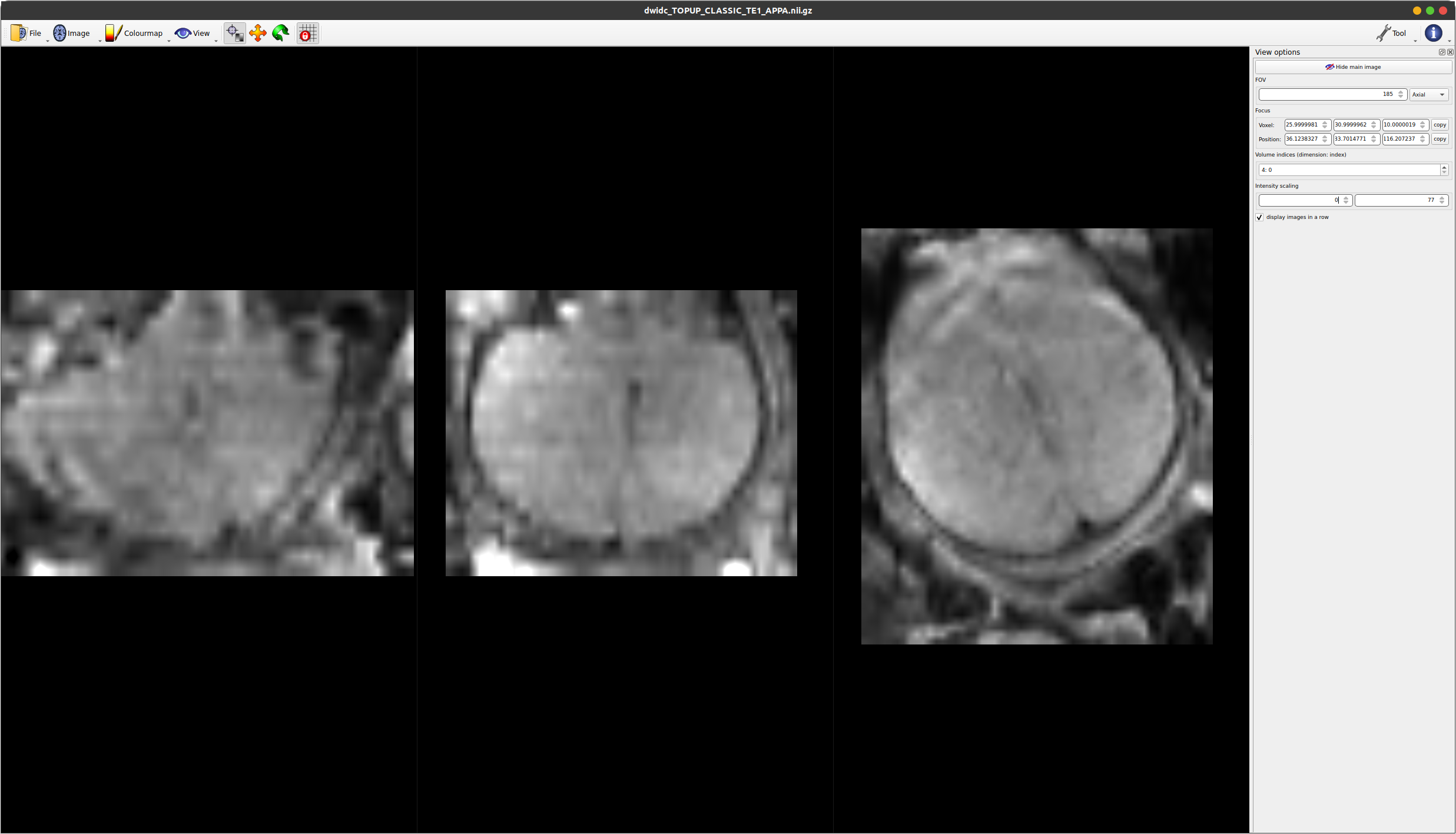} &
\includegraphics[width=0.125\textwidth,trim={0cm 16cm 64cm 17.5cm},clip]{images/distortion_FCB196_s2_dwiME_run30/raw_TE1.png} &
\includegraphics[width=0.125\textwidth,trim={0cm 16cm 64cm 17.5cm},clip]{images/distortion_FCB196_s2_dwiME_run30/voss_TE1_TE2.png} &
\includegraphics[width=0.125\textwidth,trim={0cm 16cm 64cm 17.5cm},clip]{images/distortion_FCB196_s2_dwiME_run30/volume_TE1_TE2.png} &
\includegraphics[width=0.125\textwidth,trim={0cm 16cm 64cm 17.5cm},clip]{images/distortion_FCB196_s2_dwiME_run30/static_TE1_TE2.png} \\
\includegraphics[width=0.115\textwidth,trim={52cm 12cm 16cm 13.5cm},clip]{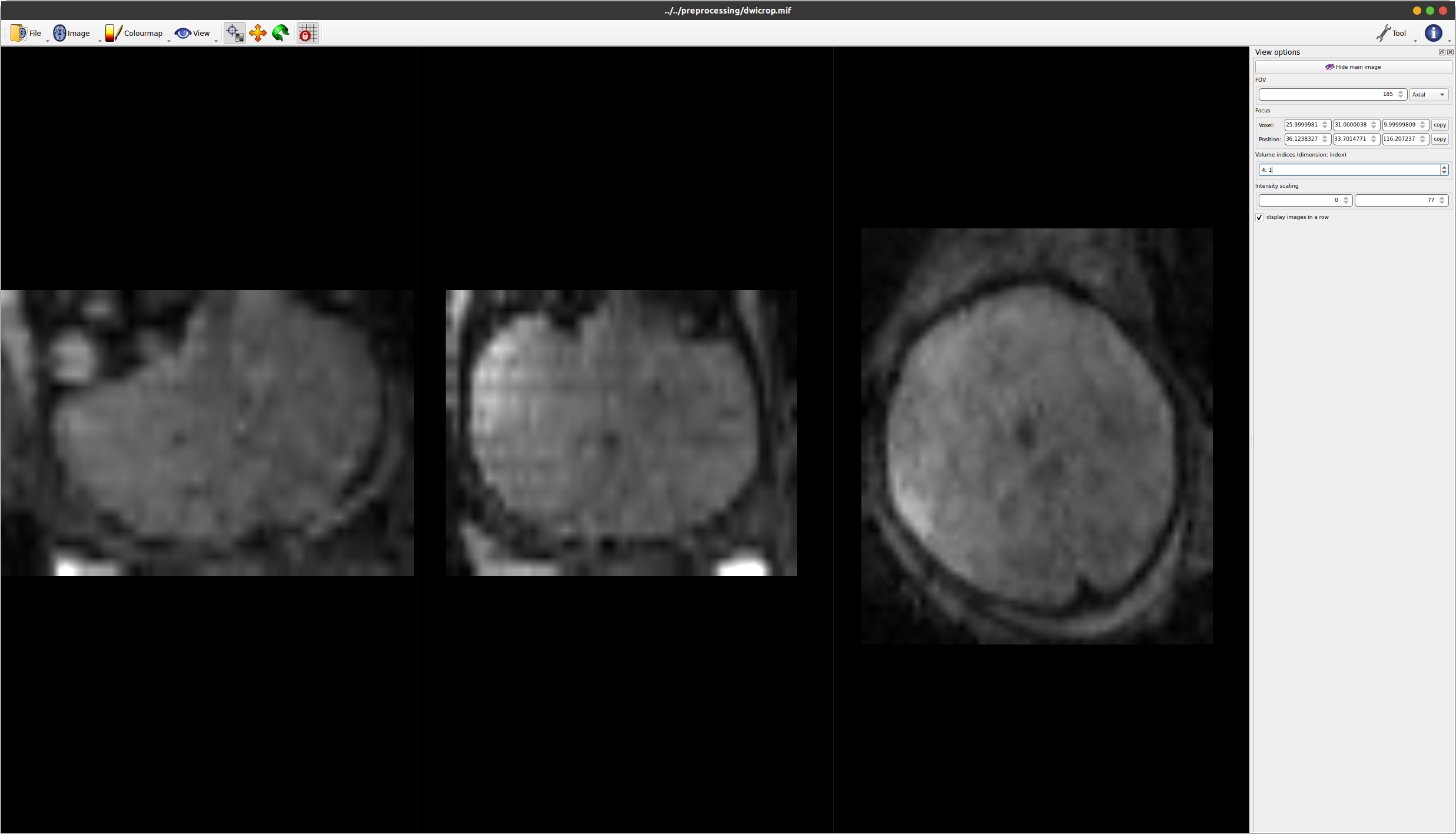} &
\includegraphics[width=0.115\textwidth,trim={52cm 12cm 16cm 13.5cm},clip]{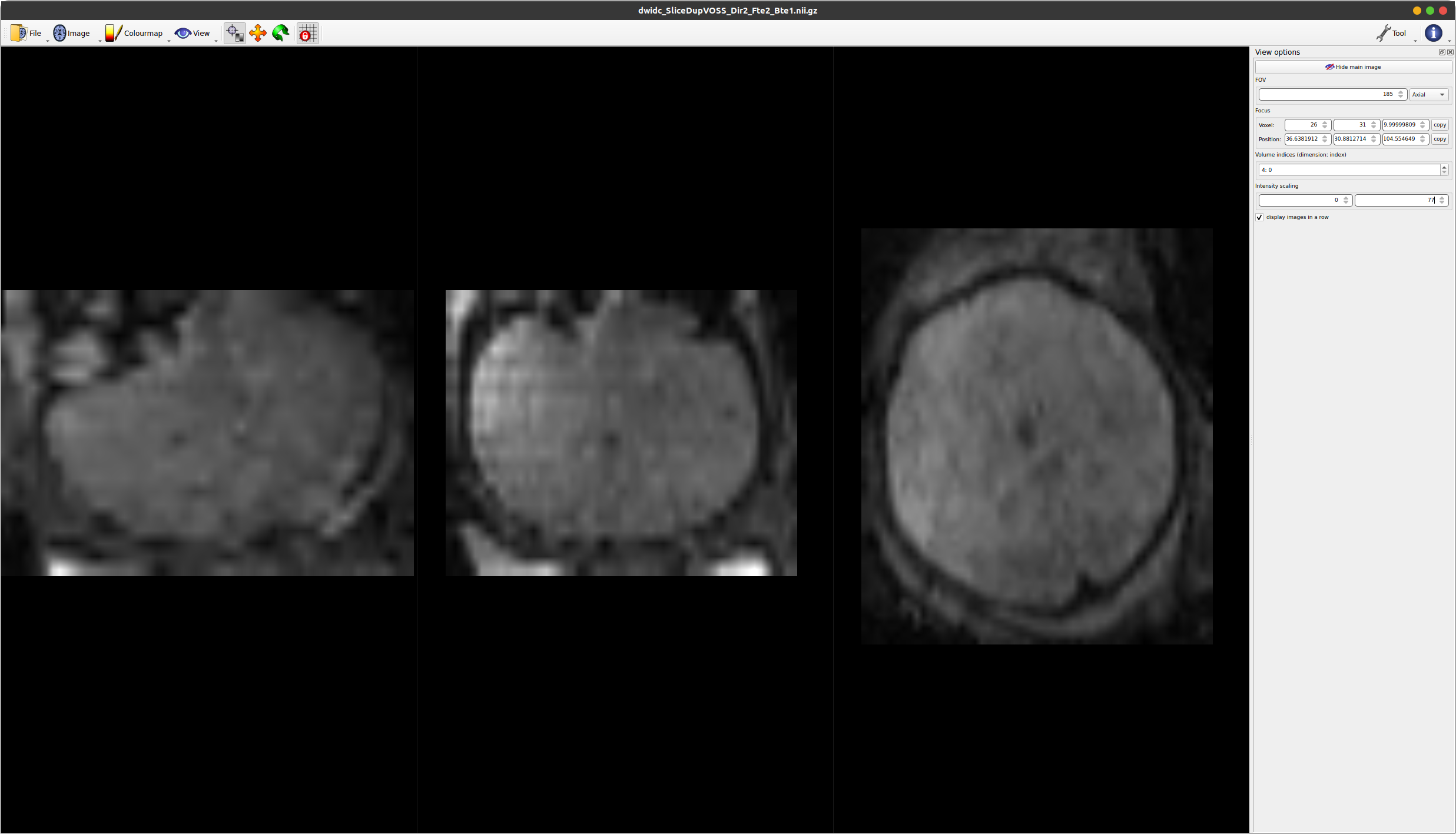} &
\includegraphics[width=0.115\textwidth,trim={52cm 12cm 16cm 13.5cm},clip]{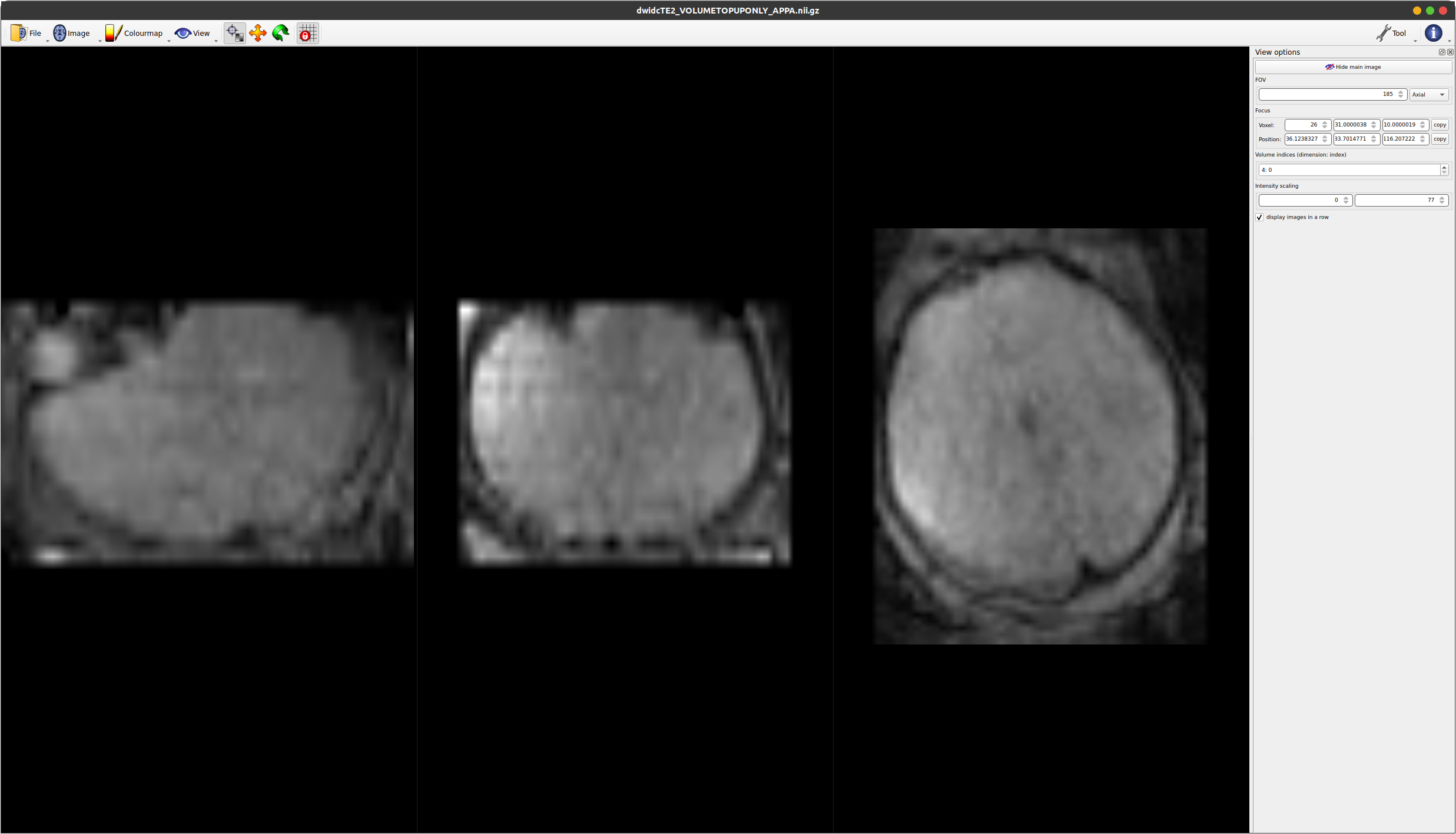} &
\includegraphics[width=0.115\textwidth,trim={52cm 12cm 16cm 13.5cm},clip]{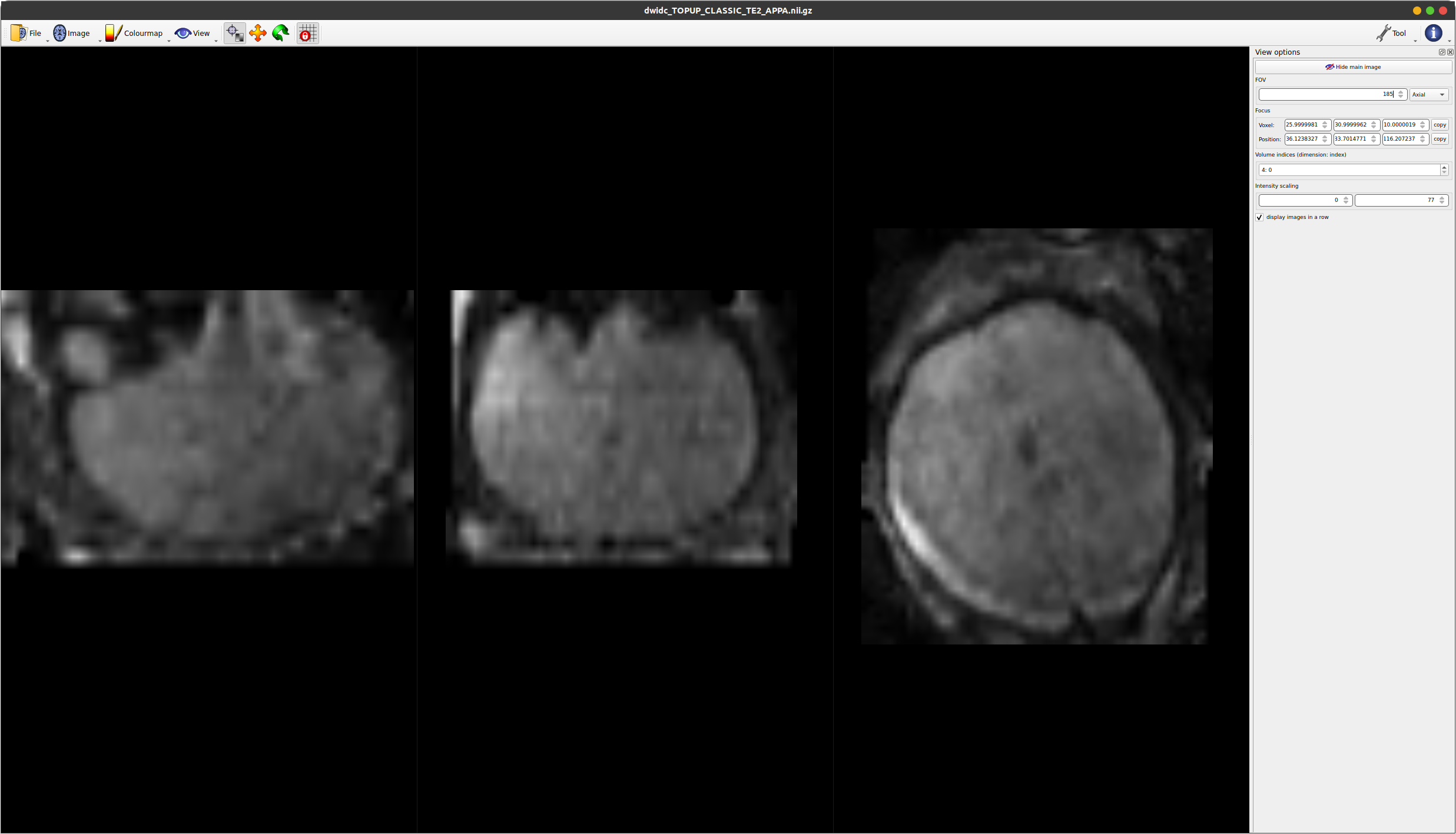} &
\raisebox{0.66\height}{\includegraphics[width=0.125\textwidth,trim={0cm 16cm 64cm 17.5cm},clip]{images/distortion_FCB196_s2_dwiME_run30/raw_TE2.png}} &
\raisebox{0.66\height}{\includegraphics[width=0.125\textwidth,trim={0cm 16cm 64cm 17.5cm},clip]{images/distortion_FCB196_s2_dwiME_run30/voss_TE2_TE1.png}} &
\raisebox{0.66\height}{\includegraphics[width=0.125\textwidth,trim={0cm 16cm 64cm 17.5cm},clip]{images/distortion_FCB196_s2_dwiME_run30/volume_TE2_TE1.png}} &
\raisebox{0.66\height}{\includegraphics[width=0.125\textwidth,trim={0cm 16cm 64cm 17.5cm},clip]{images/distortion_FCB196_s2_dwiME_run30/static_TE2_TE1.png}} \\
\textsf{\small Raw Data} & \textsf{\small Slice-wise} & \textsf{\small Volume-wise} & \textsf{\small Static} & \textsf{\small Raw Data} & \textsf{\small Slice-wise} & \textsf{\small Volume-wise} & \textsf{\small Static}
\end{tabular}
\caption{Comparison of Fetal dMRI Distortion Correction Techniques. This figure illustrates the impact of different distortion correction techniques on fetal dMRI data. The top row displays dMRI (TE1) and the bottom row shows dMRI (TE2) images. Left and right panels represent axial and coronal views, respectively. Each column showcases the results for a different dataset or correction technique: Raw Data: Uncorrected image exhibiting distortions due to susceptibility variations. Slice-wise Correction: Distortion correction is applied to individual slices independently.Volume-wise Correction: Distortion correction is simultaneously applied to the entire volume (HAITCH approach). Static Correction: Correction based on a static field map, potentially inaccurate due to fetal motion.}
\label{fig:dc_ssim}
\end{table*}

\subsection{Evaluation of the Motion Correction}
\begin{figure*}[!t]
\centering
\includegraphics[width=0.235\textwidth, trim={0cm 13.0cm 14cm 13.5cm}, clip]{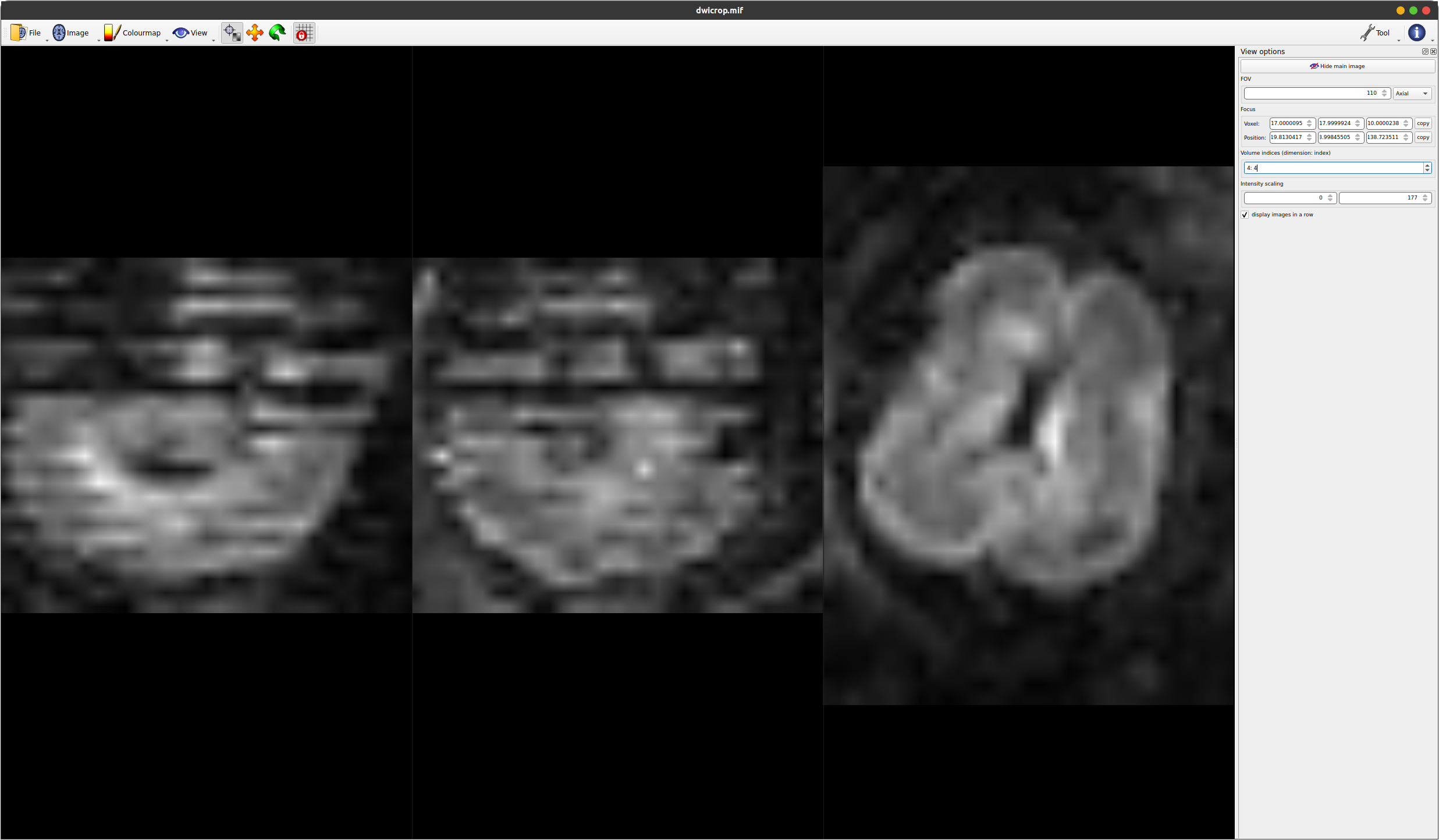}
\includegraphics[width=0.235\textwidth, trim={0cm 13.0cm 14cm 13.5cm}, clip]{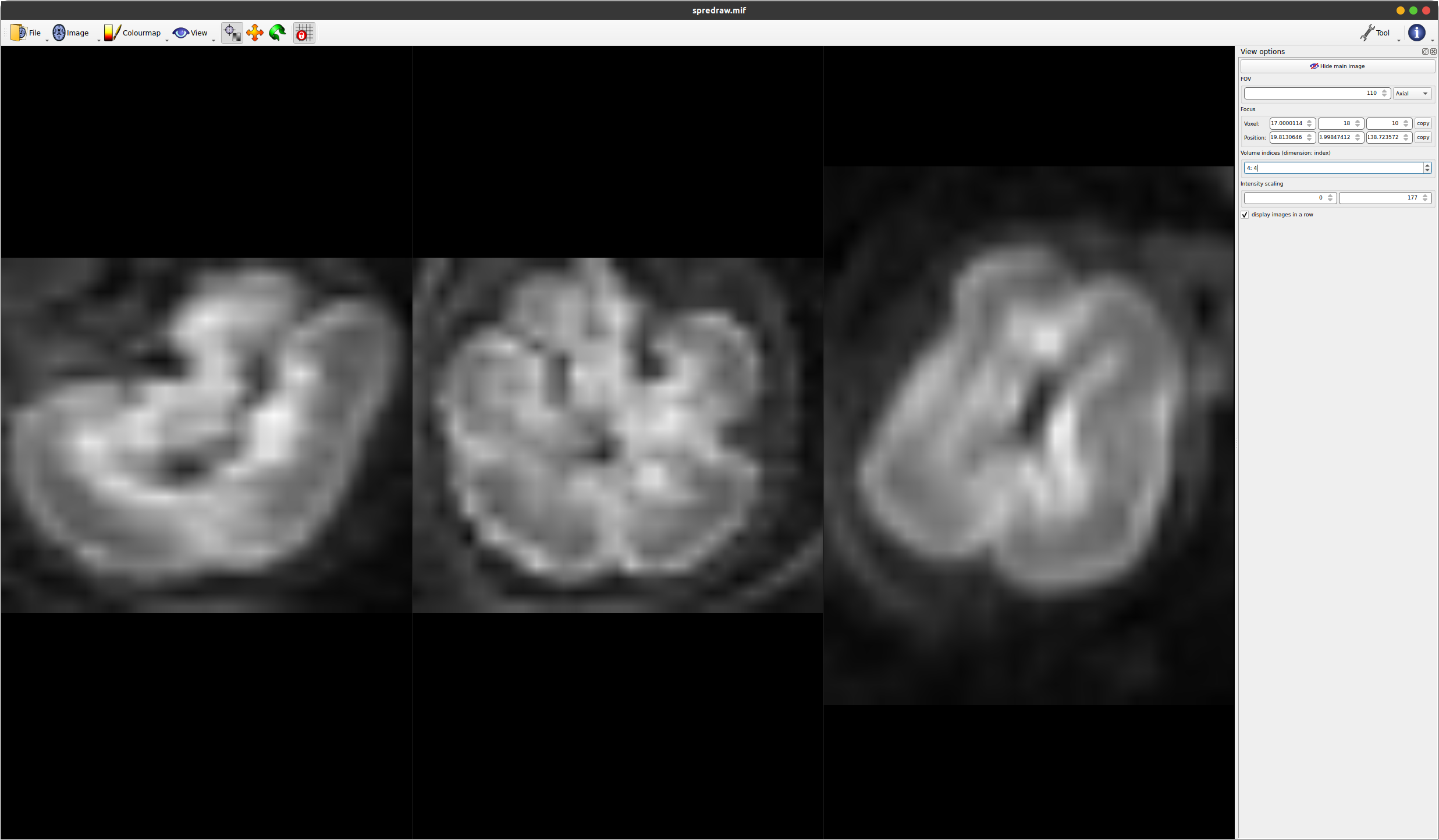}\hspace{0.1cm}
\includegraphics[width=0.235\textwidth, trim={0cm 13.0cm 14cm 13.5cm}, clip]{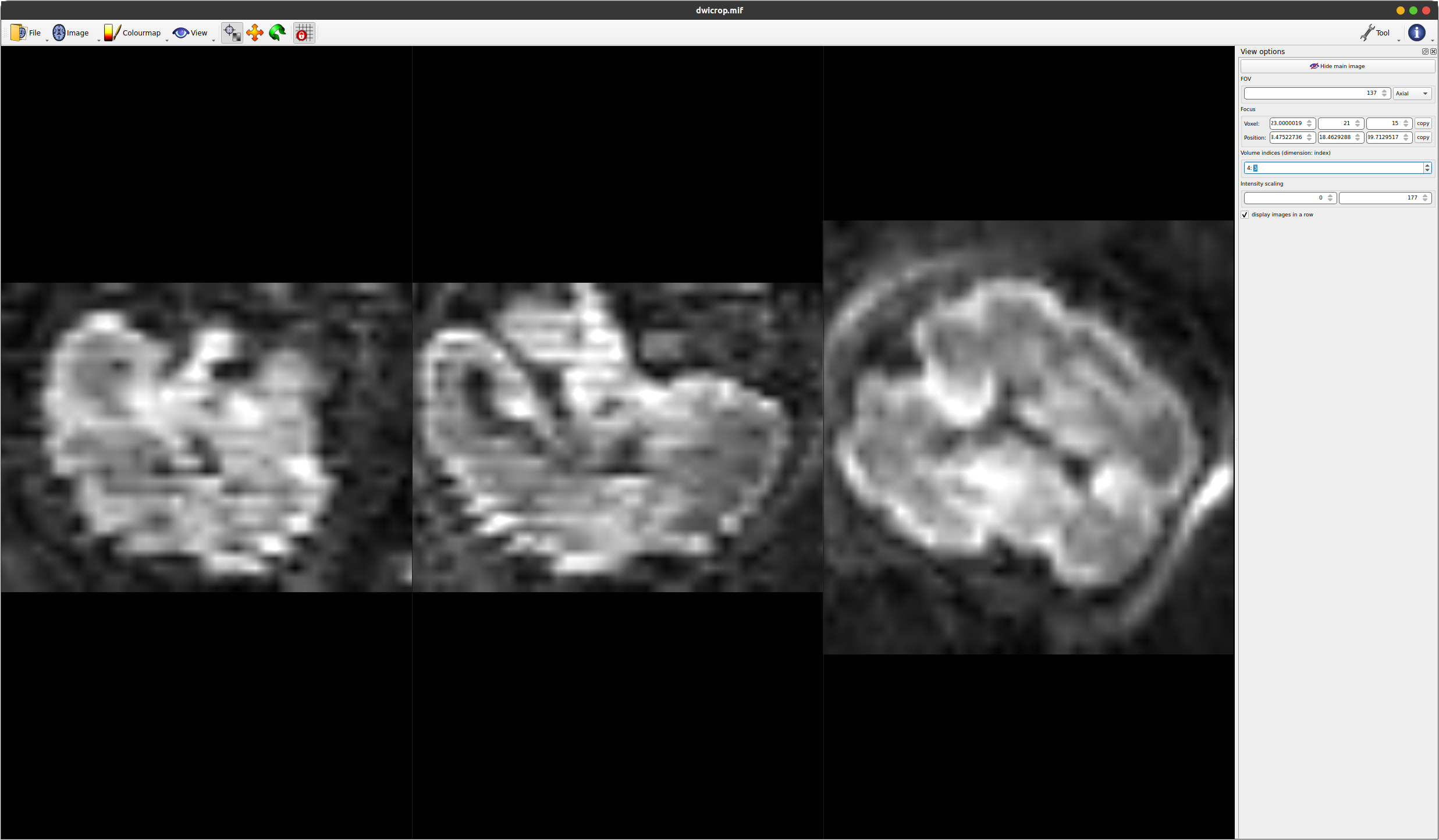}
\includegraphics[width=0.235\textwidth, trim={0cm 13.0cm 14cm 13.5cm}, clip]{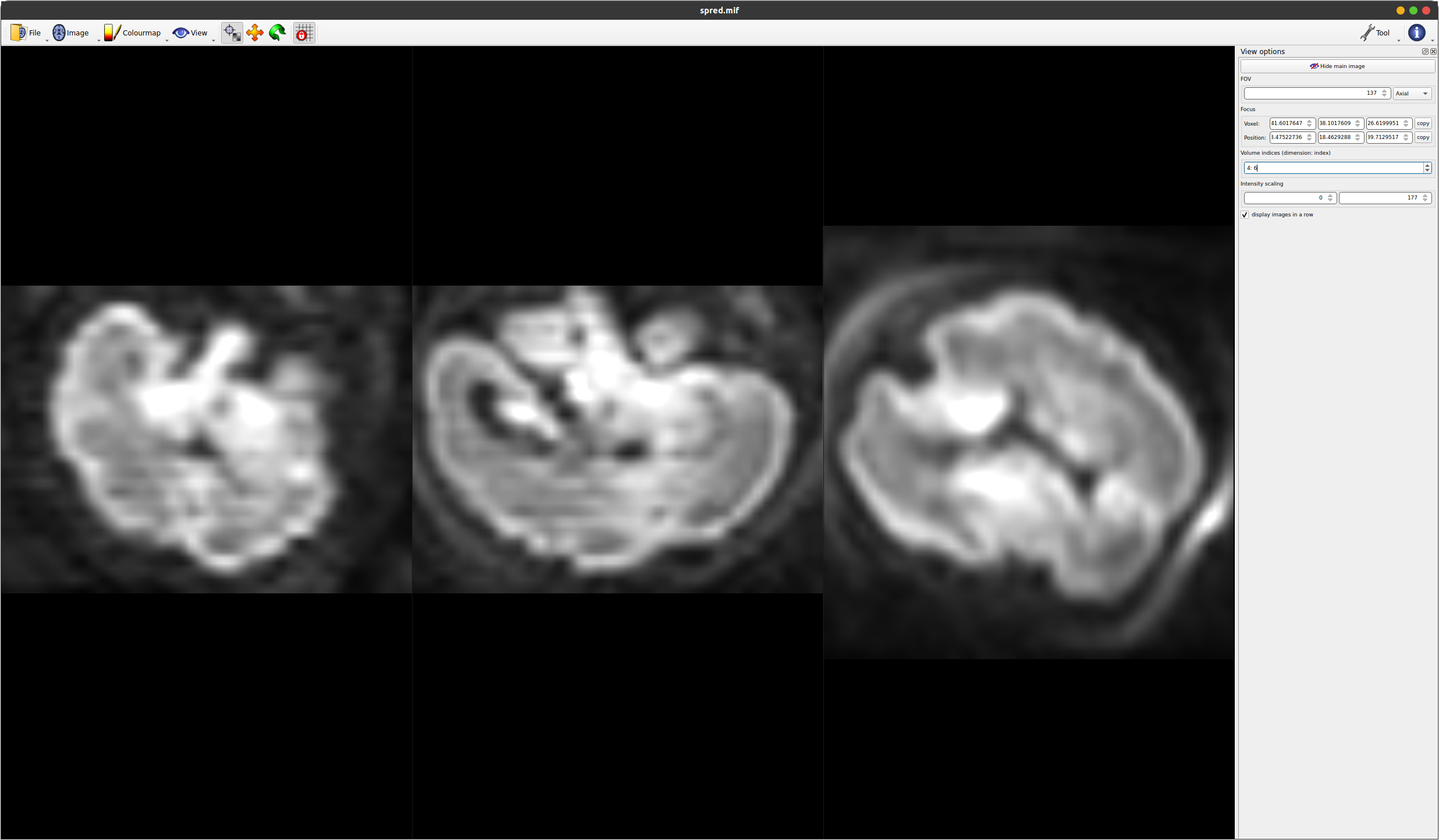}
\\
\includegraphics[width=0.235\textwidth, trim={0cm 13.0cm 14cm 13.5cm}, clip]{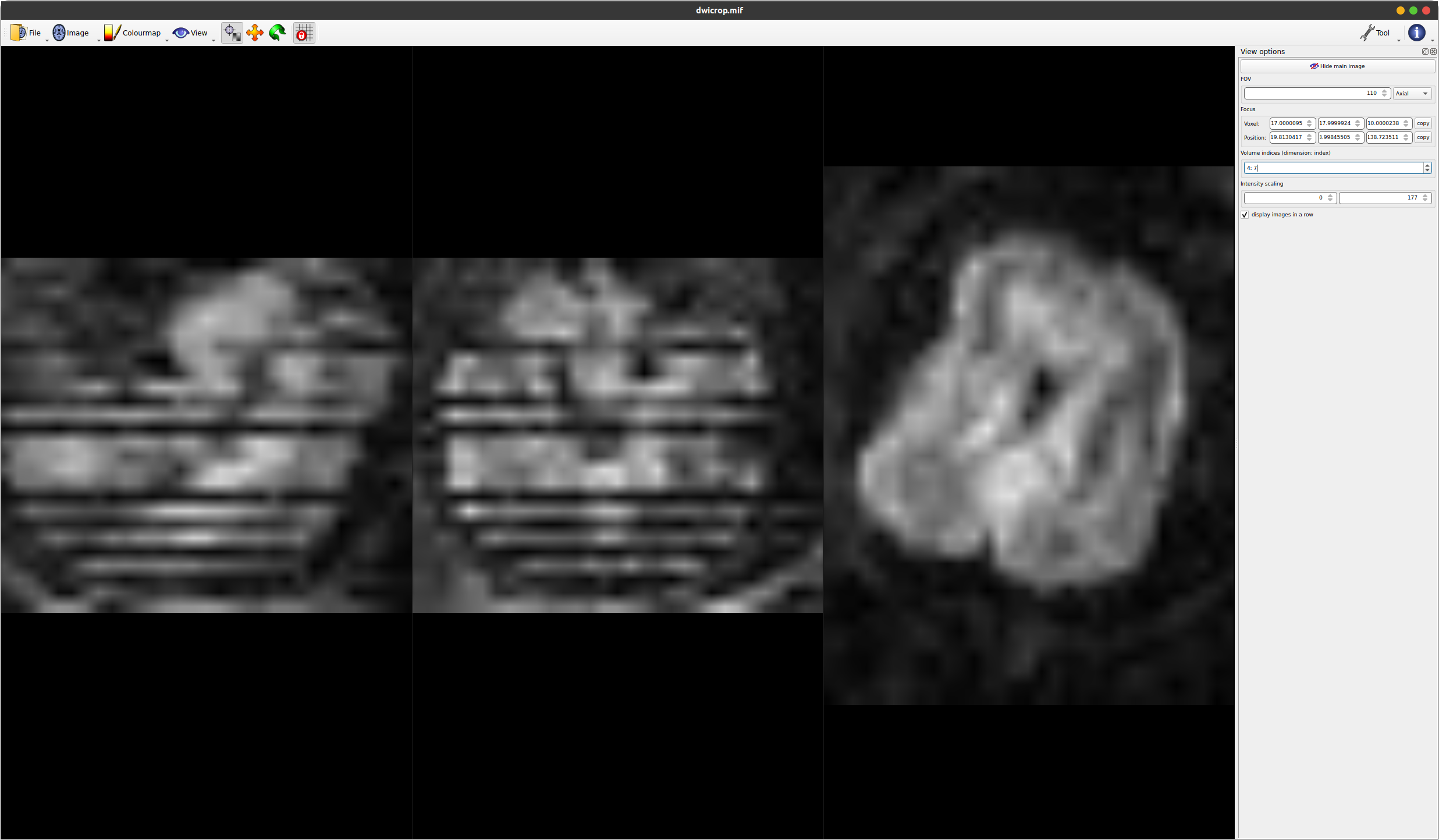}
\includegraphics[width=0.235\textwidth, trim={0cm 13.0cm 14cm 13.5cm}, clip]{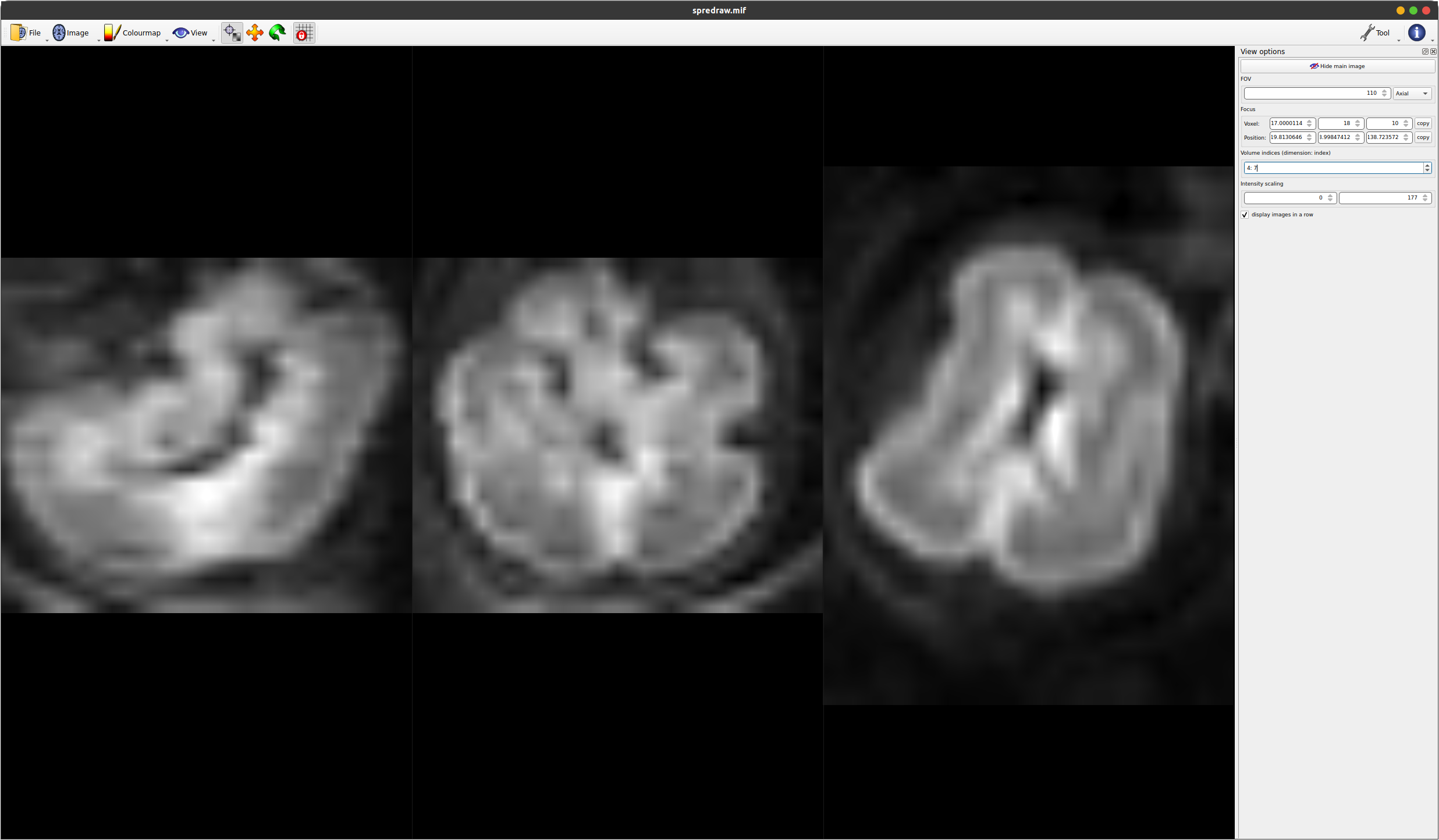}\hspace{0.1cm}
\includegraphics[width=0.235\textwidth, trim={0cm 13.0cm 14cm 13.5cm}, clip]{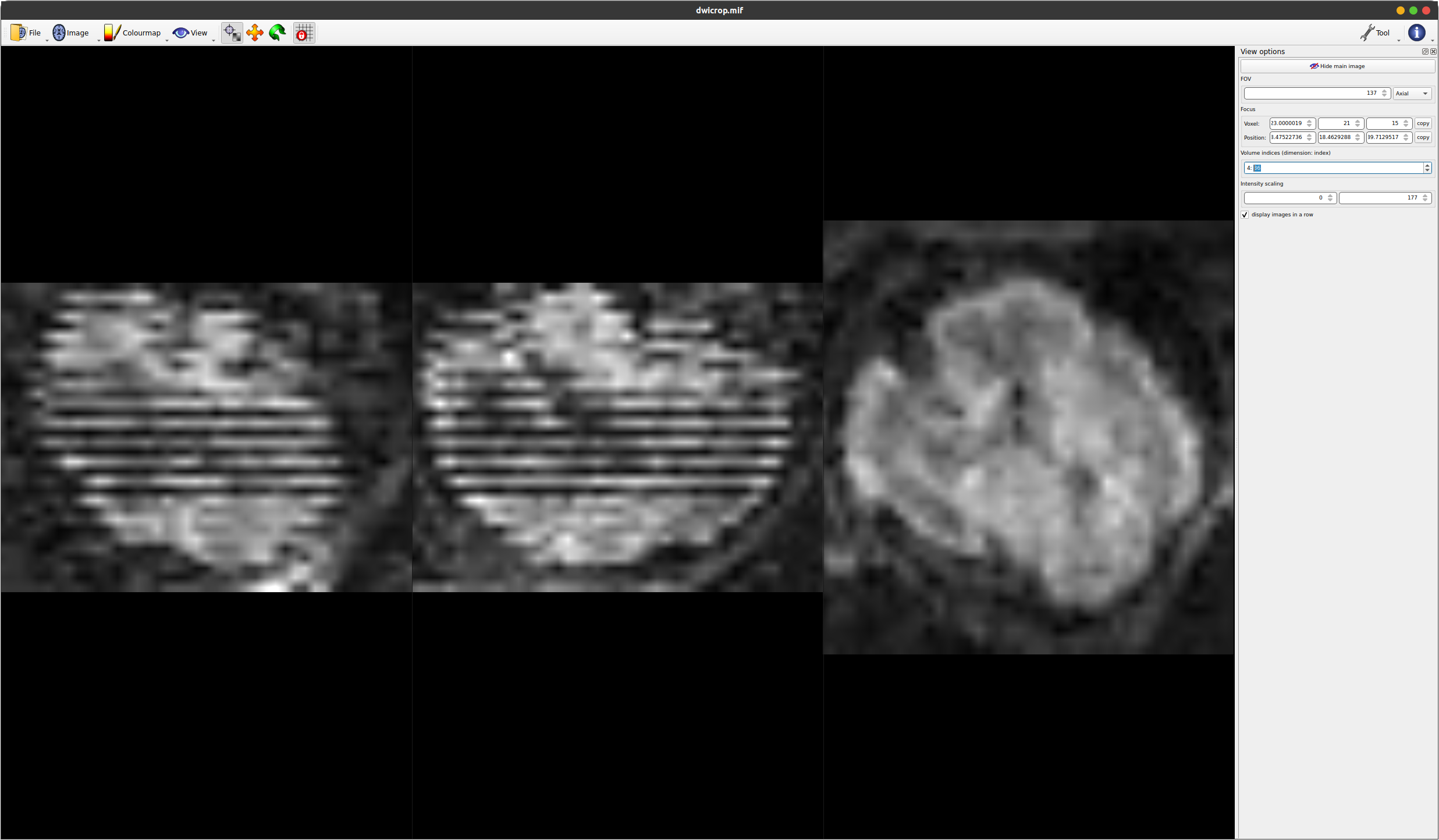}
\includegraphics[width=0.235\textwidth, trim={0cm 13.0cm 14cm 13.5cm}, clip]{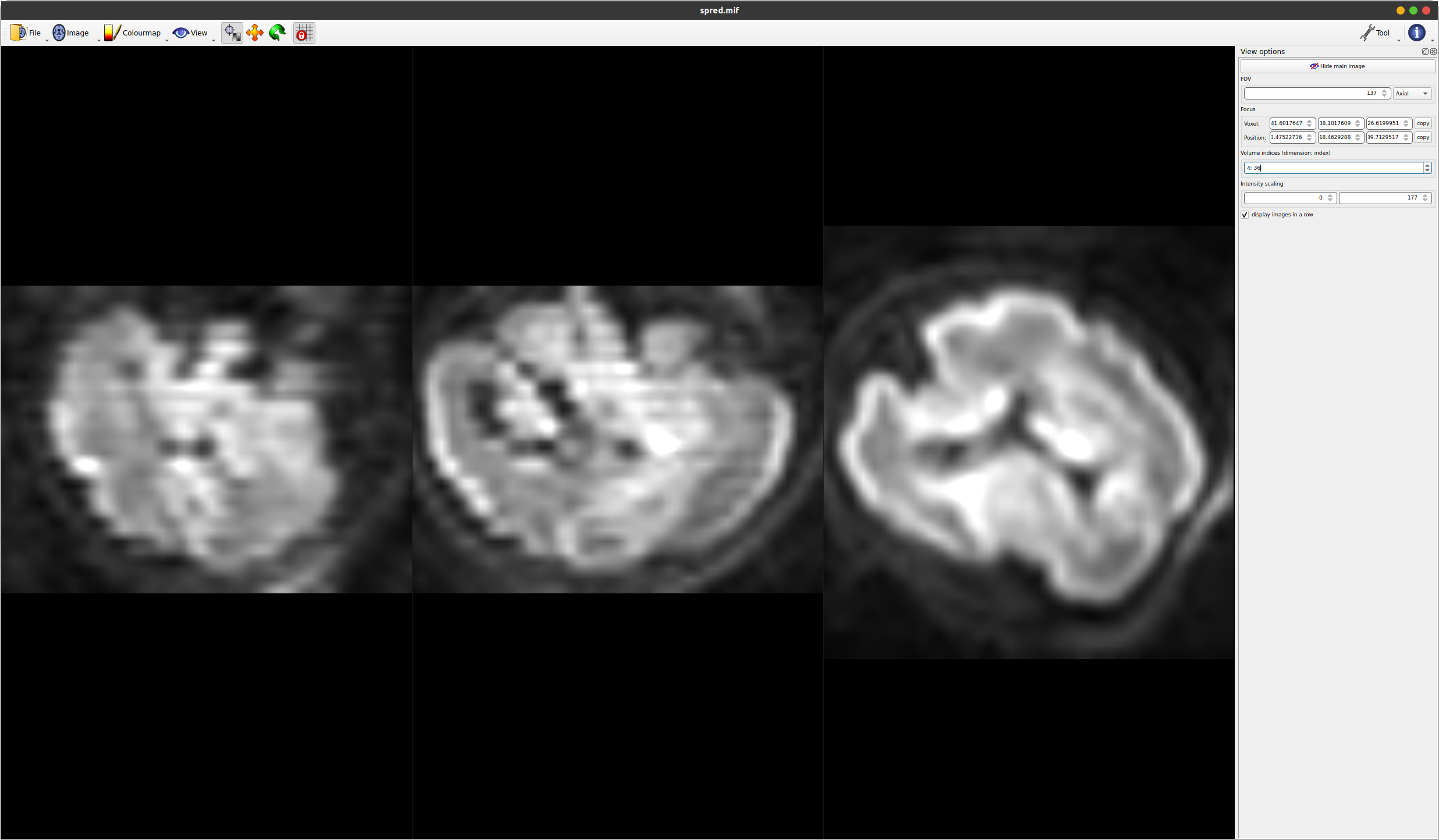}
\\
\includegraphics[width=0.235\textwidth, trim={0cm 13.0cm 14cm 13.5cm}, clip]{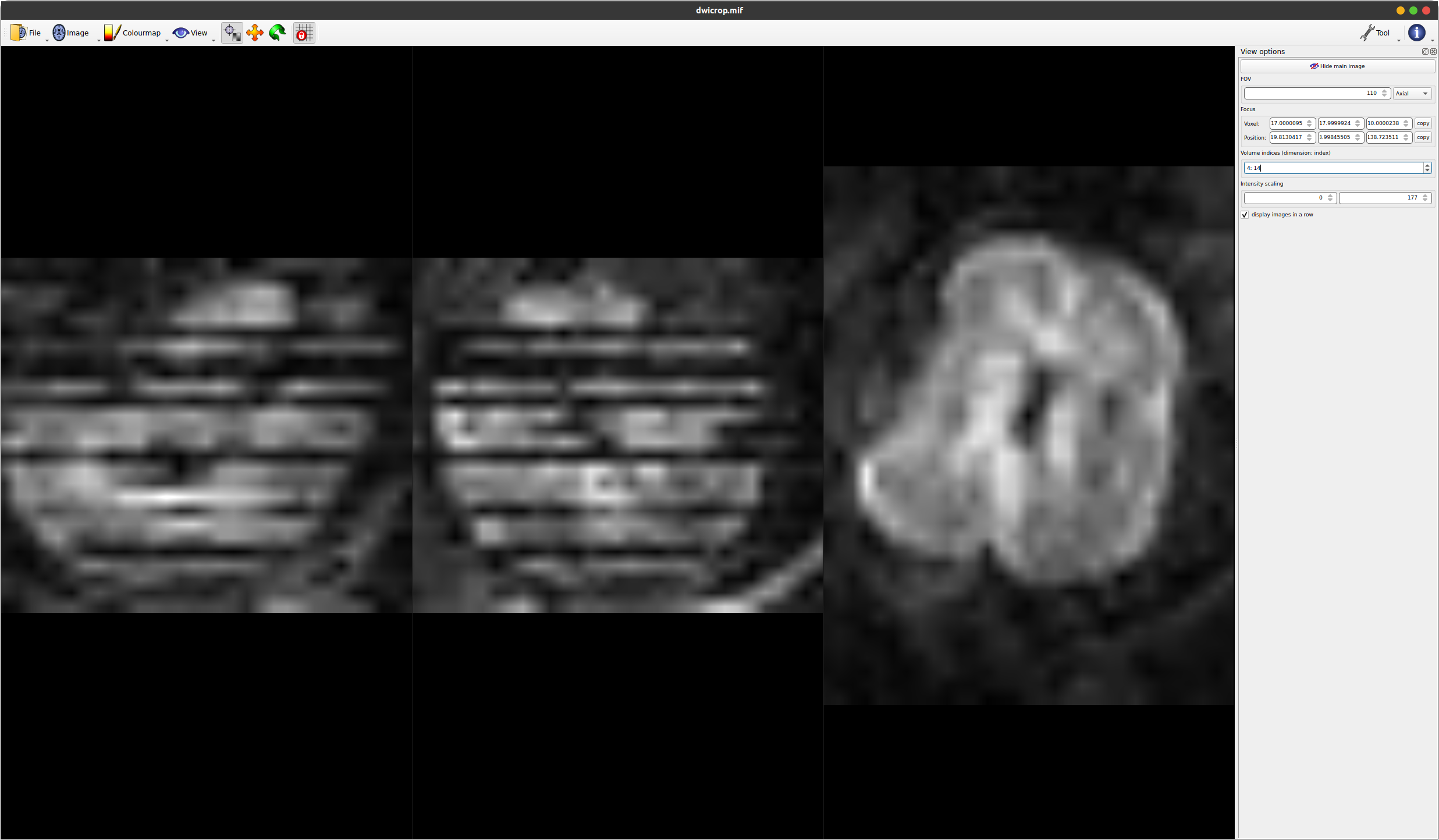}
\includegraphics[width=0.235\textwidth, trim={0cm 13.0cm 14cm 13.5cm}, clip]{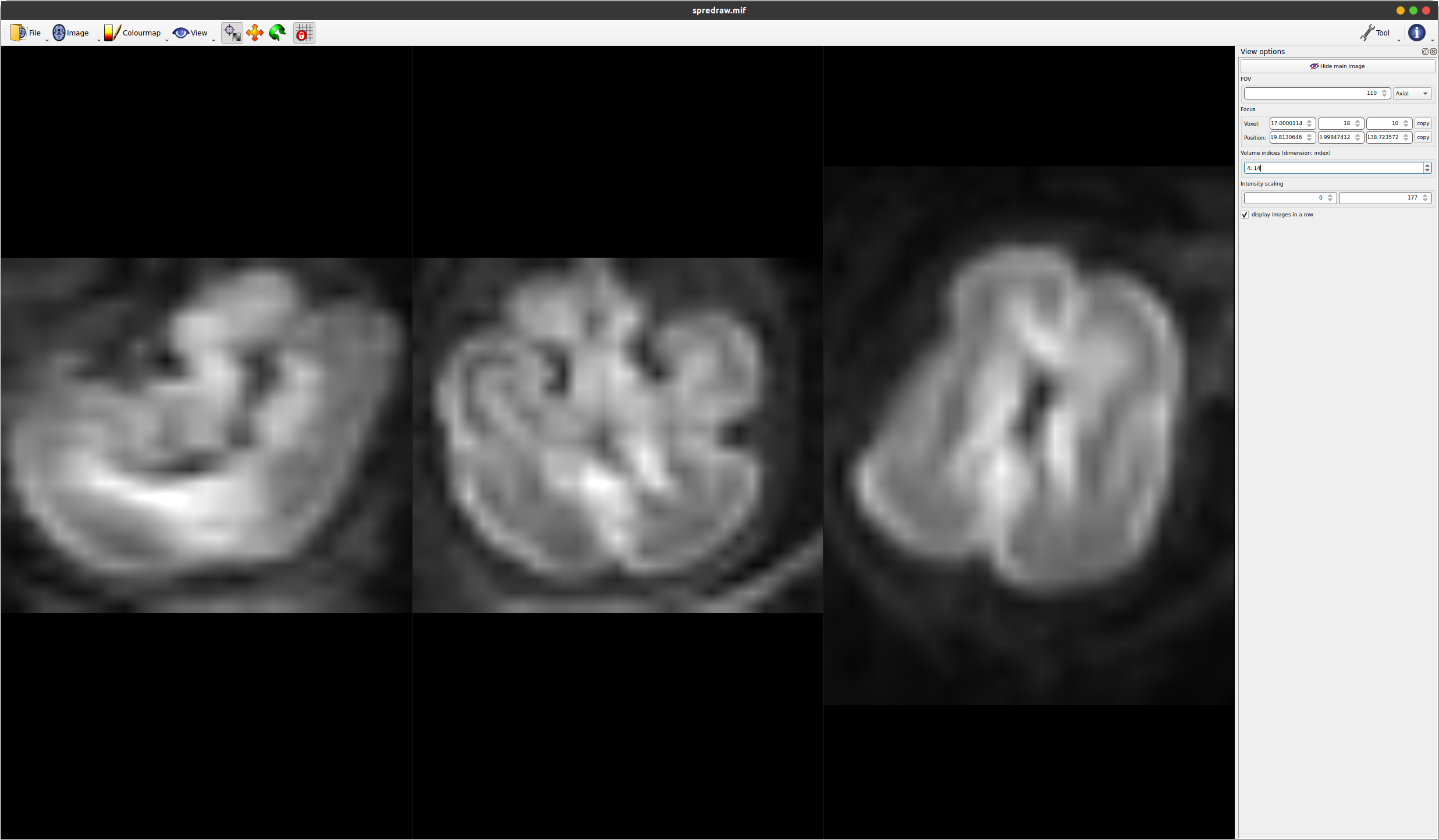}\hspace{0.1cm}
\includegraphics[width=0.235\textwidth, trim={0cm 13.0cm 14cm 13.5cm}, clip]{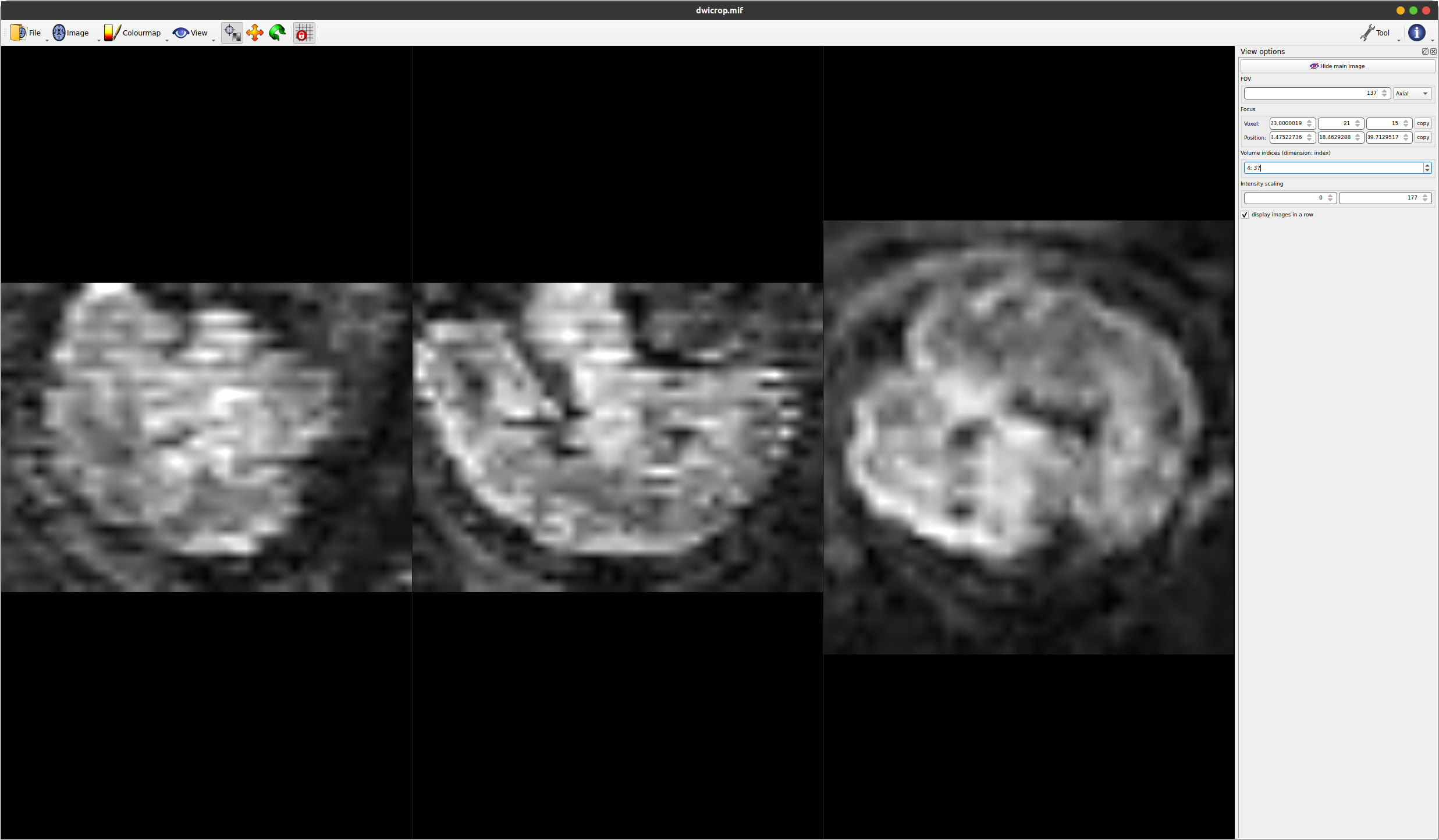}
\includegraphics[width=0.235\textwidth, trim={0cm 13.0cm 14cm 13.5cm}, clip]{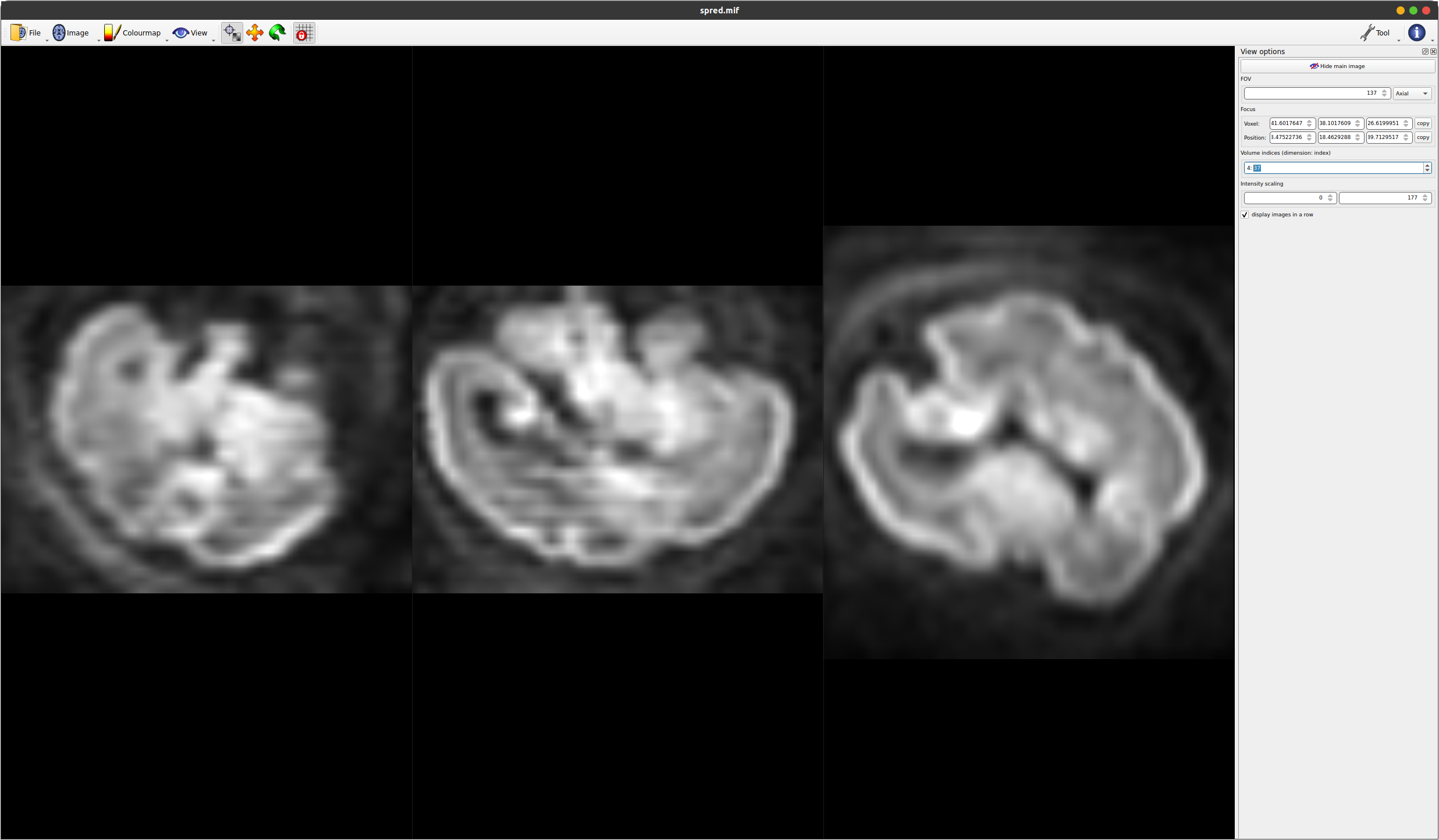}
\\
\includegraphics[width=0.235\textwidth, trim={0cm 13.0cm 14cm 13.5cm}, clip]{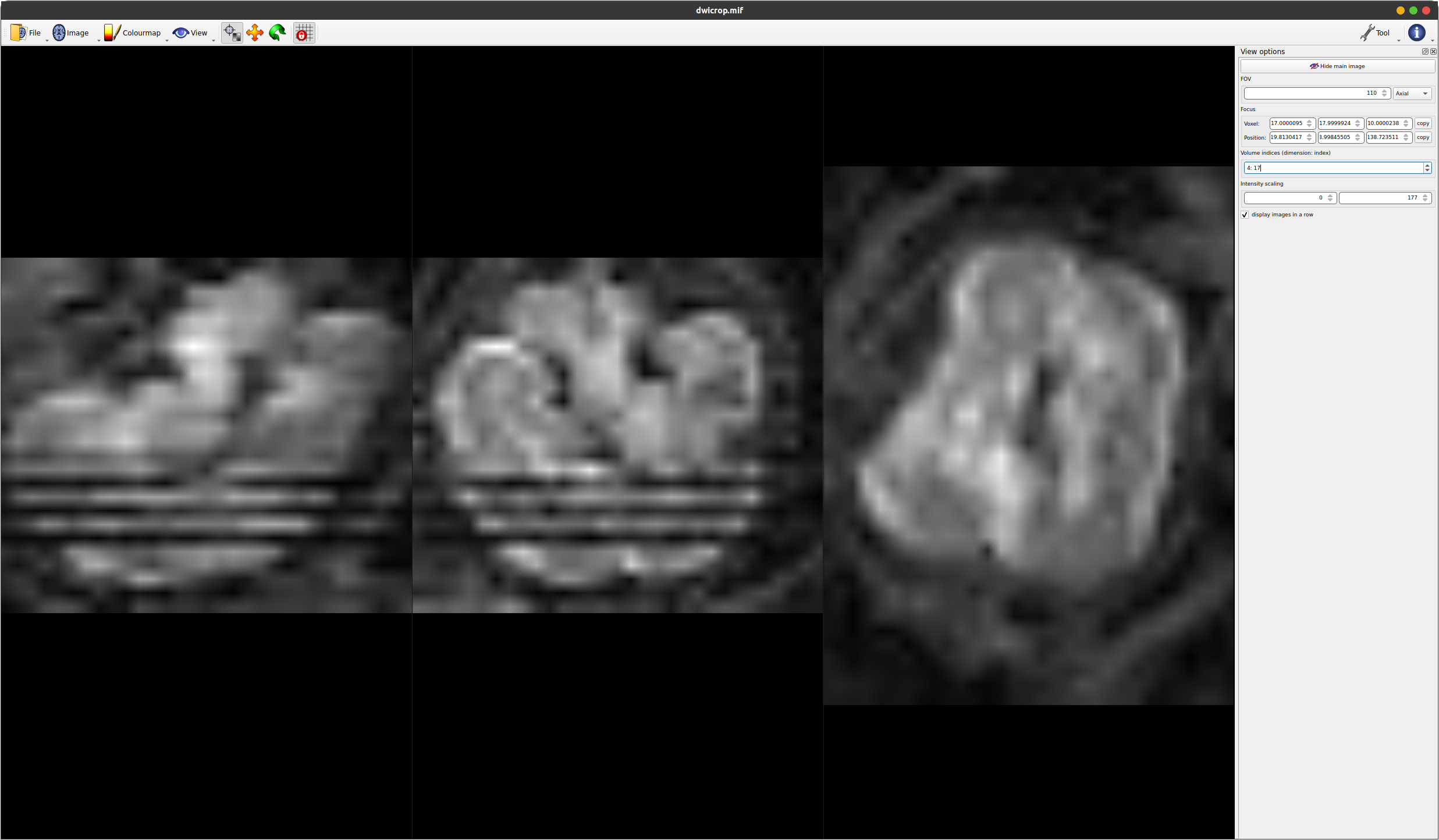}
\includegraphics[width=0.235\textwidth, trim={0cm 13.0cm 14cm 13.5cm}, clip]{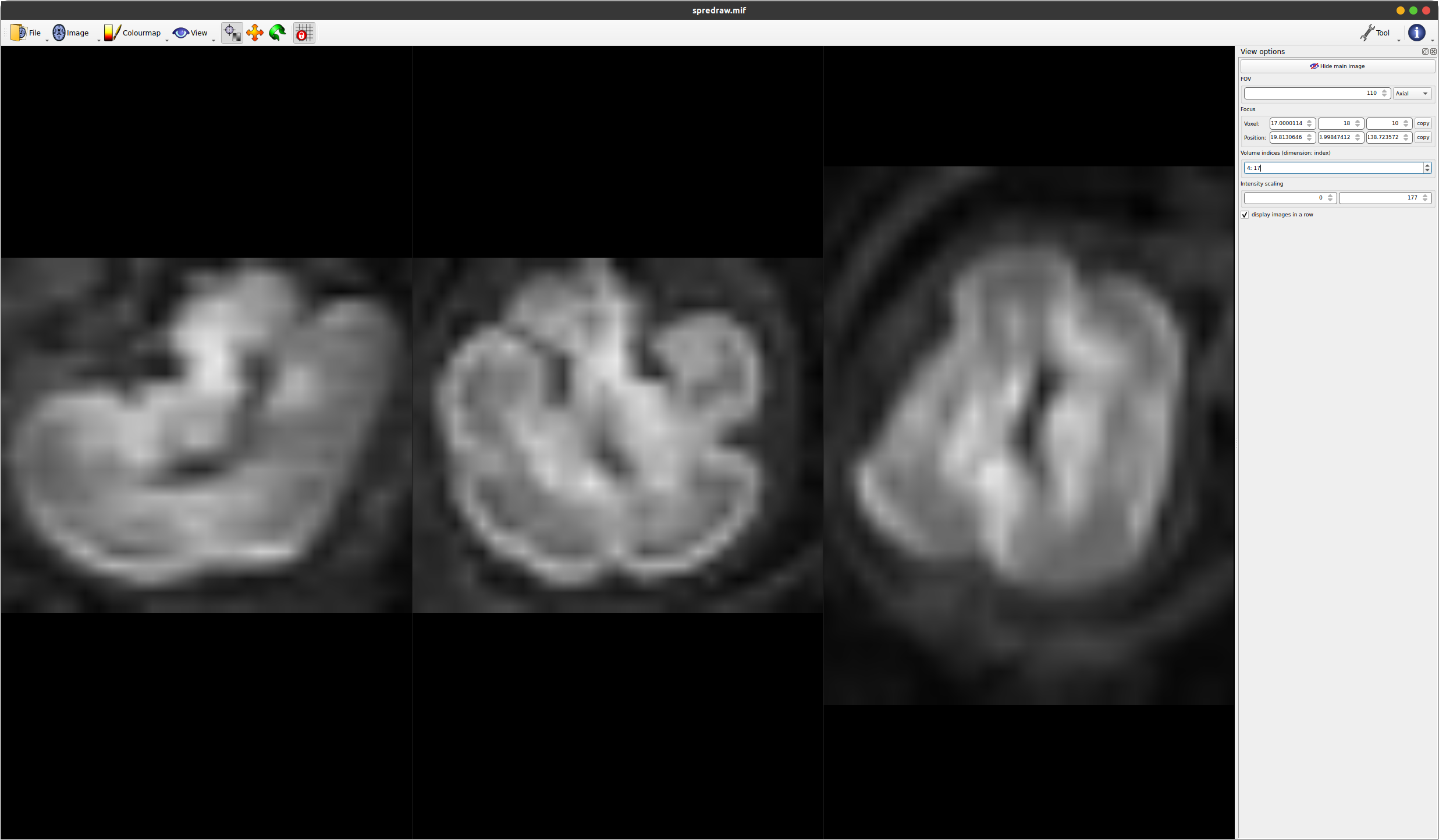}\hspace{0.1cm}
\includegraphics[width=0.235\textwidth, trim={0cm 13.0cm 14cm 13.5cm}, clip]{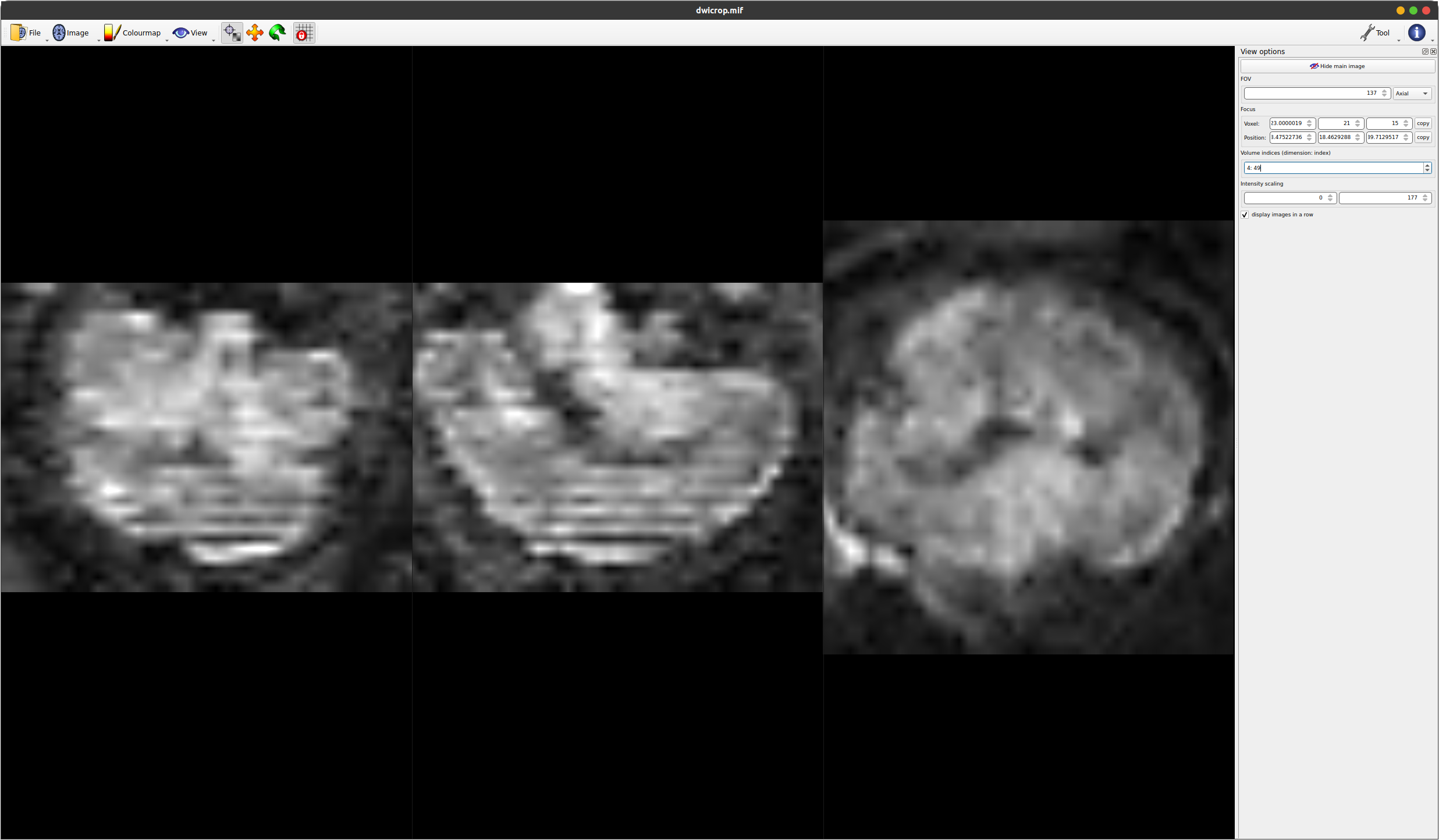}
\includegraphics[width=0.235\textwidth, trim={0cm 13.0cm 14cm 13.5cm}, clip]{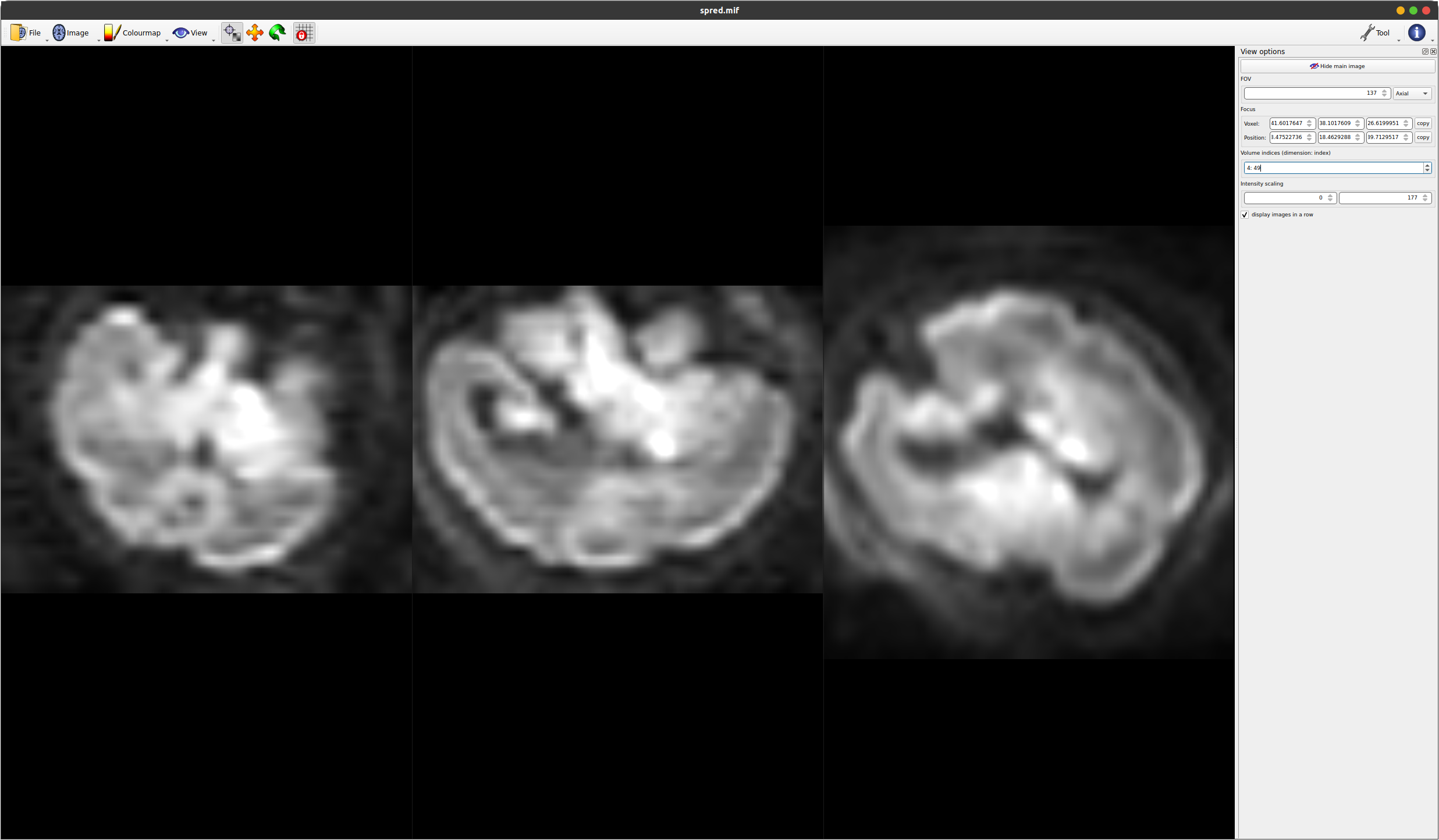}
\\
\includegraphics[width=0.235\textwidth, trim={0cm 13.0cm 14cm 13.5cm}, clip]{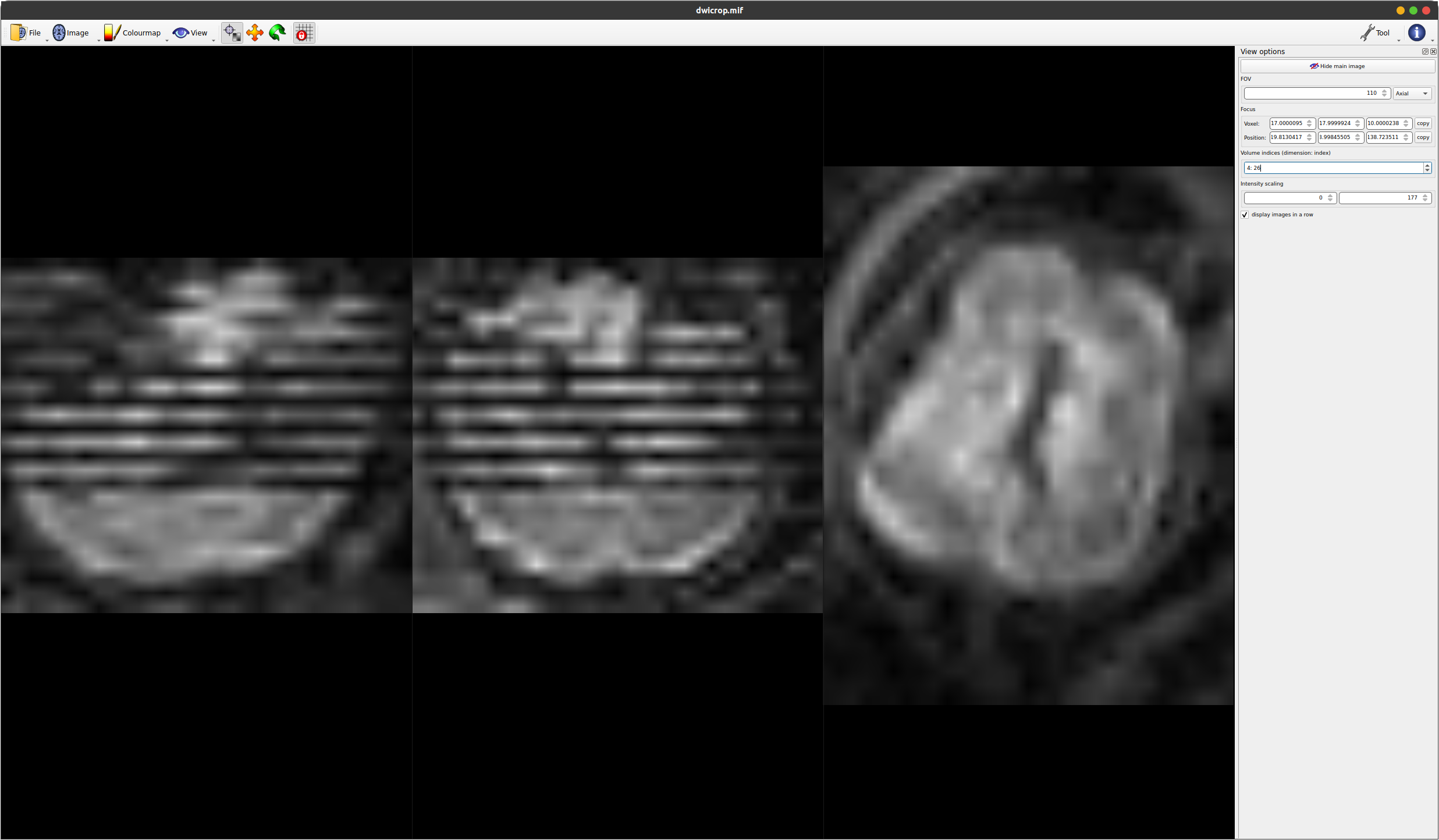}
\includegraphics[width=0.235\textwidth, trim={0cm 13.0cm 14cm 13.5cm}, clip]{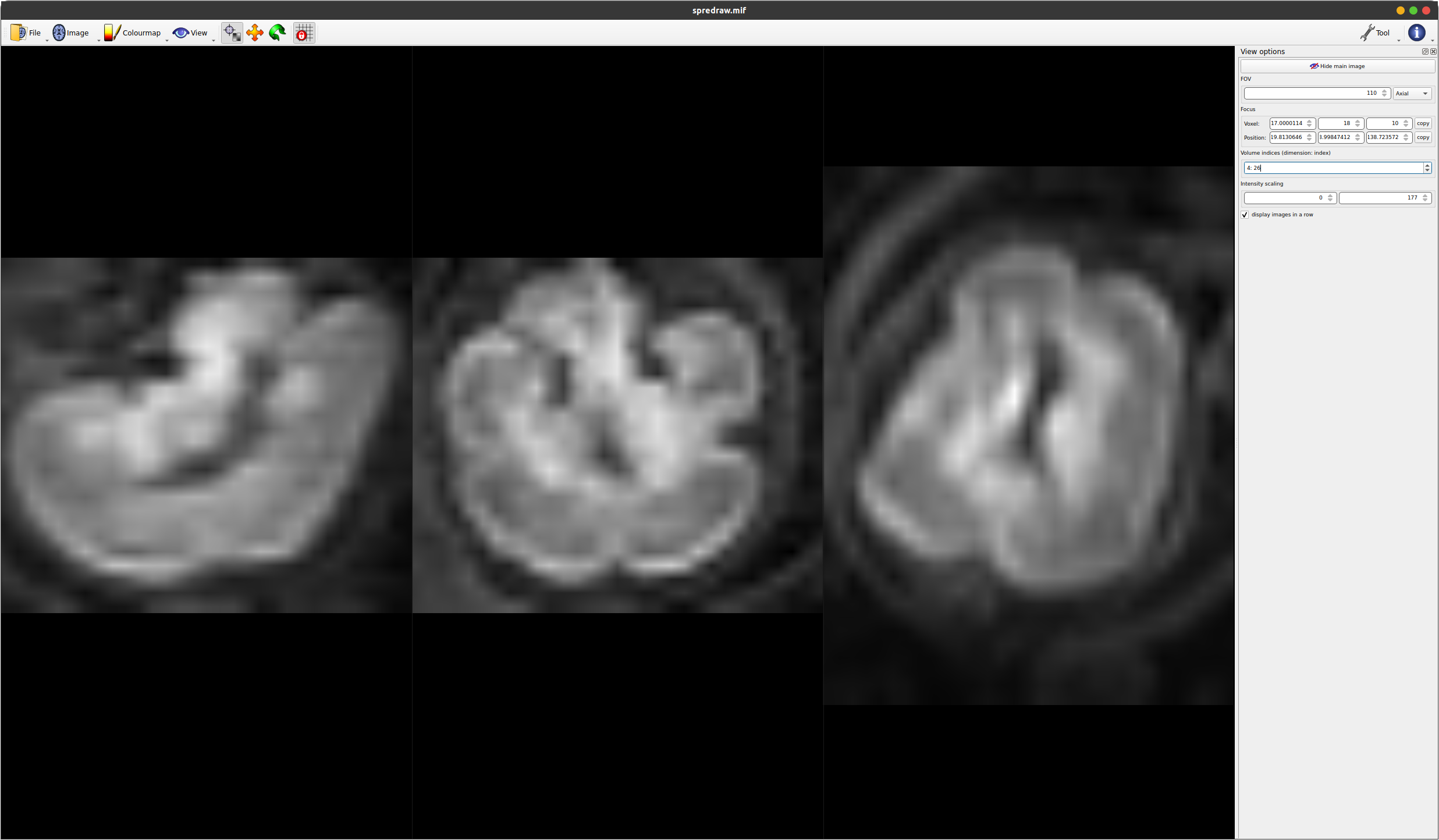}\hspace{0.1cm}
\includegraphics[width=0.235\textwidth, trim={0cm 13.0cm 14cm 13.5cm}, clip]{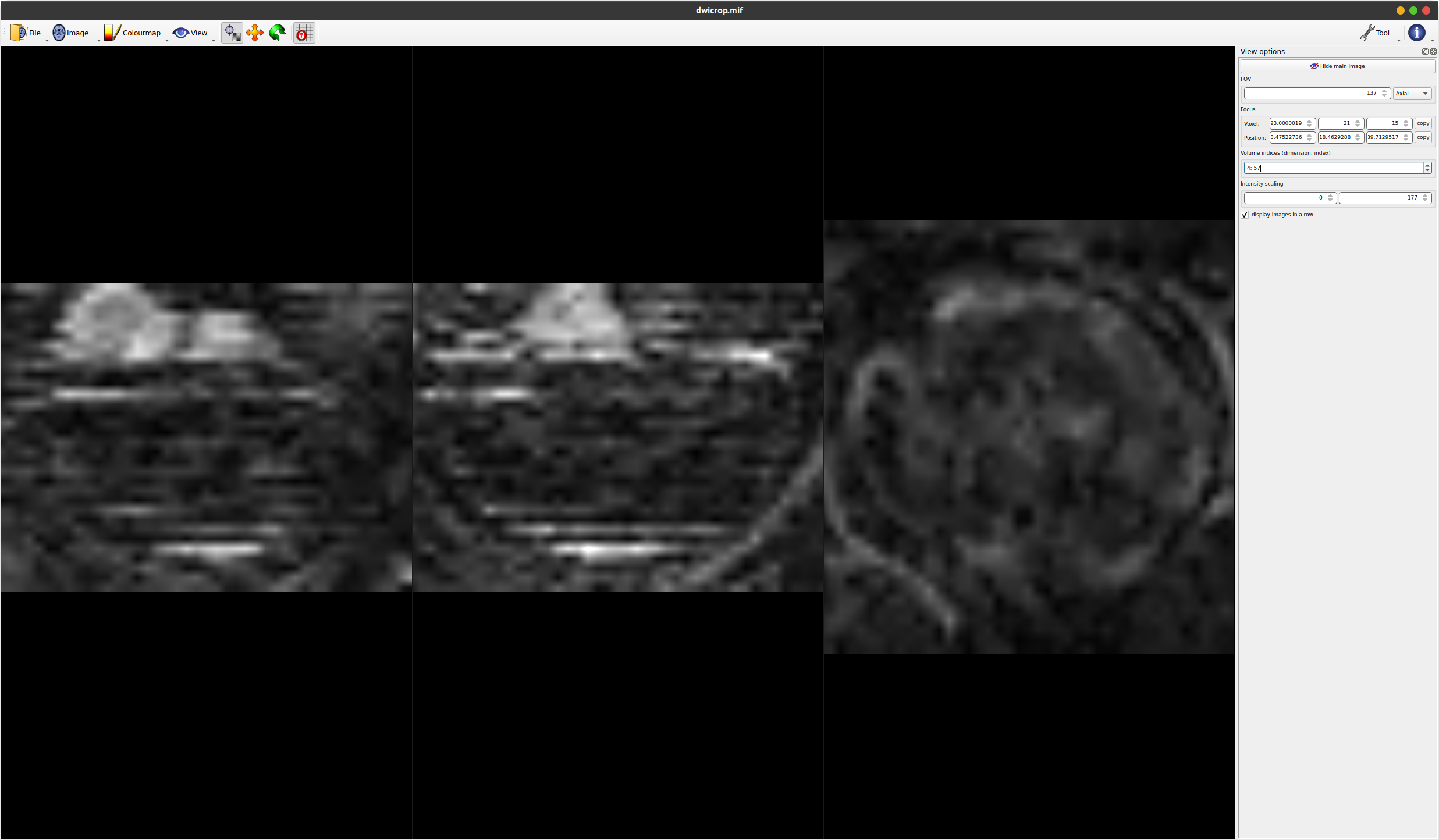}
\includegraphics[width=0.235\textwidth, trim={0cm 13.0cm 14cm 13.5cm}, clip]{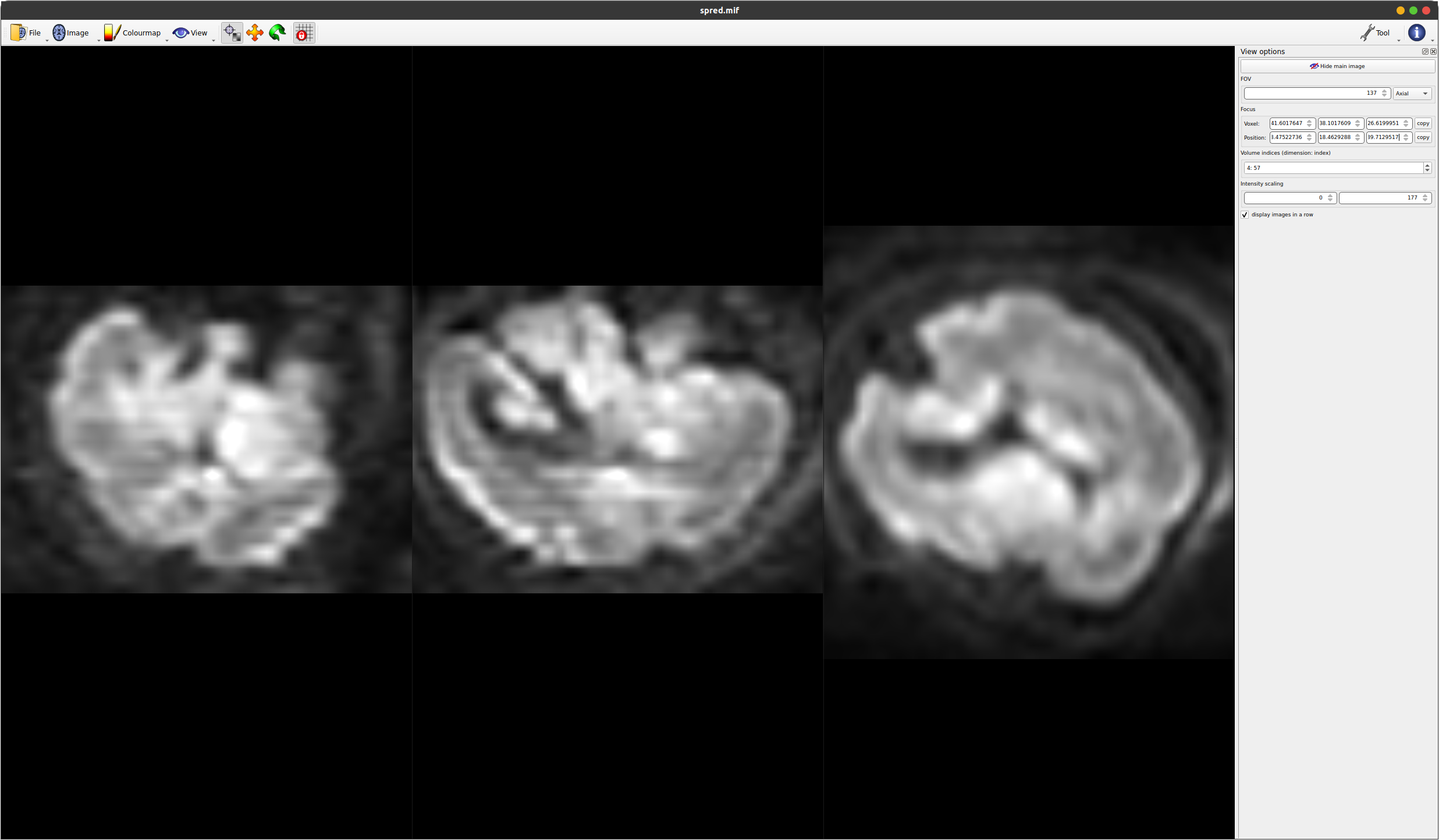}
\\
\includegraphics[width=0.235\textwidth, trim={0cm 13.0cm 14cm 13.5cm}, clip]{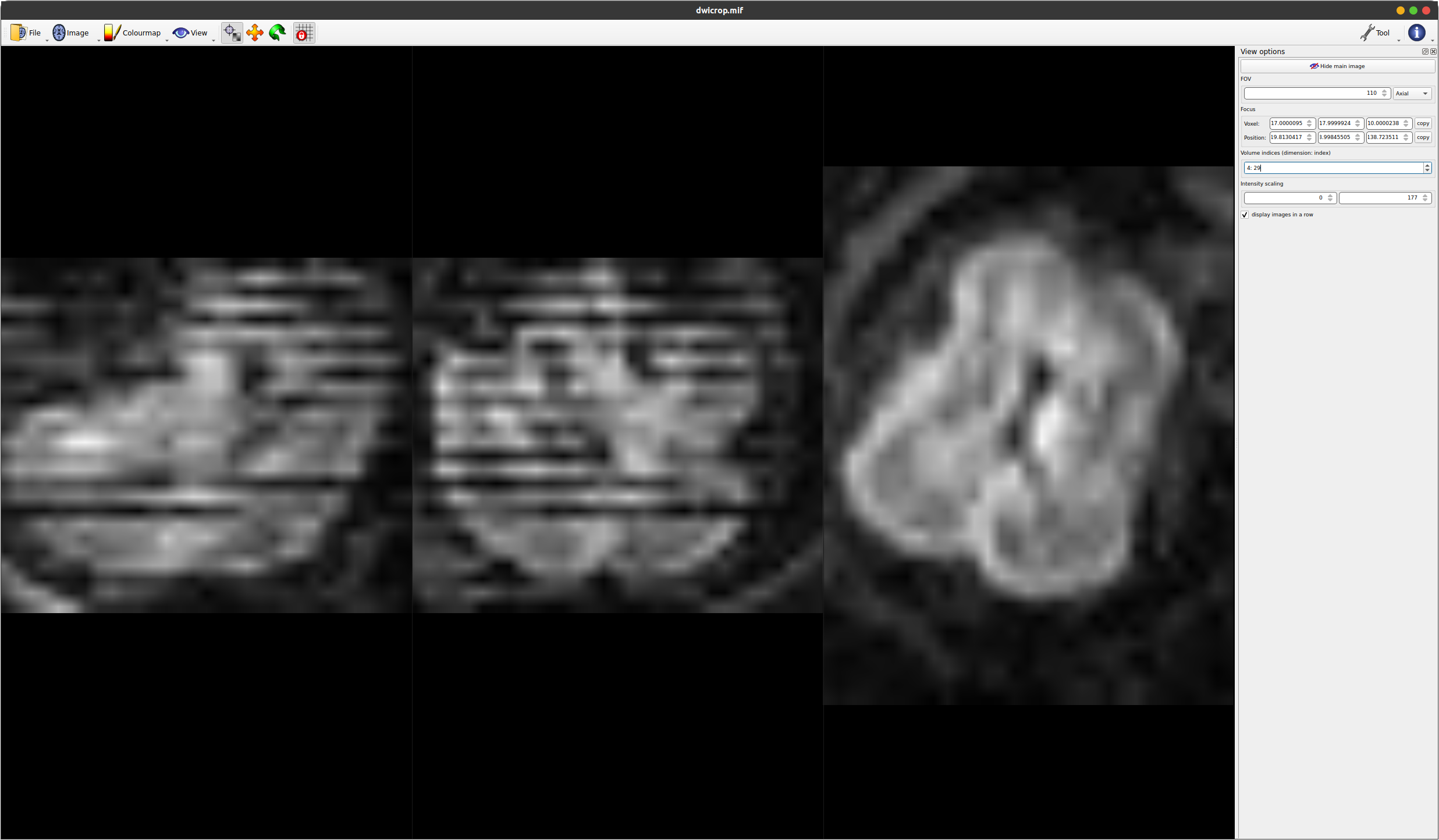}
\includegraphics[width=0.235\textwidth, trim={0cm 13.0cm 14cm 13.5cm}, clip]{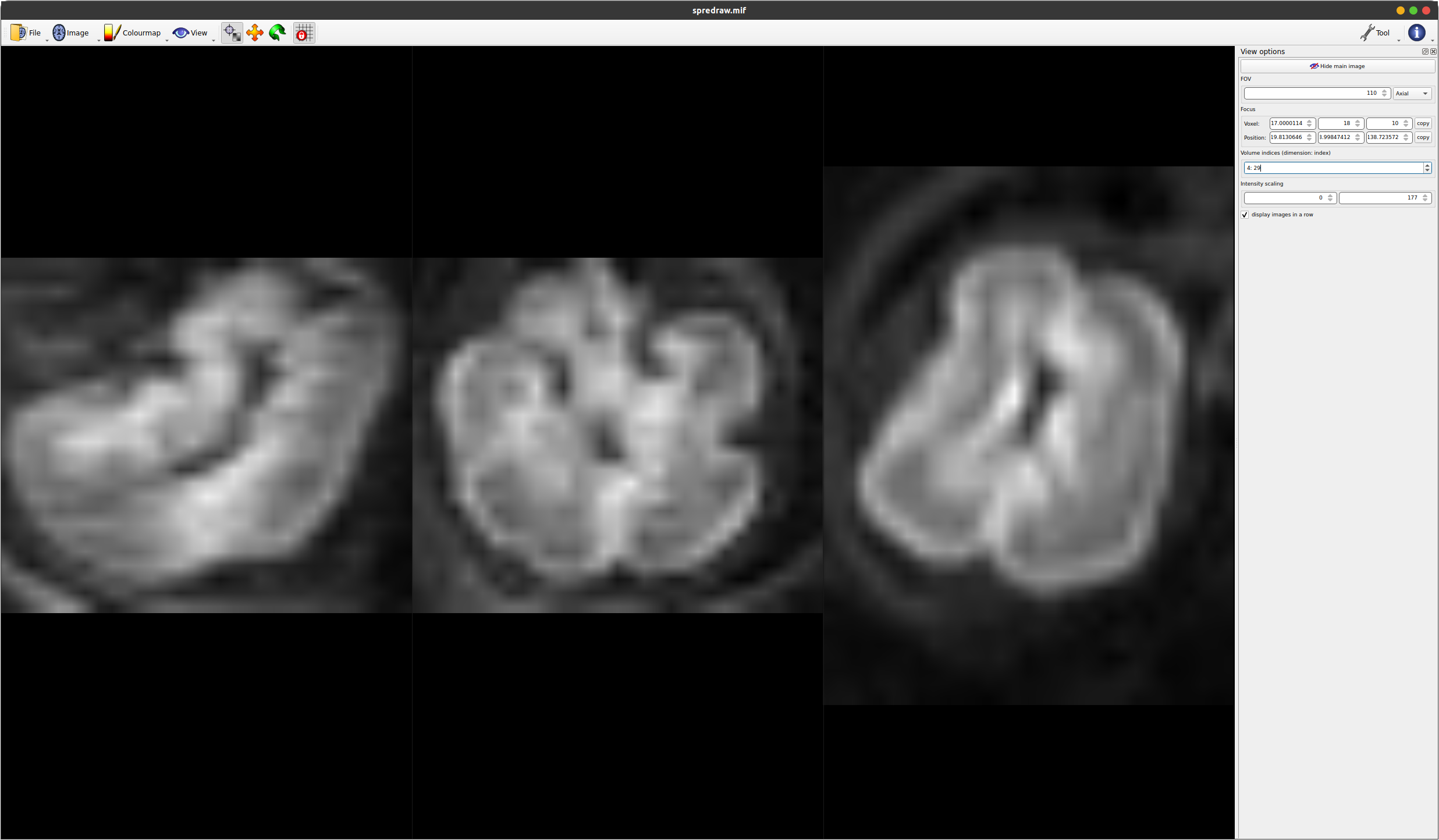}\hspace{0.1cm}
\includegraphics[width=0.235\textwidth, trim={0cm 13.0cm 14cm 13.5cm}, clip]{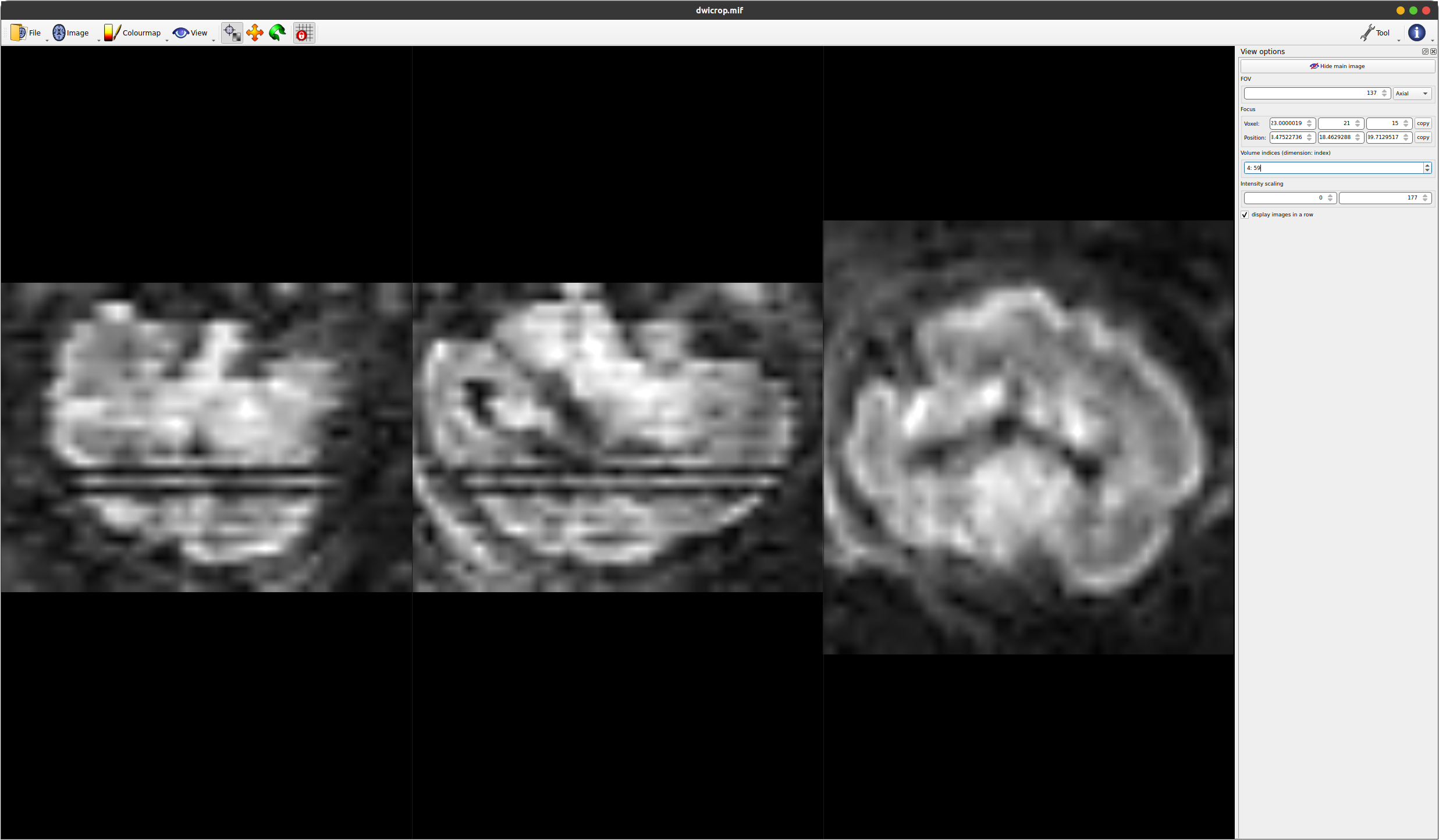}
\includegraphics[width=0.235\textwidth, trim={0cm 13.0cm 14cm 13.5cm}, clip]{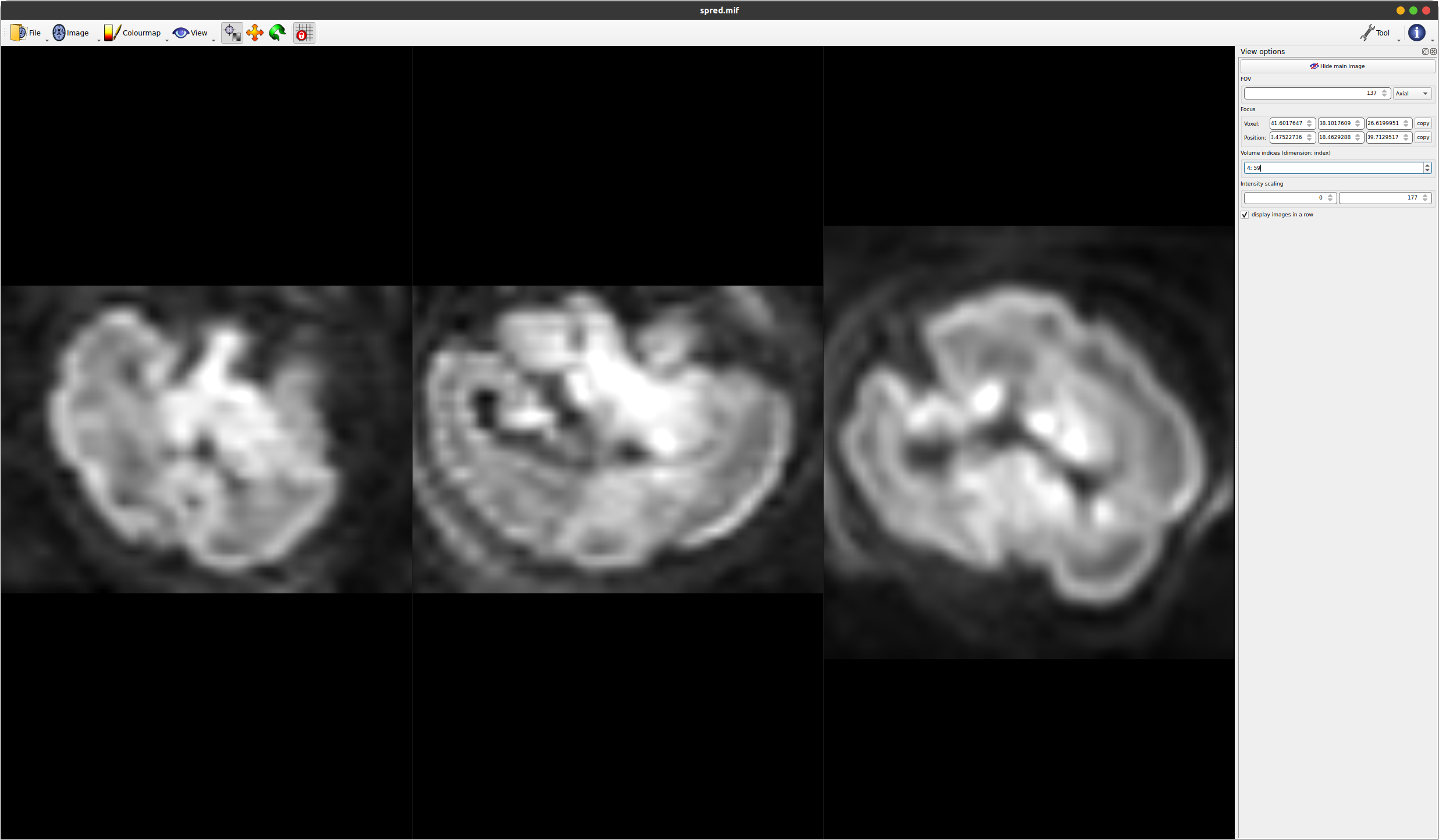}
\\
\includegraphics[width=0.235\textwidth, trim={0cm 13.0cm 14cm 13.5cm}, clip]{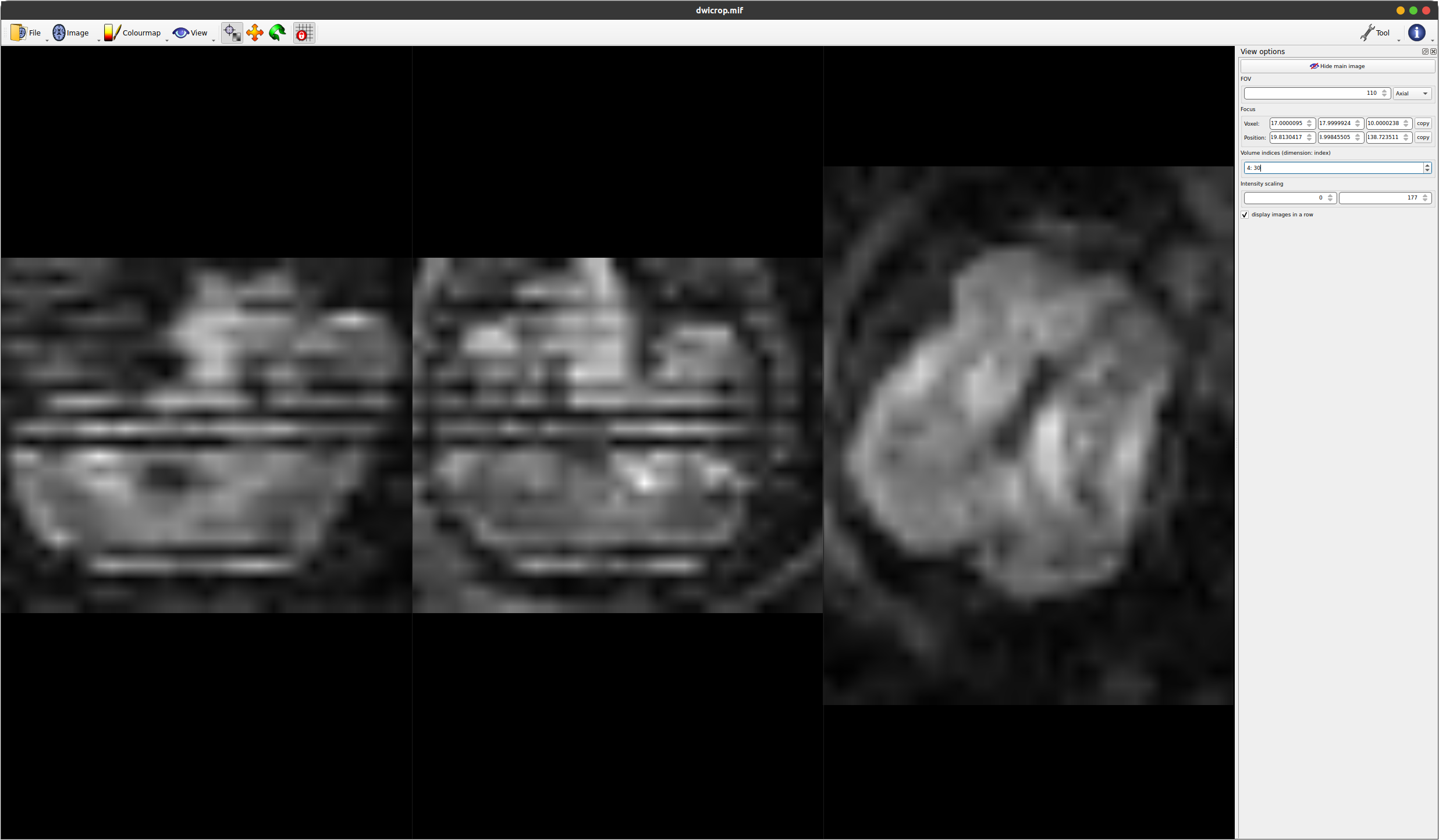}
\includegraphics[width=0.235\textwidth, trim={0cm 13.0cm 14cm 13.5cm}, clip]{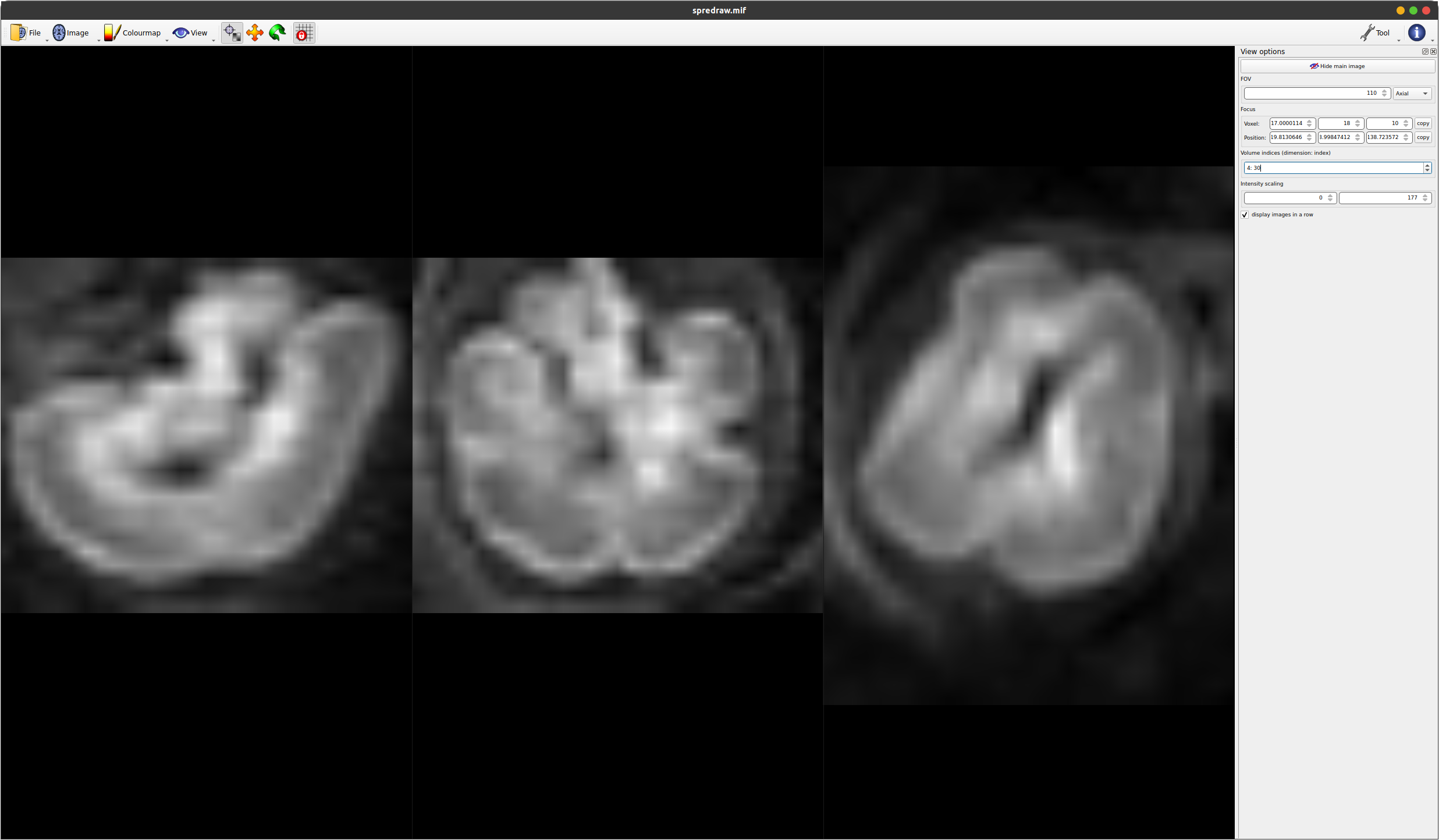}\hspace{0.1cm}
\includegraphics[width=0.235\textwidth, trim={0cm 13.0cm 14cm 13.5cm}, clip]{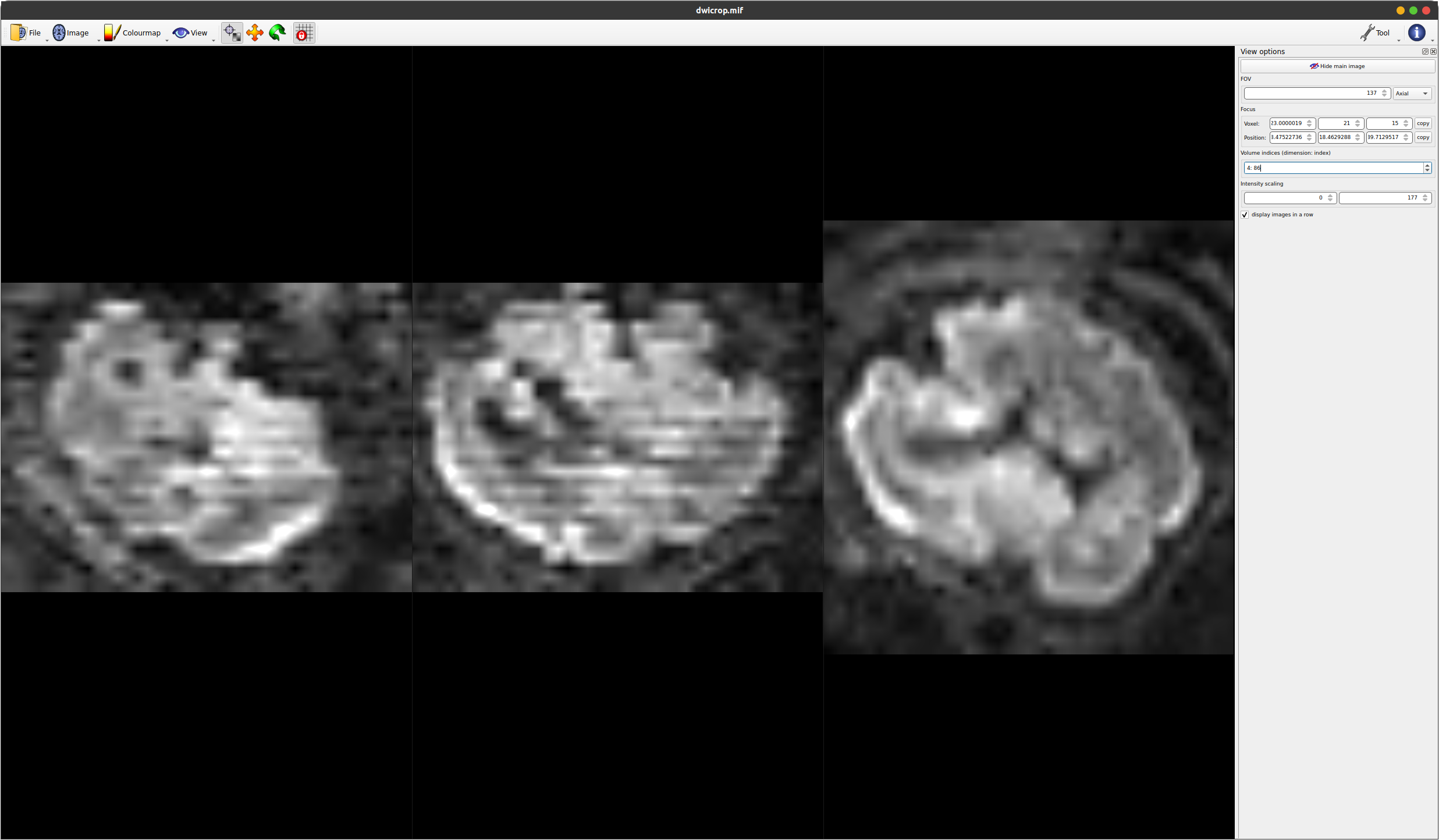}
\includegraphics[width=0.235\textwidth, trim={0cm 13.0cm 14cm 13.5cm}, clip]{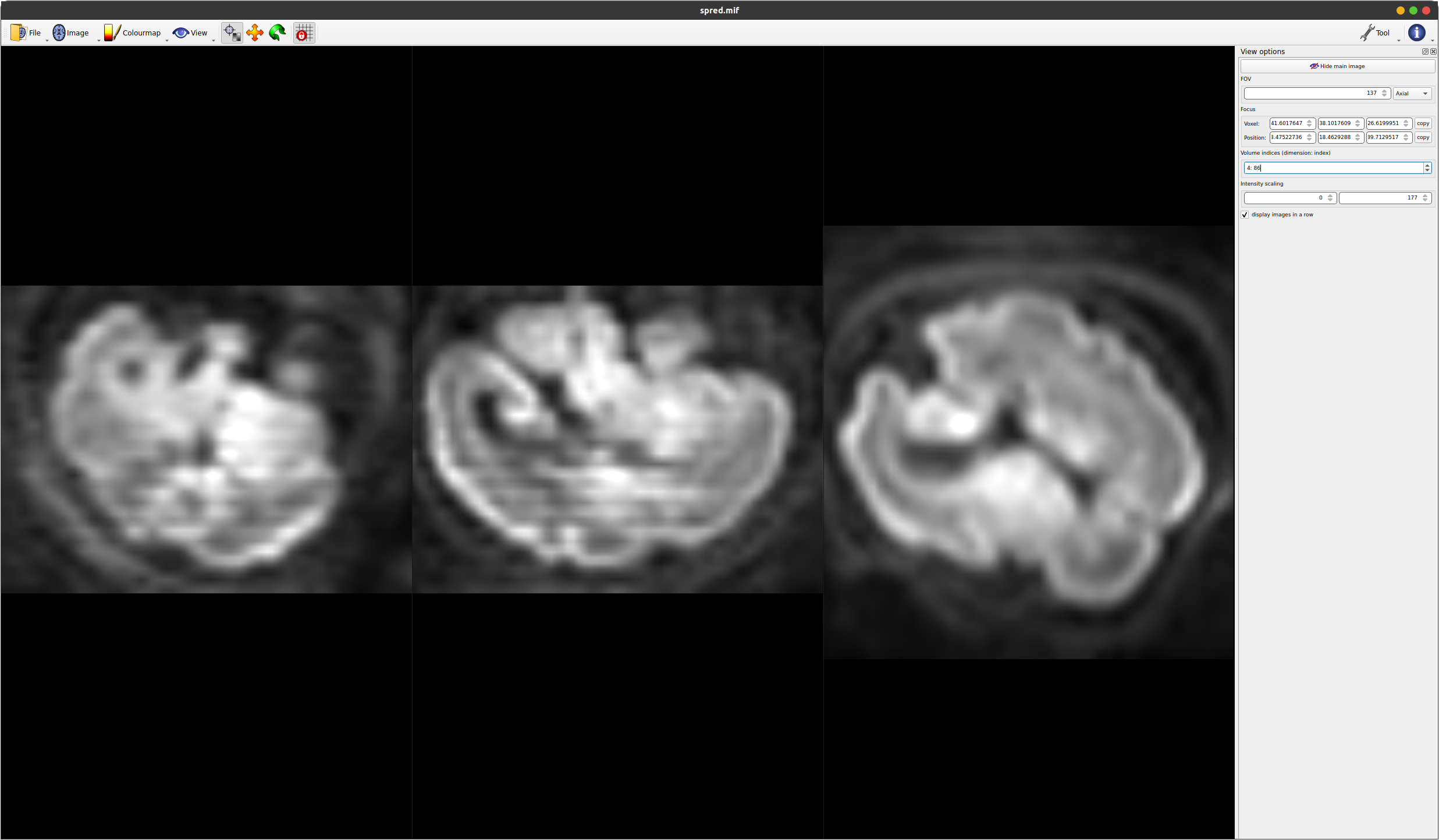}\\
\textsf{\small ~~~~~Raw Data (Subject A)~~~~~~~~~~~~~~ Corrected Data (Subject A)~~~~~~~~~Raw Data (Subject B) ~~~~~~~~~~~~Corrected Data (Subject B)~~~~~}
\vspace{0.15cm}
\caption{Example fetal dMRI scans before and after motion correction. The left two columns display axial, coronal, and sagittal views of the raw data (pre-processed up to B1 bias field correction step) (Subject A) and corresponding motion-corrected data, respectively. Each row represents a specific volume (index: 5, 8, 15, 18, 27, 30, 31 from top to bottom). The right two columns show similar data for Subject B. Each row represents a specific volume (index: 7, 37, 38, 50, 58, 60, 87 from top to bottom).}
\label{fig:mc_examples}
\end{figure*}

Fig.~\ref{fig:mc_examples} showcases the effectiveness of the motion correction stage of our framework by comparing motion-corrupted fetal dMRI data (first and third columns) with their corresponding motion-corrected reconstructions (second and fourth columns) for two subjects \textit{A} and \textit{B}. The raw data exhibits significant motion artifacts, including signal dropouts and inconsistencies in head orientation across multiple volumes. In contrast, the motion-corrected images in Fig.~\ref{fig:mc_examples} (second and fourth columns) show dramatic improvements. Signal intensity is successfully recovered, revealing clear visualization of the fetal brain anatomy across all three orthogonal views (axial, coronal, and sagittal). This visual comparison emphasizes HAITCH's capability to effectively mitigate motion artifacts, and leads to improved data quality suitable for further analysis.


\begin{figure}[!ht]
\centering
\includegraphics[width=0.5\textwidth, trim={1cm 0cm 0cm 0cm}, clip]{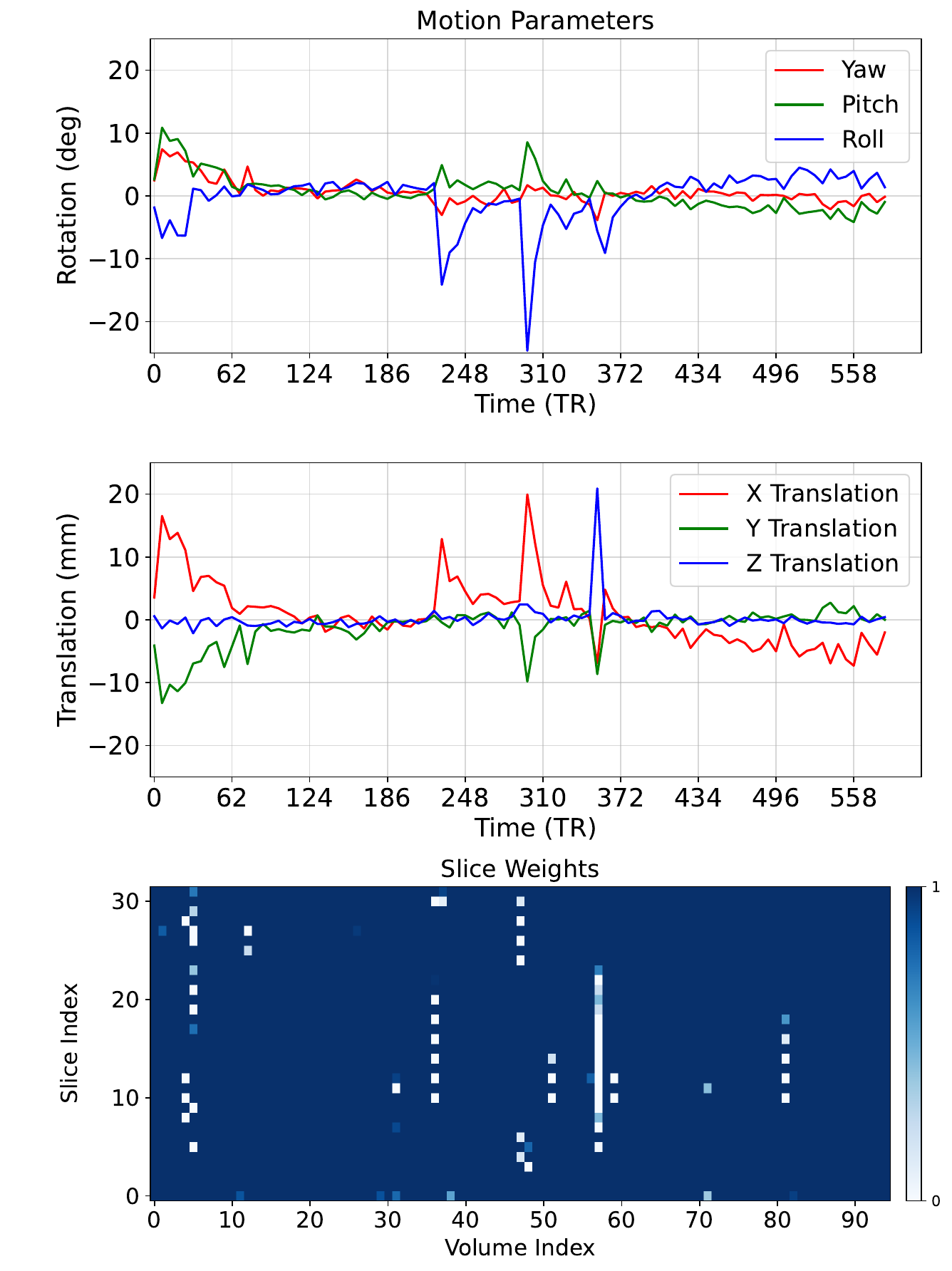}
\caption{Estimated motion parameters over time and slice weights for Subject B: The top two panels show the estimated motion parameters: translation (mm) and rotation (Euler angles). The bottom panel displays the slice weights calculated using the modified \textit{Z}-score method. Peaks in motion parameters coincide with low slice weights (outliers).}
\label{fig:mc_parmas}
\end{figure}

To gain deeper insights into the motion correction process, Fig.~\ref{fig:mc_parmas} depicts the estimated motion parameters \( \boldsymbol{\mu} \) for subject \textit{B}. The observed translations range from -12 mm to 20 mm, and rotations range from -24 to 10 degrees. The motion profile reveals a slowly varying baseline started by one extended period and punctuated by three shorter bursts of rapid motion. These transient motion events correspond to regions with low slice weights in the lower panel of Fig.~\ref{fig:mc_parmas}, as identified by Z-score.
This alignment highlights the efficacy of the HAITCH motion correction stage, as it successfully accounts for inter-volume head motion inconsistencies, corrects for intra-volume motion artifacts, and ultimately recovers signal intensity and improves slice uniformity.

\begin{figure}[ht]
\centering
\includegraphics[width=0.235\textwidth, trim={0cm 6.5cm 38.2cm 0cm}, clip]{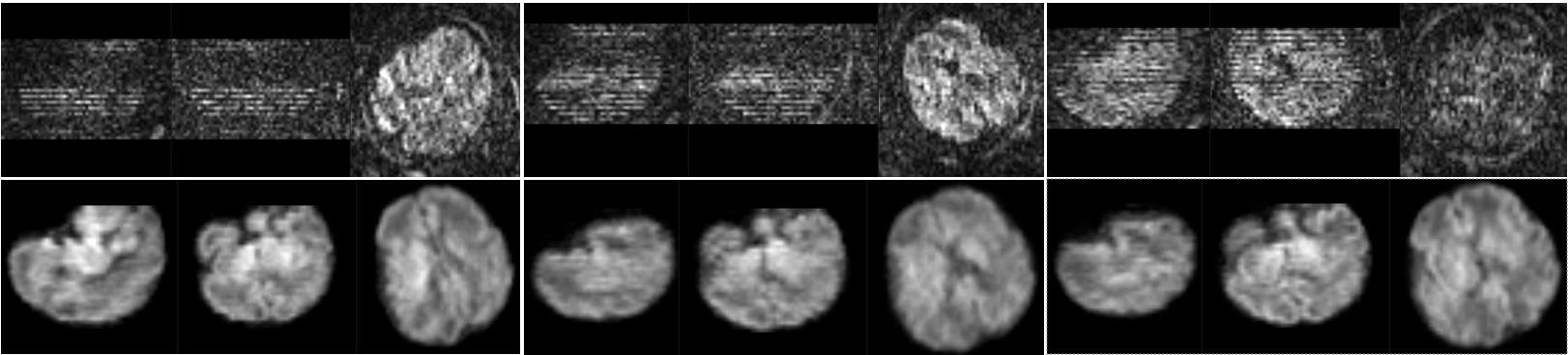}
\includegraphics[width=0.235\textwidth, trim={0cm 0cm 38.2cm 6.5cm}, clip]{images/outliers_interpolation.png}\\
\includegraphics[width=0.235\textwidth, trim={19.1cm 6.5cm 19.1cm 0cm}, clip]{images/outliers_interpolation.png}
\includegraphics[width=0.235\textwidth, trim={19.1cm 0cm 19.1cm 6.5cm}, clip]{images/outliers_interpolation.png}\\
\includegraphics[width=0.235\textwidth, trim={38.2cm 6.5cm 0cm 0cm}, clip]{images/outliers_interpolation.png}
\includegraphics[width=0.235\textwidth, trim={38.2cm 0cm 0cm 6.5cm}, clip]{images/outliers_interpolation.png}
\caption{Fetal dMRI with Motion Correction in Challenging Scenario (Subject C). The left column shows raw data from three volumes with close temporal proximity (index: 33, 46, 49 from top to bottom) with significant outliers, signal dropouts, and various brain orientations. The right column displays the corresponding masked reconstructed volumes with improved anatomical details.}
\label{fig:outliers}
\end{figure}

Fig.~\ref{fig:outliers} showcases HAITCH performance even in challenging scenarios, such as low SNR data exemplified by subject C. The figure displays three volumes with close temporal proximity and various brain orientations. Despite this, our framework successfully recovers anatomical details. The well-defined masked brain in the right panel of Fig.~\ref{fig:outliers} also demonstrates the robustness of our deep learning model in fetal brain segmentation. Note that segmentation can be performed before or after motion correction depending on the registration stage which can be sometimes better when keeping the background.

\begin{figure}[ht]
\centering
\includegraphics[width=0.45\textwidth, trim={0cm 1.4cm 0cm 0cm}, clip]{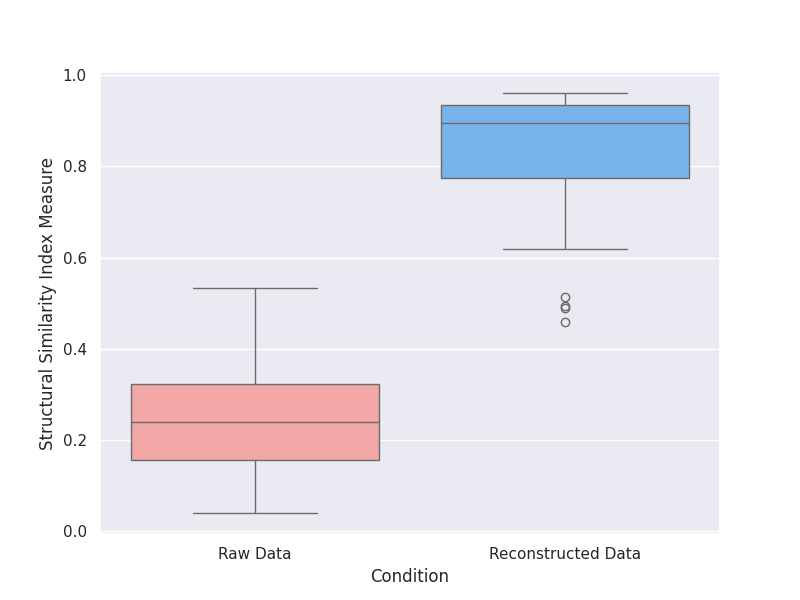}\\
\textsf{\small~~~~~~~Raw Data~~~~~~~ Corrected Data~~~~~~}
\caption{Distribution of SSIM Values for Raw and Motion-Corrected dMRI. Boxplots compare the SSIM distribution across slices of the first and last volume (b-value = 900 s/m\(^2\)) for eight subjects. Raw data (red) exhibits lower SSIM compared to motion-corrected data (blue), indicating improved data integrity after motion correction.}
\label{fig:mc_comparison}
\end{figure}

Furthermore, we employed quantitative analysis to complement the visual assessment. Fig.~\ref{fig:mc_comparison} presents the distribution of SSIM values for raw and motion-corrected data across slices between the first and last volumes (b-value = 900 s/mm²) for eight subjects. The initial and final volumes represent a significant temporal gap during the scan, making them particularly susceptible to intra-scan motion. Our analysis revealed a significant increase in SSIM values after motion correction, which indicates enhanced data integrity and reduced artifacts. This quantitative finding reinforces the visual observations of improved anatomical detail and strengthens the overall validation of the HAITCH framework.

\subsection{Anatomical Accuracy of FOD and Tractography}
\begin{figure}[ht]
\centering
\includegraphics[width=0.235\textwidth, trim={60cm 10.7cm 3cm 9.5cm}, clip]{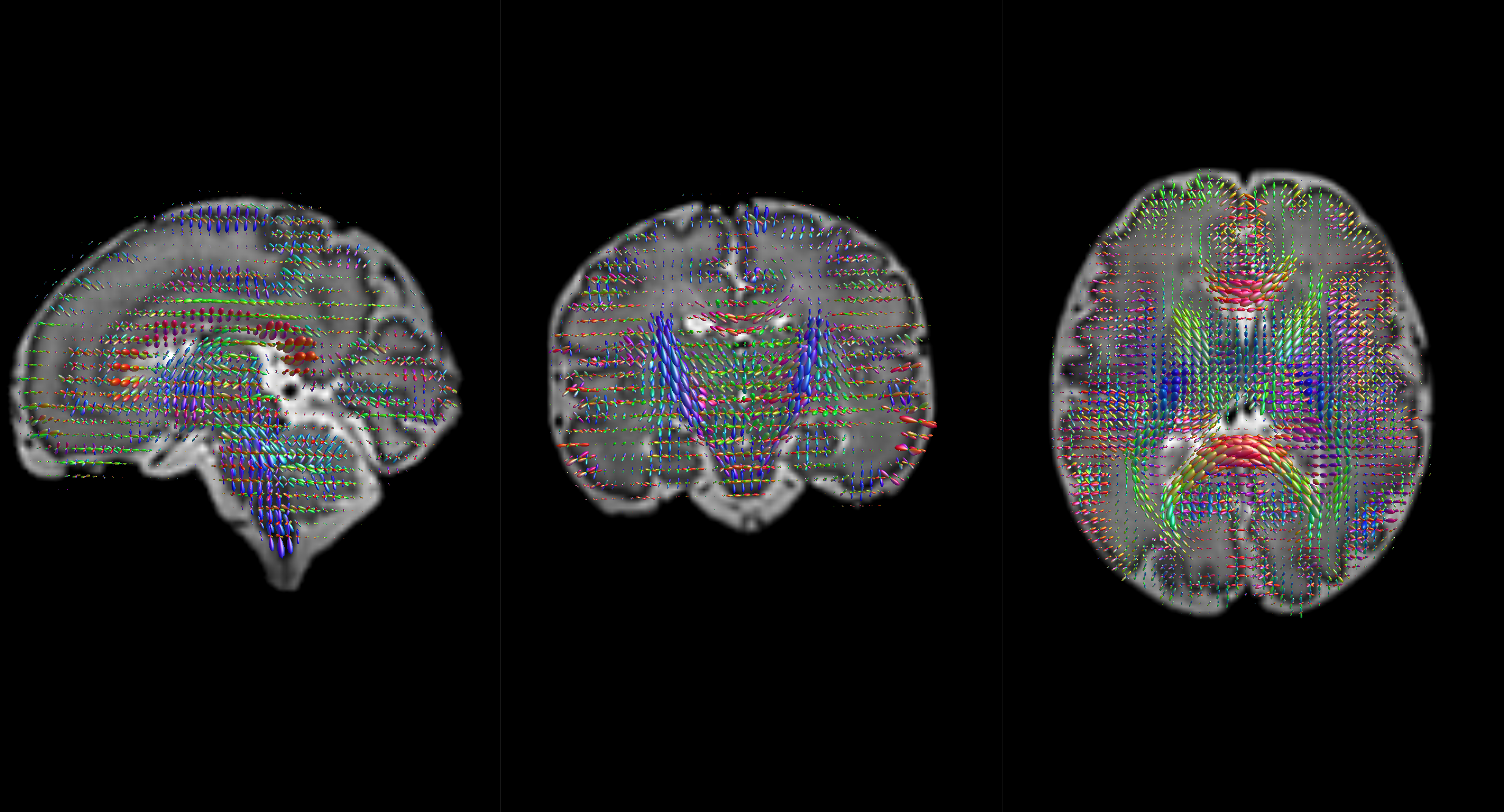}~
\includegraphics[width=0.235\textwidth, trim={60cm 10.7cm 3cm 12.25cm}, clip]{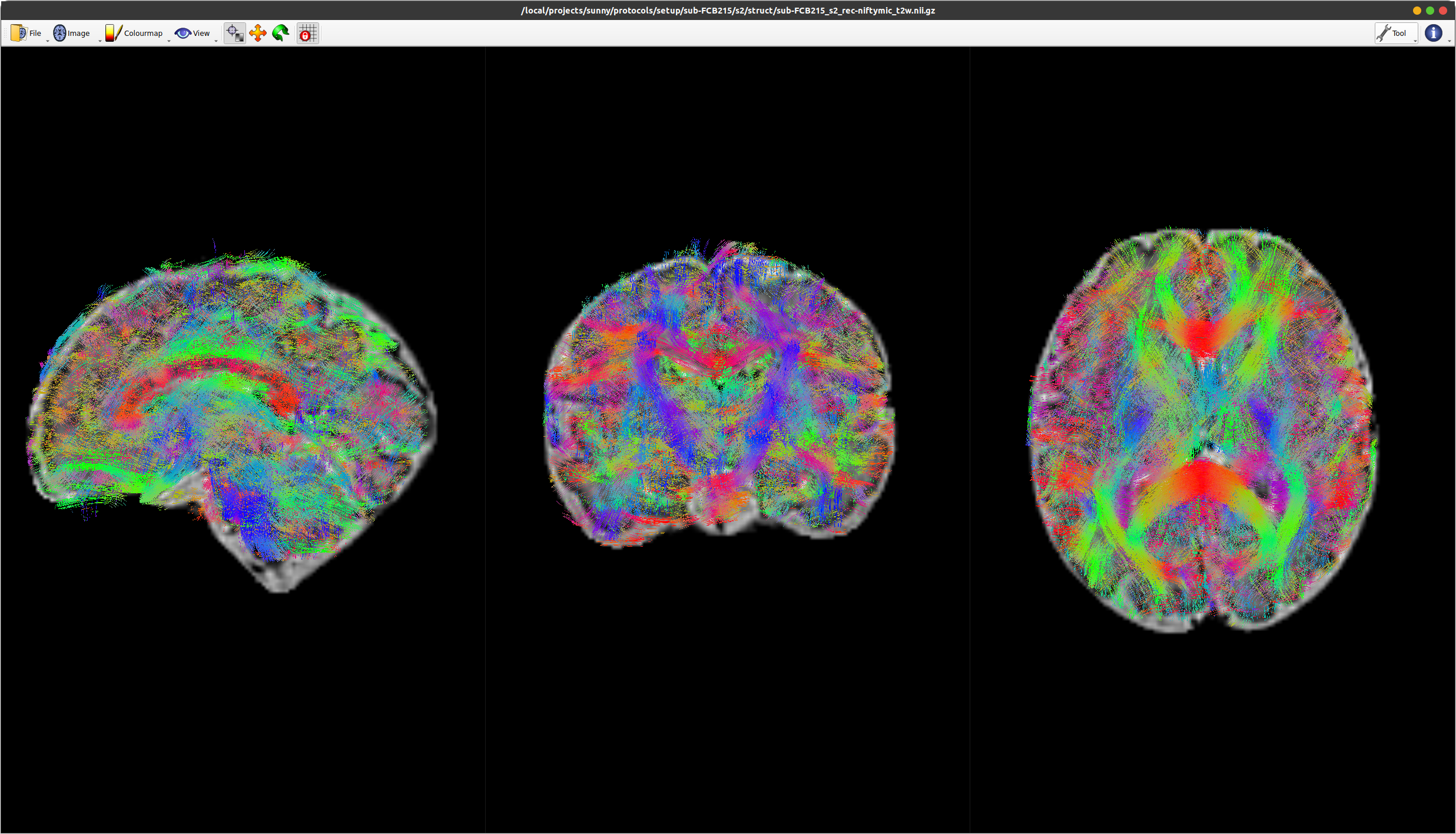}
\caption{Fetal Brain Anatomy Visualized with HAITCH Output. Left panel displays the fiber orientation distribution (FOD) overlaid on a T2-weighted image (background) of a 35-week fetus. Tractography streamlines, colored according to diffusion directions, are shown in the right panel.}
\label{fig:FOD}
\end{figure}

While visual inspection of the dMRI data itself provides valuable insights, an additional assessment of the framework's effectiveness in motion correction can be achieved by analyzing and comparing the derived white matter pathways with structural scan. Fig.~\ref{fig:FOD} demonstrates good alignment of the FODs with the developing white matter structures, as visualized by the tractography streamlines overlaid on the T2w image. This qualitative finding suggests the potential of the framework for generating reliable dMRI data suitable for studying fetal brain development.

\section{Discussion}
Fetal dMRI faces unique challenges due to inherent motion and susceptibility-induced distortions. These factors lead to scattered data in the q-space domain and result in inconsistencies within and across volumes regarding the diffusion encoding direction and the corresponding diffusion-weighted contrast. Additionally, fetal motion and maternal breathing cause significant fluctuations in the B0 inhomogeneity during scans and render static field map correction techniques ineffective. These limitations necessitate dynamic field map estimation and multi-dimensional q-space reconstruction for reliable analysis of the fetal brain microstructure. HAITCH addresses these challenges by incorporating several strategies. First, the recommended multi-shell HARDI sampling scheme in HAITCH enriches the information content of the acquired data while enhancing its tolerance to fetal motion. Second, HAITCH leverages a dual-echo EPI sequence for dynamic field map estimation. Classic static field map correction shows inaccurate results due to motion occurring between the two unweighted (b=0) scans. Third, HAITCH builds on a data-driven representation that spans both the angular and the radial q-space domain, using SH alongside generalized Laguerre polynomials. The volume-level registration proved to be effective across hundreds of processed datasets. This feature is attributed to the effectiveness of the new slice and voxel-level weighting and reconstruction steps in HAITCH to handle motion-corrupted data. 
The reconstructed fetal dMRI data obtained from HAITCH allows fitting various diffusion models, which paves the way for subsequent processing steps and enables a deeper understanding of tissue microstructure and white matter connectivity within the developing brain.

HAITCH is the first and the only publicly available tool, to-date, to process multi-shell and/or multi-echo fetal dMRI data. While validating the accuracy of HAITCH, it was not feasible to quantitatively compare it with the existing methods due to their limitations - limitations that HAITCH sought to address. The SVRTK~\cite{deprez2019higher} toolkit includes modules for fetal dMRI processing, but is restricted to single-shell data and cannot readily process multi-shell or multi-echo dMRI data. 
The Spherical Harmonics And Radial Decomposition (SHARD)~\cite{christiaens2021scattered}, which was originally developed for neonatal dMRI data, requires an interleaved phase encoding scheme, which is based on a specific sequence described in~\cite{bhushan2014improved}. SHARD utilizes an invariant B0 field map, neglecting dynamic field variations. An extended version of SHARD for fetal dMRI, explained by~\cite{cordero2018spin}, utilizes the spin and field echo (SAFE) sequence for phase-based dynamic B0 estimation following~\cite{dymerska2018method}. While promising, this method relies on MRI sequences that may not be accessible to many researchers, and involves complex phase subtraction, which can be challenging. \cite{dymerska2018method} reported limited accuracy of phase-based B0 estimation in the presence of motion exceeding 8.1 degrees, which may not be sufficient for all fetal MRI applications. \cite{hutter2018slice} proposed a slice-level diffusion encoding and interleaved double spin-echo sequence for distortion and motion correction. We did not have access to this sequence and did not re-implement it on our scanner platform. Future work warrants a detailed comparison between various sequences and their effectiveness in the presence of fetal motion and geometric and intensity distortions that vary by fetal and maternal motion.

The information gained by fetal diffusion MRI cannot be obtained by any other imaging modality. Therefore, there has been significant interest in studying the development of the fetal brain microstructure and structural connectivity using dMRI. Early \textit{ex-vivo} studies ~\cite{huang2009anatomical,vasung2010development,takahashi2012emerging,takahashi2014development,huang2014gaining,vasung2017spatiotemporal} have shed light on the capacity of dMRI to study the microstructure of the fetal brain. These studies were compared to histology and have also been used to indirectly validate \textit{in-vivo} studies, e.g.~\cite{kebiri2024deep}. However, postmortem fetal brain samples are scarce, and may not be available in sufficient numbers to study any specific disorder. \textit{In-vivo} fetal dMRI is challenging due to fetal motion, geometric distortion, and the extremely low SNR available from the small fetal brain anatomy. Despite these impediments, significant progress has been made in the field: various motion-robust fetal dMRI techniques have been proposed~\cite{oubel2012reconstruction,fogtmann2013unified,marami2017temporal,deprez2019higher}, atlases have been built and released~\cite{khan2019fetal,chen2022deciphering,uus2023multi,calixto2024advances,calixto2024detailed}, and several studies characterized normal and abnormal development of the fetal brain based on dMRI~\cite{jakab2015disrupted,jakab2019developmental,jaimes2020vivo,machado2021spatiotemporal,calixto2023characterizing,calixto2023population}. Almost all of these works, however, were based on the diffusion tensor model, which is known to have significant limitations in depicting complex fiber structures and connections. Advanced diffusion models that are needed to characterize complex brain connections and study subtle differences between groups of normal and abnormal fetuses, require multi-shell high-angular resolution dMRI, which cannot be processed reliably with any of the existing tool for fetal dMRI processing. By providing an open-source multi-stage framework for multi-shell high-angular resolution fetal dMRI, HAITCH fills in this critical gap.


\section{Conclusion}
We have developed HAITCH, a novel framework and an open-source public toolkit for acquiring and reconstructing high-quality fetal diffusion-weighted MRI. HAITCH utilizes a dual-echo EPI sequence which enables the dynamic field map estimation and reduces motion-related distortions. Our framework employs an advanced scheme that enriches the informational depth of the data and improves its motion tolerance. For motion correction and reconstruction, HAITCH leverages neighborhood directional information and a sophisticated data-driven \textit{model-free} representation in conjunction with slice weighting and registration, effectively addressing signal dropouts and scattered data. Rigorous validation experiments on fetal dMRI scans have demonstrated HAITCH's ability to significantly improve the integrity and reliability of fetal brain diffusion MRI data, paving the way for more accurate analyses of fetal brain development. The enhanced data quality facilitated by HAITCH has the potential to unlock new insights into the complex processes of brain development \textit{in-utero}.

\section*{Acknowledgment}
\noindent This research was supported in part by the National Institute of Biomedical Imaging and Bioengineering, the National Institute of Neurological Disorders and Stroke, and Eunice Kennedy Shriver National Institute of Child Health and Human Development of the National Institutes of Health (NIH) under award numbers R01NS106030, R01EB031849, R01EB032366, R01HD109395, R01HD110772, R01NS128281, and R01NS121657; in part by the Office of the Director of the NIH under award number S10OD025111; and in part by the National Science Foundation (NSF) under grant number 212306. This research was also partly supported by an award from NVIDIA Corporation and utilized NVIDIA RTX A6000 and RTX A5000 GPUs. The content of this publication is solely the responsibility of the authors and does not necessarily represent the official views of the NIH, NSF, or NVIDIA.

\subsection*{Code Availability}
The code is publicly available in \url {https://github.com/bchimagine} and \url{https://github.com/FEDIToolbox}.



\bibliographystyle{unsrt}

\end{document}